\documentclass{article}

\usepackage[T1]{fontenc}    % use 8-bit T1 fonts
\usepackage{hyperref}       % hyperlinks
\usepackage{url}            % simple URL typesetting       % professional-quality tables
\usepackage{amsfonts}       % blackboard math symbols
\usepackage{nicefrac}       % compact symbols for 1/2, etc.
\usepackage{microtype}      % microtypography        % colors
\usepackage{framed}
\usepackage{algorithm}
\usepackage{algorithmicx}
\usepackage{algpseudocode}
\usepackage[utf8]{inputenc}
\usepackage{amsmath}
\usepackage{amssymb}
\usepackage[export]{adjustbox}
\usepackage{comment}
\usepackage{setspace}
\usepackage{multirow}
\usepackage{mathtools}
\usepackage[dvipsnames]{xcolor}
\usepackage{makecell}
\usepackage{booktabs,enumitem,ragged2e}
\usepackage{graphicx}
\usepackage{subcaption}
\usepackage{longtable}
\usepackage{amsmath}
\usepackage{caption}
\usepackage{float}
\usepackage{tabularx}
\usepackage[switch]{lineno} 
\usepackage[utf8]{inputenc}
\usepackage{amsmath,amssymb,amsthm,upref,bbold}
\usepackage{float}
\usepackage{graphicx}
\usepackage{subcaption}
\usepackage[a4paper, total={6in, 10in}]{geometry}
\usepackage{lscape}
\usepackage[utf8]{inputenc}
\usepackage[english]{babel}

\usepackage{verbatim}
\usepackage{glossaries}
\usepackage{pgf, import, layouts}
\usepackage{tikz}
\usepackage{tikz-cd}
\usepackage{pgfplots}
\usepackage[capitalize]{cleveref}
\usepackage{algorithm}
\usepackage{multirow}
\usepackage{xcolor}
\usepackage{wrapfig}

\usepackage{url}

\usepackage{authblk} % For author and affiliation 

\DeclareMathOperator*{\argmax}{argmax}

\usepackage[round, sort, authoryear]{natbib}
\usepackage{bibunits}

\defaultbibliography{main}  % or whatever your .bib filename is

\DeclareMathOperator*{\E}{\mathbb{E}}
\usepackage{fancyhdr}

\usepackage{xcolor}

\title{Mitigating Metropolitan Carbon Emissions with Dynamic Eco-driving at Scale}

\author[1]{Vindula Jayawardana}
\author[2]{Baptiste Freydt}
\author[1]{Ao Qu}
\author[1]{Cameron Hickert}
\author[1]{Edgar Sanchez}
\author[1]{Catherine Tang}
\author[3]{Mark Taylor}
\author[3]{Blaine Leonard}
\author[1]{Cathy Wu}

\affil[1]{Massachusetts Institute of Technology \protect\\ \{vindula, qua, chickert, edgarrs, cattang, cathywu\}@mit.edu}
\affil[2]{ETH Zürich \protect\\ bfreydt@student.ethz.ch}
\affil[3]{Utah Department of Transportation \protect\\ \{marktaylor, bleonard\}@utah.gov}

\date{}
\begin{document}

\newcommand{\vj}[1]{{\color{blue} {[VJ: #1]}}}

\maketitle


%Ride Sharing

\section{Abstract}

The sheer scale and diversity of transportation make it a formidable sector to decarbonize. Here, we consider an emerging opportunity to reduce carbon emissions: the growing adoption of semi-autonomous vehicles, which can be programmed to mitigate stop-and-go traffic through intelligent speed commands and, thus, reduce emissions. But would such \textit{dynamic eco-driving} move the needle on climate change? A comprehensive impact analysis has been out of reach due to the vast array of traffic scenarios and the complexity of vehicle emissions. Such an analysis would require careful modeling of many traffic scenarios and solving an eco-driving problem at each one of them - a challenge that has been out of reach for previous studies. We address this challenge with large-scale scenario modeling efforts and by using multi-task deep reinforcement learning with a carefully designed network decomposition strategy. We perform an in-depth prospective impact assessment of dynamic eco-driving at 6,011 signalized intersections across three major US metropolitan cities, simulating a million traffic scenarios. Overall, we find that vehicle trajectories optimized for emissions can cut city-wide intersection carbon emissions by 11-22\%, without harming throughput or safety, and with reasonable assumptions, equivalent to the national emissions of Israel and Nigeria, respectively. We find that 10\% eco-driving adoption yields 25\%-50\% of total reduction, and nearly 70\% of the benefits come from 20\% of intersections, suggesting near-term implementation pathways. However, the composition of this high-impact subset of intersections varies considerably across different adoption levels, with minimal overlap, calling for careful strategic planning for eco-driving deployments. Moreover, the impact of eco-driving, when considered jointly with projections of vehicle electrification, hybrid vehicle adoption, and travel growth, remains significant. More broadly, this work paves the way for large-scale analysis of traffic externalities, such as time, safety, and air quality, and the potential impact of solution strategies. 
Visual details can be found on the project page \href{https://vindulamj.github.io/eco-drive}{https://vindulamj.github.io/eco-drive}.

\section{Introduction}
\label{introduction}

\begin{bibunit}[plainnat]

\begin{figure*}[!h]
\centering
  \includegraphics[width=0.7\textwidth]{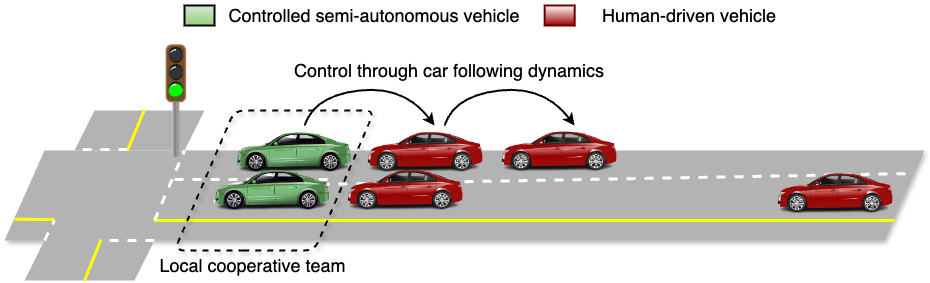}
\caption{A schematic illustration of eco-driving at signalized intersections. Controlled vehicles are driven in a way that reduces their carbon emissions. Through car following dynamics and implicitly forming local cooperative teams, controlled vehicle behaviors can affect human-driven vehicle behaviors, leading to a reduction in carbon emissions in human-driven vehicles as well. }
\label{fig:eco-driving-example}
\end{figure*}

For interrupted traffic flow dynamics, dynamic eco-driving has attracted significant attention in recent decades~\citep{Mintsis2020survey}. Dynamic eco-driving is concerned with reducing carbon emissions and/or fuel consumption of vehicles individually and at the fleet level by strategically modifying driving patterns and behaviors. (Figure~\ref{fig:eco-driving-example}). Owing to advances in vehicle-to-infrastructure (V2I) communication technologies and the standardization of the signal phase and timing (SPaT) messages, these technologies have facilitated the research of dynamic eco-driving, especially near signalized intersections, where traffic flow is persistently interrupted~\citep{Mintsis2020survey}. Based on prior studies, we crudely estimate that persistent yet unproductive idling and excess accelerations at signalized intersections could contribute up to 14.6\% of land transportation CO$_2$ emissions in the United States (US) (refer to Appendix \ref{back-of-the-envelop}). Highlighting its significance, this is comparable to half of the well-scrutinized US airline sector emissions \citep{lee2009aviation}. Reducing emissions at signalized intersections thus emerges as a promising avenue for achieving climate change mitigation goals.

However, despite nearly four decades of research, large-scale implementation of dynamic eco-driving is yet to be seen. A potential reason and a notable observation is that existing studies vary widely in assumptions and, thus, reported outcomes. In particular, previous studies report a significant variation in emission benefits ranging from 2\% to 56\%~\citep{Mintsis2020survey}. Furthermore, our observation is that a majority of studies are small-scale (e.g., one or two intersections or a few intersections in a route) and mainly focus on technical questions that concern \textit{how to eco-drive}~\citep{yang2016eco, yang2020eco, sun2020optimal}. But the fundamental socio-technical question of \textit{whether to eco-drive}, which requires going beyond a few intersections-based analyses, remains unanswered~\citep{barkenbus2010eco}. Our aim of this work is to contribute a large-scale prospective analysis of eco-driving at signalized intersections, highlighting city-scale emission benefits and uncovering insights that may influence widespread adoption—factors that are often elusive in smaller-scale studies. Such an analysis provides a scientific basis for informed decision-making, strategic planning of infrastructure, and technology development~\citep{barkenbus2010eco}.

However, conducting a large-scale analysis of eco-driving presents significant challenges, a complexity that has largely eluded in previous studies. Diverse eco-driving scenarios stemming from factors such as intersection topologies, vehicle-related factors (e.g., fuel type, age, model, etc.), weather conditions, and seasonal changes must be considered to ensure that the prospective eco-driving impact assessments are representative~\citep{falkenauer1998method, jayawardana2022impact}. It not only ensures that insights are generalizable but also aligns with industry best practices for impact assessments~\citep{ieee-standards, ROSS201747, DUINKER2007206}. While field studies are considered a gold standard, as they reflect real-world conditions, it is prohibitively expensive to conduct traffic field tests at scale~\citep{gloudemans202324}. Simulation-based analyses offer greater scalability, and uncertainties stemming from the sim-to-real gap can be addressed~\citep{vinitsky2022smoothing, jang2019simulation}. 

Simulation-based analysis consists of three key steps. First, each scenario is modeled within a high-fidelity agent-based traffic simulator, replicating real-world intersections and calibrating for real-world conditions. This step is termed \textit{scenario modeling}. Second, a set of eco-driving strategies is devised for each scenario to control the vehicles based on a given eco-driving adoption level, referred to as \textit{eco-driving control}. Lastly, the impact of eco-driving using the devised strategies on the modeled scenarios is assessed and compared against naturalistic human-like driving as a baseline. It is termed \textit{impact assessment}. 

We identified 33 influential factors for eco-driving based on previous studies and expert opinion. An eco-driving scenario manifests as a combination of these factor valuations. Each factor can take multiple valuations, creating a vast scenario space of at least $2^{33}$ scenarios. A comprehensive eco-driving analysis thus suffers from the \textit{curse of dimensionality}~\citep{donoho2000high}. The high-dimensional scenario space grows exponentially with each additional factor, amplifying computational complexity in scenario modeling and eco-driving control. Furthermore, in eco-driving control, each scenario involves solving an optimal eco-driving control problem, making a comprehensive large-scale analysis challenging due to the need to solve a vast number of them—a scale far out of reach of previous studies.

This article addresses these challenges to comprehensively analyze eco-driving at signalized intersections at scale. We consider semi-autonomous vehicles, such as vehicles equipped with advanced driver assistance systems (ADAS), which we can be leveraged to intelligently manage stop-and-go traffic, effectively reducing carbon emissions. To navigate the complexity of this undertaking, we take a data-driven approach to scenario modeling and model the scenario space with representative scenarios instead of exhausting all scenarios with inputs from domain experts. We then optimize eco-driving control across this––still significantly large––spectrum of representative scenarios by employing model-free multi-task deep reinforcement learning and zero-shot transfer learning instead of often used model-based methods. 
To make the optimization of eco-driving control tractable, we employ a carefully designed network decomposition strategy. This approach focuses on optimizing eco-driving at each intersection individually, instead of network scope optimizations, which is a significantly hard problem and intractable even with advances in modern computing and thus remains an open optimization challenge~\citep{qadri2020state}. While network decomposition strategies are commonly used in eco-driving research, many studies fail to analyze the impact of eco-driving optimized on individual intersections when deployed at the network scale. For example, if the throughput is increased due to eco-driving at an optimized intersection, that may lead to a queue spillback event in the following intersection at the network level. This oversight can render the previous analyses less reliable, potentially leading to network-level traffic overflows when such strategies are implemented. In contrast, we ensure that such impacts are accounted for in the assessment to ensure that the benefits observed in our analysis are consistent and reliable at the network level.

Analyzing nearly \textit{one million} traffic scenarios stemming at \textit{6,110} intersections across three major high-emissions US metropolitan cities~\citep{gately2015cities}- Atlanta, Los Angeles, and San Francisco- our analysis reveals insights that were elusive at small scale analyses and strengthen the existing knowledge with high confidence. We find that implementing eco-driving could reduce US intersection CO$_2$ emissions by 11-22\%, which under reasonable extrapolating assumptions is equivalent to the national carbon emissions of Israel and Nigeria (Refer to Appendix \ref{equal-benefit-estimation}). Such an implementation does not affect intersection throughput and could even assist in increasing throughput due to smoothed traffic flow. Furthermore, we observe that eco-driving is as safe as conventional driving, measured by safety surrogate measures~\citep{wang2021review}. With just 10\% of eco-driving adoption, 25\%-50\% of the total emission benefits (benefits if the adoption is 100\%) can be achieved, underscoring its significant near-term potential as a climate change mitigation strategy. However, when eco-driving adoption increases, the impact level of each factor affecting eco-driving benefits changes. For instance, while longer green signal times may negatively affect benefits at low adoption levels, they can positively contribute at higher adoption rates. This dynamic relationship necessitates frequent adjustments in traffic engineering practices and a reevaluation of long-term infrastructure investments at intersections. Furthermore, approximately 70\% of the total benefits are concentrated in just 20\% of intersections, even at lower adoption levels. However, the composition of this high-impact subset of intersections varies considerably across different adoption levels, with minimal overlap. This finding underscores the need for a more nuanced and adaptive approach to eco-driving deployment, emphasizing the importance of careful planning and strategic implementation to maximize benefits as adoption rates evolve over time.

Diverging from the sole technical focus often found in eco-driving research, this article adopts a socio-technical perspective, seeking to contextualize eco-driving within the framework of societal priorities. In a broader context, this work paves the way for scalable analyses of traffic externalities and solution strategies—a critical opportunity in light of the expanding adoption of roadway automation. Below, we summarize our main contributions, 

\begin{itemize}
    \item We present, to the best of our knowledge, the first prospective impact assessment of metropolitan city-scale eco-driving at signalized intersections, considering three main cities in the United States - Atlanta, Los Angeles, and San Francisco.
    \item We perform data-driven modeling of one million traffic scenarios at 6,110 intersections in the three cities in an agent-based microscopic traffic simulator. 
    \item We devise generalizable eco-driving policies for vehicles using multi-task deep reinforcement learning and zero-shot transfer learning across these scenarios with a carefully defined network decomposition strategy. 
    \item We conduct a comprehensive assessment of eco-driving at signalized intersections using the modeled scenarios and devised control policies and show insights that are elusive in small-scale analyses. 
\end{itemize}

\section{Related Work}
\label{ralated-work}

In the following sections, we present a summary of related work by categorizing them into four sub-sections that are closely related to our work. 

\subsection{Eco-driving at signalized intersections}

Over the past decade, numerous studies have investigated the challenge of dynamic eco-driving at signalized intersections. While they often study different traffic settings,~\cite{Mintsis2020survey} show an average eco-driving benefit of 17\%, with a standard deviation of 11\% by considering 21 past studies. While this highlights the potential of eco-driving, these studies often consider only a few factors that are known to have an impact on eco-driving benefits, making the shown benefits scenario/traffic setting specific. Based on the literature, we identify nearly 33 factors that affect eco-driving benefits. We refer the reader to Appendix~\ref{eco_driving_factors_related_work} for more details on these factors and their known impact. %However, to the best of our knowledge, there are no works that explicitly consider major factors that affect eco-driving, which this work aims to address. 

Moreover, most eco-driving studies focus on devising better eco-driving strategies (e.g., how to eco-drive). Such studies often take three forms: 1) analytical approaches, 2) simulation approaches, and 3) field experiments~\citep{Mintsis2020survey}. Analytical approaches often focus on analyzing the eco-driving problem from a theoretical point of view by analyzing properties such as the stability of the system under certain eco-driving strategies~\citep{coppola2022eco}. On the other hand, simulation-based approaches are used to test the potential of new eco-driving strategies and also to conduct impact assessments. However, one of the major limitations of such work is the lack of consideration of many factors that affect eco-driving benefits\citep{barkenbus2010eco, Mintsis2020survey, jayawardana2022learning, meng2020eco}. Furthermore, for simplicity and tractability of eco-driving control optimization scope, most of these studies analyze a few intersections under simplified traffic conditions (e.g., uniform inflow rates)~\citep{yang2016eco, yang2020eco, sun2020optimal}. A few studies that focus on multiple intersections often solely focus on the generalization of eco-driving strategies and rely on synthetically generated traffic scenarios while not considering the impact assessment requirements~\citep{jayawardana2024generalizing}. Finally, field studies are also conducted to demonstrate the benefits of eco-driving and to validate the applicability of eco-driving strategies~\citep{hao2018eco, stahlmann2018exploring, mintsis2017evaluation}. However, these studies are often confined to small regions (e.g., a few intersections) due to the logistical difficulties of conducting such experiments at scale~\citep{Mintsis2020survey}.

\subsection{Traffic intersection scenario modeling}

Numerous studies have focused on modeling intersections for various purposes, such as traffic signal control~\citep{ault2021reinforcement} and autonomous driving~\citep{highway-env}. However, they often remain limited to specific scenarios~\citep{wang2021multi, ault2021reinforcement}, primarily due to a focus on proof of concepts rather than comprehensive assessments of proposed methods. Additionally, modeling multiple intersections can be prohibitively expensive, requiring data fusion from various sources and handling missing data.

However, our requirements for modeling intersection scenarios are objectively and strategically different from previous work, necessitating us to develop our own datasets. First, we are neither presenting an eco-driving proof of concept nor attempting to answer how to do eco-driving. Our goal is to analyze the impact of eco-driving to answer the question of whether to eco-drive. Such questions can not be answered by analyzing merely a few intersection scenarios. Unless we have a representative set of intersection scenarios, merely relying on a select intersection could lead to evaluation overfitting where the insights extracted from a few scenarios do not generalize to the class of scenarios~\citep{jayawardana2022impact, whiteson2011protecting}. 

Furthermore, we must avoid overfitting to a specific region in our impact assessment~\citep{gately2015cities}. While there are popular intersection datasets available for Manhattan Island in New York~\citep{wei2019survey}, using them could lead to regional overfitting. Manhattan Island exhibits distinct spatial and temporal traffic dynamics due to its block-oriented road network. Insights drawn from an impact assessment solely based on these intersections would not be effective in generalizing to other cities or regions.

Another option is to use synthetic datasets, which are easier to generate~\citep{li2022metadrive}. However, impact assessments based on these datasets are not particularly meaningful as they may not reflect the real-world intersection distributions and may contain oversimplified scenarios~\citep{feng2023trafficgen}.

Thus, the fundamental requirement for modeling intersection scenarios lies in their diversity, encompassing variations at both the intersection and regional levels and making sure they are drawn from the real-world distribution. However, it is crucial to exercise caution when creating datasets, ensuring that the fused data originate from the same distribution. For instance, the IntD dataset~\citep{inDdataset} containing a set of intersections in Germany is popular within the research community, but combining it with U.S.-specific data like vehicle inflows may result in a distribution mismatch. %These datasets could be influenced by different confounding factors, undermining the reliability of the analysis.

Thus, to ensure the accuracy and validity of the modeling process, we must meticulously consider the sources and characteristics of the data we fuse, striving to maintain consistency within the datasets to avoid any potential biases or inaccuracies and be explicit about the assumptions we make. In this work, we follow these steps to carefully approach the scenario modeling while incorporating feedback from domain experts, as will be explained later. 

\subsection{Methods for eco-driving control}

Earlier studies on eco-driving have predominantly explored five control strategies for controlling vehicles. They include 1) dynamic programming, 2) heuristics, 3) trajectory optimization, 4) model-based reinforcement learning (model-based RL), and 5) model-free reinforcement learning (model-free RL). Each of these approaches exhibits distinct advantages and drawbacks. In Table~\ref{table:eco-driving-method-comparison}, we offer a comparative analysis of these methods, focusing on five prevalent shortcomings.

\begin{table*}[!ht]
\centering
\small
\begin{tabular}{|l|c|c|c|c|c|}
\hline
\multirow{2}{*}{\textbf{Method}} & \textbf{Sub opt-} & \textbf{Rely on} & \textbf{Model-based} & \textbf{High comp-} & \textbf{Small} \\
 & \textbf{imality} & \textbf{estimations} & \textbf{assumptions} & \textbf{uting load} & \textbf{Scale} \\
\hline
Dynamic programming\textsuperscript{[1,2]} & x & \checkmark & \checkmark & \checkmark & \checkmark \\
Heuristics\textsuperscript{[3]} & \checkmark & x & \checkmark & x & x \\
Trajectory optimization\textsuperscript{[4,5]} & x & \checkmark & \checkmark & x & $*$ \\
Model-based RL\textsuperscript{[6,7]} & \checkmark & $*$ & \checkmark & x & x \\
Model-free RL\textsuperscript{[8-10]} & \checkmark & x & x & x & x \\
\hline
\end{tabular}
\caption{A comparison of the shortcomings of five known methodologies for eco-driving control policy synthesis.  The 'x' indicates that the method does not demonstrate the shortcoming, a '\checkmark' indicates that the method demonstrates the shortcoming, and a '$*$' indicates that the shortcoming may or may not present.
\textsuperscript{[1]}\citep{maamria2016use}, \textsuperscript{[2]}\citep{doan2018iterative}, \textsuperscript{[3]}\citep{katsaros2011performance}, \textsuperscript{[4]}\citep{kamal2010board}, \textsuperscript{[5]}\citep{wang2019model}, \textsuperscript{[6]}\citep{lee2020model}, \textsuperscript{[7]}\citep{zhu2022safe}, \textsuperscript{[8]}\citep{jayawardana2022learning}, \textsuperscript{[9]}\citep{wegener2021automated}, \textsuperscript{[10]}\citep{bai2022hybrid}}
\label{table:eco-driving-method-comparison}
\end{table*}

\vspace{-0.2cm}

\begin{itemize}
\item \textbf{Optimality:} Refers to the method's ability to produce optimal results.
\item \textbf{Rely on estimations:} Refers to having access to additional information, like intersection queue lengths derived from fundamental diagrams during the optimization stage. This shortcoming encompasses both data acquisition and communication of those data-related estimations.
\item \textbf{Model-based assumptions:} Involves relying on a potentially complex model of vehicle dynamics, especially challenging when dealing with inter-vehicle dynamics and lane changes. Simplified models can lead to sub-optimal solutions.
\item \textbf{Computational load:} Indicates the inference time computational burden needed for obtaining solutions. This is crucial, considering that vehicles may not have powerful onboard computers. We note that low inference time computational burden does not necessarily mean low training time computational burden, as in the case of deep reinforcement learning. Given that training happens only once and does not involve the vehicle onboard computers, and inference happens continuously in the vehicle onboard computers, here we only focus on inference time computational burden. However, we note that our work focuses on not coming up with the most efficient inference time algorithm but rather assessing the impact of eco-driving.

\item \textbf{Scale:} Refers to the optimization problem's scale, e.g., one intersection or two. Here, scale does not consider deployment scale directly, although some synergies may exist.
\end{itemize}

As can be seen from Table~\ref{table:eco-driving-method-comparison}, model-free deep reinforcement learning appears as a favorable choice among the alternatives. Thus, in this work, we leverage model-free deep reinforcement learning to devise eco-driving strategies. 

\subsection{Reinforcement learning for vehicle-based traffic control}

Contrary to conventional traffic control mechanisms such as traffic signals and stop signs, vehicle-based traffic control is increasingly being explored in recent research. Such vehicle-based control is referred to as Lagrangian traffic control and involves controlling a fleet of vehicles to achieve certain fleet-level (including both controlled and uncontrolled vehicles) objectives such as congestion management~\citep{bottleneck}, reducing emissions~\citep{jayawardana2022learning}, and fostering smoother traffic flow~\citep{flow, yan2022unified}. While various studies have presented proofs-of-concept, they still fall short of accurately modeling real-world scenarios~\citep{yan2022unified, wu2017emergent, di2021survey} or go beyond simplified traffic problems~\citep{yan2022unified}.

Existing research often relies on simplified traffic models, such as simplified traffic blocks~\citep{yan2022unified}, to assess the benefits of Lagrangian traffic control methods. However, none of these studies explore the potential of large-scale implementations that could offer meaningful insights for practical decision-making. While simplified tasks are valuable for gaining preliminary insights into a method, they might not resonate strongly with policymakers and stakeholders due to the lack of rigorous analysis of the benefits at a scale that holds significance~\citep{esser2000large}.

In order to create more impactful outcomes that can drive policy changes and engage stakeholders, future research should focus on conducting comprehensive analyses that consider real-world complexities and practical scenarios. By doing so, the potential benefits of Lagrangian traffic control can be better understood, and its implementation may gain the attention and support needed to truly transform transportation systems, and make a substantial positive impact on traffic congestion and emission reduction, and improve overall traffic flow dynamics. In this work, we take a step towards this direction by analyzing the impact of Lagrangian eco-driving at signalized intersections in the metropolitan city scale. 

\section{Preliminaries}

\subsection{Intelligent Driver Model}
\label{idm}

In our simulations, we use the Intelligent Driver Model (IDM)~\citep{treiber2013traffic} to simulate the human drivers. IDM defines the instantaneous acceleration $a(t)$ of a vehicle at time $t$. It is given by the Equation~\ref{IDM-eq} where $v_0, s_0$ and $T$ denote desired velocity, space headway, and time headway, respectively. Maximum acceleration is given by $a$, and comfortable braking deceleration is given by $b$. $v(t)$ is the ego-vehicle velocity, $\delta$ is a constant, and $\Delta v(t)$ denotes the velocity difference between the ego vehicle and the vehicle in front of it.

\begin{equation}
a(t)=\frac{d v(t)}{d t}=a\left[1-\left(\frac{v(t)}{v_{0}}\right)^{\delta}-\left(\frac{H(v(t), \Delta v(t)}{h(t)}\right)^{2}\right]
\label{IDM-eq}
\end{equation}

\begin{equation}
H(v(t), \Delta v(t))=s_{0}+\max \left(0, v(t) T+\frac{v(t) \Delta v(t)}{2 \sqrt{a b}}\right)
\end{equation}

\subsection{Reinforcement Learning}
\label{rl}
In reinforcement learning, an agent learns a control policy by interacting with its environment, typically modeled as a Markov Decision Process (MDP) denoted as $M = \left\langle\mathcal{S}, \mathcal{A}, p, r, \rho, \gamma \right\rangle$. Here, $\mathcal{S}$ represents the set of states, $\mathcal{A}$ denotes the possible actions, $p(s_{t+1}|s_t,a_t)$ denotes the transition probability from the current state $s_t$ to the next state $s_{t+1}$ upon taking action $a_t$,  the reward received for action $a_t$ at state $s_t$ is $r(s_t, a_t) \in \mathbb{R}$, a distribution over the initial states is $\rho$, and $\gamma \in [0,1]$ is a discounting factor that balances immediate and future rewards.

Given the MDP, we seek to find an optimal policy $\pi^*: \mathcal{S} \rightarrow \mathcal{A}$ over the horizon $H$ that maximizes the expected cumulative discounted reward over the MDP. 
\vspace*{-0.02cm}
\begin{equation}
\pi^* = \argmax_{\pi} \mathop{\mathbb{E}} \left[\sum_{t=0}^{H}{\gamma^t r (s_t,a_t)| s_0, \pi}\right]
\end{equation}

\subsection{Contextual Reinforcement Learning}
\label{CMDPs}

A Contextual Markov Decision Process (CMDP) is an extension to the definition of an MDP introduced in Section~\ref{rl} by incorporating contextual information. The context parameterizes environmental variations, like changes in speed limits or lane lengths at different intersections in eco-driving. A CMDP is denoted as $\mathcal{M} = \left\langle\mathcal{S}, \mathcal{A}, \mathcal{C}, p_c, r_c, \rho_c, \gamma \right\rangle$. Different from MDPs, a context space $\mathcal{C}$ is introduced, but the action space $\textit{A}$ and state space $\textit{S}$ stay the same. The transition dynamics $\textit{p}_c$, rewards $\textit{r}_c$, and initial state distribution $\rho_c$ are conditioned on the context $c \in \mathcal{C}$. In other words, a CMDP $\mathcal{M}$ is a set of MDPs, each defined based on a context, and $\mathcal{M} = \{M_c\}_{c \sim \mathcal{C}}$ where $M_c$ is an MDP.

Solving a CMDP requires solving an algorithmic generalization problem within the task. This can be achieved by finding a policy that performs well in the CMDP overall, as denoted by Equation~\ref{eq-mrtl}. 

\begin{equation}
\label{eq-mrtl}
\pi^* = \argmax_{\pi} \mathop{\mathbb{E}} \left[\sum_{c \in \mathcal{C}}\sum_{t=0}^{H}{\gamma^t r_{c}(s_t,a_t)| s_0^c, \pi}\right]
\end{equation}

\subsection{Multi-task Learning}
\label{multi-task-learning}

In multi-task reinforcement learning, we extend the single-MDP (single task) reinforcement learning in Section~\ref{rl} to multiple MDPs (multiple tasks). Accordingly, our objective in finding optimal policy thus becomes, 
\vspace*{-0.1cm}
\begin{equation}
\label{eq_mtl}
\pi^* = \argmax_{\pi} \mathop{\mathbb{E}} \left[\sum_{\omega \in \mathcal{T}}\sum_{t=0}^{H}{\gamma^t r_{\omega}(s_t,a_t)| s_0^\omega, \pi}\right]
\end{equation}

where $\mathcal{T}$ is the set of MDPs (tasks). When this set of MDPs is defined based on a context, multi-task learning solves a CMDP. Also, note that we seek a unified policy in multi-task reinforcement learning that is performant over all MDPs (tasks). 

\subsection{Zero-shot Transfer Learning}
\label{zero_shot_transfer_rl}

Zero-shot transfer learning extends the concept of learning across multiple tasks, as in CMDPs, where the agent must perform well on new, unseen tasks without additional training. The goal is to leverage knowledge from a set of source tasks $\mathcal{T}_{\text{source}}$, which the agent has experienced during training, to generalize effectively to a set of target tasks $\mathcal{T}_{\text{target}}$. These target tasks are typically drawn from a distribution that may encompass tasks not encountered during training. In particular, a policy $\pi$ trained with the $\mathcal{T}_{\text{source}}$ is evaluated on the target tasks $\mathcal{T}_{\text{target}}$ without further training on the target tasks. Such a transfer learning strategy reduces the training overhead and can be more economical than training a policy from scratch on the target tasks~\citep{kirk2111survey}. 

\vspace{-0.2cm}
\section{Problem Formulation}

\subsection{Regional eco-driving problem}
\label{problem-formulation}

Optimizing eco-driving at the traffic network scale can yield the highest benefits, but it is a significantly hard problem and intractable even with advances in modern computing, and thus remains an open optimization challenge~\citep{qadri2020state}. Therefore, many studies adopt a network decomposition from the full network to the intersection level and optimize vehicle trajectories at individual intersections. While this has been a standard practice, it has to be performed carefully as isolated optimization of intersection-level behaviors can cause adverse events (e.g., intersection queue spillback from one intersection to another) at the network level. In this work, we adopt the same network decomposition strategy and take careful precautions to isolate the effects of the changed vehicle behaviors on each individual intersection. 

In light of this, consider a region comprised of traffic scenario variations $\Phi$ where $|\Phi|$ is very large (e.g., millions of individual signalized intersection scenarios). With the network decomposition strategy, the traffic scenario variations are stemming at the intersection level (e.g., changes in traffic flow, temperature, etc.). Let $\pi \coloneqq (\pi_{\phi})_{\phi \in \Phi}$ denotes a set of driving control laws. Similarly, let $\pi^b \coloneqq (\pi^b_{\phi})_{\phi \in \Phi}$ denotes the baseline (status quo driving). Let $f(\pi^{\prime}, \phi)$ denote a function that captures the CO$_2$ emission for the scenario $\phi$ under a control law $\pi^{\prime} \in \{\pi, \pi^b\}
$. We then define the \textit{regional eco-driving effectiveness} as follows. Note that each scenario $\phi$ implicitly encapsulates its own dynamics, which are abstracted within $\phi$ in this formulation.

\begin{equation}
E_{\Phi}(\pi, \pi^b) \coloneqq 1 - \frac{\E_{\phi \in \Phi} [f(\pi_{\phi}, \phi)]}{\E_{\phi \in \Phi} [f(\pi^b_{\phi}, \phi)]}
\label{regional_assemenent-eq}
\end{equation}
We thus seek to solve the \textit{regional eco-driving problem} for the best eco-driving control laws $\pi^e$ while avoiding negative system impacts to throughput, queue length, and waiting time. That is, we find:
\begin{equation}
\label{regional-policy-obj}
\pi^e = \arg\max_{\pi}E_{\Phi}(\pi, \pi^b) = \arg\min_{\pi} \E_{\phi \in \Phi} [f(\pi_{\phi}, \phi)]
\end{equation}
subject to, 
\begin{equation}
\label{regional-policy-throughput}
n(\pi_{\phi}, \phi) \geq n(\pi^b_{\phi}, \phi) \;  \;  \; \; \forall  \; \phi \in \Phi
\end{equation}

\begin{equation}
\label{regional-policy-next-int-throughput}
m(\pi_{\phi}, \phi) \leq \tau   \;  \;  \; \; \forall  \; \phi \in \Phi
\end{equation}

\begin{equation}
\label{regional-policy-gliding}
w(\pi_{\phi}, \phi) \leq w(\pi^b_{\phi}, \phi)  \;  \;  \; \; \forall  \; \phi \in \Phi
\end{equation}

Here, $n(\cdot)$ defines the intersection throughput dynamics function that takes the scenario $\phi$ and the control laws $\pi_{\phi}$ and $\pi^b_{\phi}$ and outputs the intersection throughput when the given control policy is applied to the scenario. Similarly, $m(\cdot)$ defines a function that captures the average observed queue length to lane length ratio of the immediately following intersection to the intersection where the eco-driving is assessed, and $\tau$ is a hyperparameter. Finally, $w(\cdot)$ defines a function that captures the average waiting time per vehicle in the immediate preceding intersection to the intersection where the eco-driving is assessed under control laws of $\pi_{\phi}$ and $\pi^b_{\phi}$. The functions $n(\cdot)$, $m(\cdot)$, and $w(\cdot)$ serve as global abstractions of the underlying physical traffic phenomena and takes in scenario $\phi$ and the relevant control policy to make it scenario specific (e.g., to capture the effects of each scenario dynamics) to capture how vehicle behaviors and interactions evolve under different control laws in a compact and measurable form.

The Equations \ref{regional-policy-throughput}, \ref{regional-policy-next-int-throughput}, and \ref{regional-policy-gliding} ensure the network decomposition strategy would not induce traffic events that would adversely affect the traffic flow at the network level. In particular, Equations \ref{regional-policy-throughput} ensures that by eco-driving, we do not reduce the intersection throughout, as a reduction in throughput could create traffic queues that would adversely affect the network traffic flow. Equations \ref{regional-policy-next-int-throughput} ensure that if there is an increase in the throughput at an intersection with eco-driving, it will not lead to a saturated incoming approach in the immediately following intersection, which ensures that the immediately following intersection has the capacity to absorb the increased throughput. Finally, Equation \ref{regional-policy-gliding} ensures that if the eco-driving affects the inflow of vehicles to the intersection being assessed, it would not be propagated back to the intersection preceding it. These constraints ensure that changes in the traffic flow due to eco-driving are either localized to the assessed intersection or implemented with high confidence that they will not adversely affect overall network traffic flow. While this conservative approach may yield a lower bound of potential benefits, it provides a reliable assessment of the impact of eco-driving, considering the network scale traffic flow dynamics.  

\subsection{Eco-driving control problem at an intersection}
\label{optimal-control-problem}

Finding the policy $\pi^{e}$ as described in Section~\ref{problem-formulation}, leads to solving an eco-driving control problem for every traffic scenario. Here, we offer a high-level overview of this eco-driving control problem for a generic scenario. It is important to note that every scenario demands a customized formulation to address its unique instantiations of the eco-driving factors.

The main objective of the optimal eco-driving control problem for a given scenario at a signalized intersection is to minimize the emissions of a fleet of vehicles (both controlled semi-autonomous vehicles and human-driven vehicles). However, this should be done with minimal impact on the travel times of vehicles. Given a vehicle dynamics model $x_i$ for guided semi-autonomous vehicle $i$ and $y_j$ for human-driven vehicle $j$ and the traffic signal timing plan $p$ of the signalized intersection being assessed, we formulate the optimal-control problem in discrete time-space as follows, 

\begin{equation}
\min J = \sum_{i=1}^{n} \sum_{t=0}^{T_i} E\left(a_i(t), v_i(t)\right) + \alpha T_i
\label{obj}
\end{equation}

such that for every vehicle $i$, 

\begin{equation}
a_i(t) =
\begin{cases}
x_i(s_i(t), \dot{s}_i(t), v_i(t), p) & \quad \text{if vehicle $i$ is controlled to eco-drive}  \\
y_i(s_i(t), \dot{s}_i(t), v_i(t), p) & \quad \text{otherwise} 
\end{cases}
\end{equation}

\begin{equation}
\sum_{0}^{T_i} v_i(t)\Delta t = d
\end{equation}
\begin{equation}
s_{min} \leq s_i(t) \leq s_{max} \quad \forall t \in [0, T_i]
\end{equation}
\begin{equation}
v_{min} \leq v_i(t) \leq v_{max} \quad \forall t \in [0, T_i]
\end{equation}
\begin{equation}
a_{min} \leq a_i(t) \leq a_{max} \quad \forall t \in [0, T_i]
\end{equation}

Here, $n$ represents the total number of vehicles in the fleet, $a_i(t)$, $v_i(t)$, $s_i(t)$, and $\dot{s}(t)$ denote the acceleration, velocity, headway and relative velocity of vehicle $i$ at time $t$, $E(\cdot)$ denotes the emission function that measures instantaneous emission of vehicle $i$ at time $t$, the multi-objective trade-off hyperparameter is $\alpha$, and $T_i$ denotes the travel time of vehicle $i$. The vehicle dynamics are coupled through relative position (headway) and relative velocity. The total required travel distance is denoted by $d$, and includes both arrival to and departure from the intersection. Furthermore, $\Delta t$ defines the discretized time step, usually in $[0.5, 1]$. 

In Eq~\ref{obj}, we aim to capture the total emission objective value of each vehicle (first summation term) over the entire planning horizon of each vehicle (second summation term). This objective has to be achieved given the minimum and maximum headway values denoted as $s_{min}$ and $s_{max}$, while $v_{min}$, $v_{max}$ and $a_{min}$, $a_{max}$ indicating the same for velocities and accelerations. These variables and their corresponding constraints ensure safety, connectivity via V2V communication (if applicable), and passenger comfort.

We use IDM~\citep{Treiber2000CongestedTS} as the dynamics model for human drivers. Thus, $y_j(\cdot)$ entails the ordinary differential equations defined by the calibrated parameterization of IDM for human-driven vehicle $j$. The $x_i(\cdot)$ for controlled vehicle $i$ is learned using multi-task deep reinforcement learning. 

\begin{figure}[t]
    \centering
    \includegraphics[width=1\linewidth]{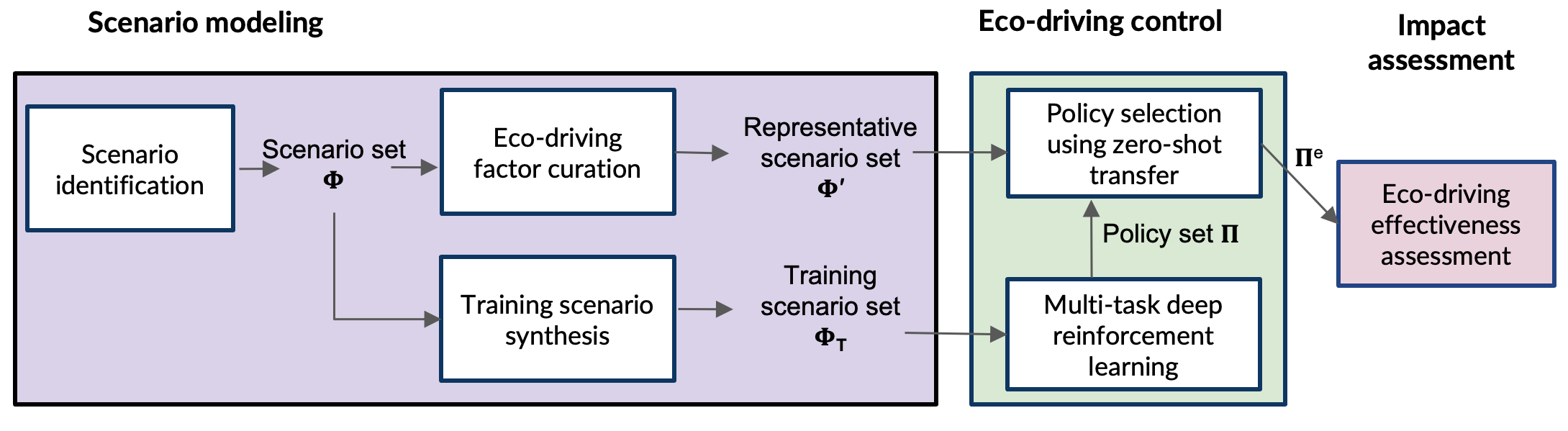}
    \caption{An overview of all steps included in assessing the impact of eco-driving. The high-level steps include scenario modeling, eco-driving control, and impact assessment. In scenario modeling, a set $\Phi$ of traffic scenarios is identified based on the factors that affect eco-driving benefits. These scenarios are pruned to select a subset $\Phi'$ of representative scenarios. Another subset $\phi_T$ is selected for training eco-driving policies. By using the subset $\Phi_T$, we then train a set of eco-driving policies $\Pi$ using multi-task deep reinforcement learning. These policies are then used for zero-shot transfer learning-based policy selection to select the best policy for each scenario in $\Phi'$. This final scenario to policy assignment is denoted by $\Pi^e$. Finally, the impact assessment is performed using $\Pi^e$ on $\Phi'$.   }
    \label{fig:overall_method}
\end{figure}
\vspace{-0.3cm}

\section{Methodology}
\vspace{-0.2cm}

\begin{figure*}[!t]
\centering
  \includegraphics[width=0.99\textwidth]{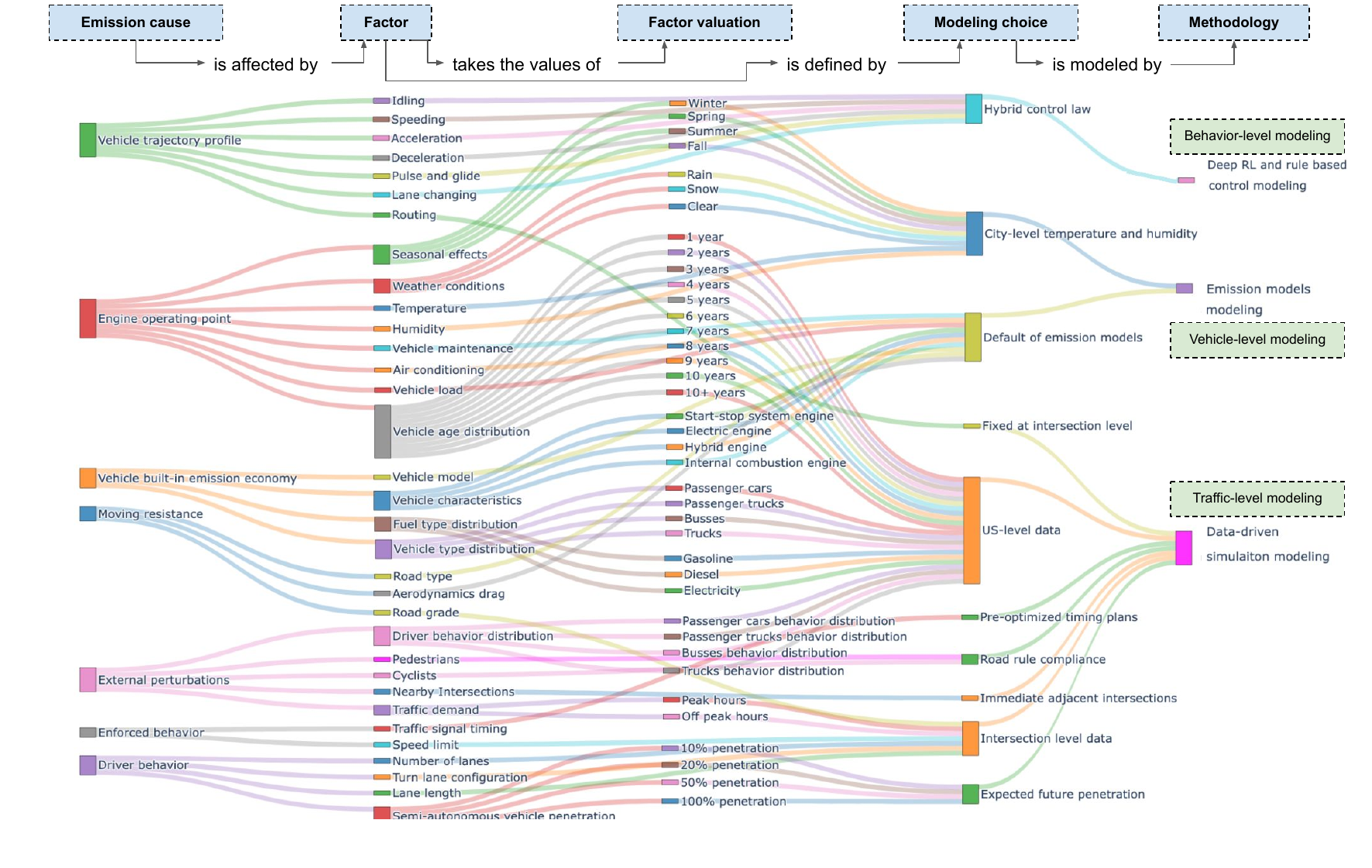}
  \caption{A visual illustration of eco-driving factors and how they affect emissions modeling. The legend on the top indicates how each column is connected. \textit{Emission cause}: Factors stem from different emission causes. A single factor can manifest due to multiple causes, but we illustrate a known predominant cause for simplicity. \textbf{Factor}: there are around 33 major factors affecting emissions of vehicles. \textit{Factor valuations}: A list of factor values we consider as each factor can take multiple values. For simplicity of the figure, we omit factor values for continuous variable factors. \textit{Modeling choice}: Indicates what knowledge or data sources inform each factor/factor value. \textit{Methodology}: Indicates how we translate the modeling choice into our simulation-based analysis. We leverage three levels of modeling: traffic-level, vehicle-level, and behavior-level. Each modeling level is captured by one modeling methodology.} 
  \label{fig:modeling_overview_main}
\end{figure*}

In the following subsections, we detail our approach to assessing the impact of eco-driving at signalized intersections by describing our approach to scenario modeling, solving eco-driving control, and how we perform impact assessment. A visual illustration of the overall methodology and how each of the three steps are connected is given in Figure~\ref{fig:overall_method}.

\vspace{-0.15cm}
\subsection{Scenario Modeling}
\vspace{-0.1cm}

\begin{wrapfigure}{r}{0.37\textwidth}
  \centering
  \includegraphics[width=0.37\textwidth]{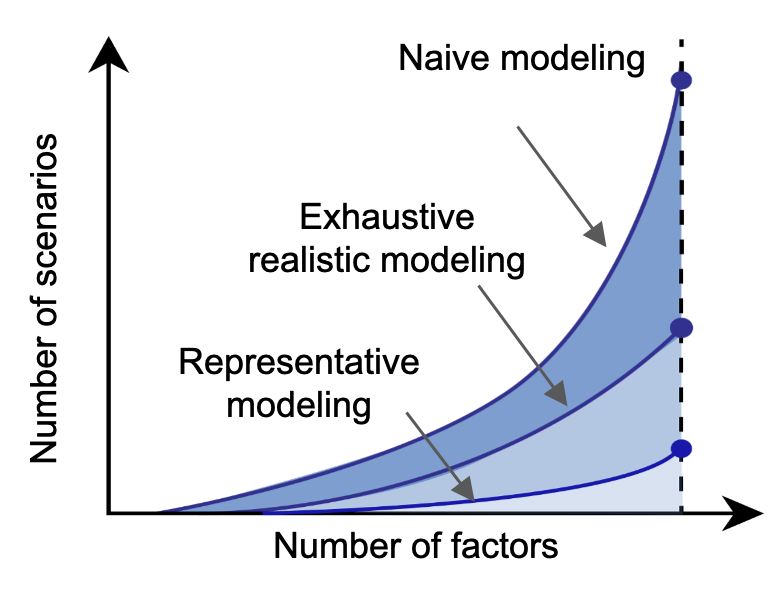} 
  \caption{Diverse scenario modeling choices and their relative comparison in terms of how many scenarios each modeling choice would require to model (not drawn to a scale). \vspace{-0.3cm}}
  \label{fig:types_of_scenario_modeling}
\end{wrapfigure}

Scenario modeling involves modeling each eco-driving scenario within a high-fidelity agent-based traffic simulator, replicating intersections as digital intersections, and calibrating for real-world conditions. Naive scenario modeling, which exhaustively models all possible scenarios, lacks practicality due to the inclusion of unrealistic scenarios (e.g., snow in Phoenix, Arizona). On the other hand, exhaustive yet realistic modeling offers superfluous detail that is redundant in decision-making (e.g., temperature impact on eco-driving at a fine-grained scale). Our approach seeks a representative subset of scenarios ($\Phi^{'} \subset \Phi$), ensuring realism, computational feasibility, and comprehensive coverage, as shown in Figure~\ref{fig:types_of_scenario_modeling}.

\begin{wrapfigure}{l}{0.37\textwidth}
  \centering
  \includegraphics[width=0.32\textwidth]{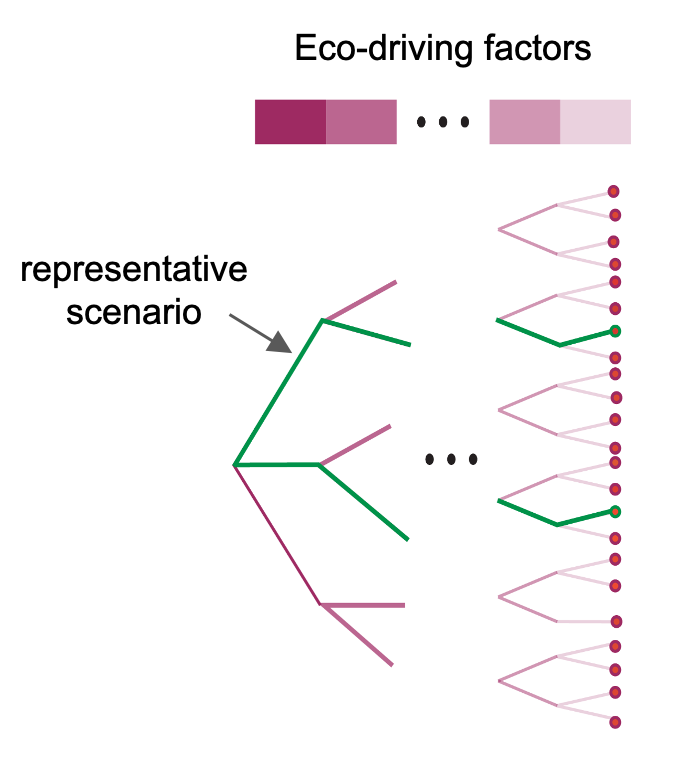} 
  \caption{Eco-driving scenarios tree using eco-driving factors. Each path from the root to a leaf node indicates a scenario, whereas green paths indicate representative scenarios. }
  \label{fig:scenario_tree}
\end{wrapfigure}

As illustrated in Figure~\ref{fig:modeling_overview_main}, related literature and expert opinions identify 33 major factors that impact eco-driving benefit levels. We omit the details of each factor here and refer the reader to Section~\ref{eco_driving_factors_related_work} in the Appendix for further details on each of these 33 factors and their known effects. Given the diverse spectrum of valuations each factor can take (e.g., seasons spanning summer, fall, winter, or spring), an eco-driving scenario manifests as a combination derived from the potential values across all factors. This results in an exponentially expanding scenario tree, as in Figure~\ref{fig:scenario_tree}. 

The scenario tree yields an exponential number of scenarios (at least $2^{33}$ scenarios). Our first goal is to simplify this scenario tree, balancing representativeness and the tractability of analysis. For this purpose, we prune the potential values each factor can take based on their significance, as known in the literature and based on the authors' assessment of how much additional information we would obtain if we were to assess a given scenario in the presence of other already selected scenarios. For instance, we take average seasonal temperatures per city instead of every temperature variation, which drastically reduces the tree branching involving the temperature factor. This is motivated by the observation that assessing all temperature variations will result in redundant information and little to no additional knowledge beyond analyzing a selected few temperatures. This curation involves expert consultation and extensive literature review, resulting in a subset $\Phi^{'} \subset \Phi$ of one million scenarios ($2^{20}$), representing a scenario space reduction by orders of magnitude. Due to the scale of this curation, we omit details here and refer the reader to Appendix Tables~\ref{tab:factor_curation_1} \ref{tab:factor_curation_2}, \ref{tab:factor_curation_3}, and \ref{tab:factor_curation_4} and Section~\ref{extra_modeling} for detailed curated modeling of each factor and Appendix~\ref{sceanrio_count} for a high-level breakdown of the one million scenarios. Below, we provide a high-level overview of our approach in scenario modeling.

For any given scenario, we capture the emissions effects of these factors through three interrelated levels of modeling: 1) traffic-level, 2) vehicle-level, and 3) behavior-level as shown in Figure~\ref{fig:modeling_overview_main}. All simulation scenario modeling is performed in the open-source agent-based SUMO microscopic traffic simulator~\citep{SUMO2018}.

\begin{wrapfigure}{r}{0.40\textwidth}
\centering
  \includegraphics[width=0.40\textwidth]{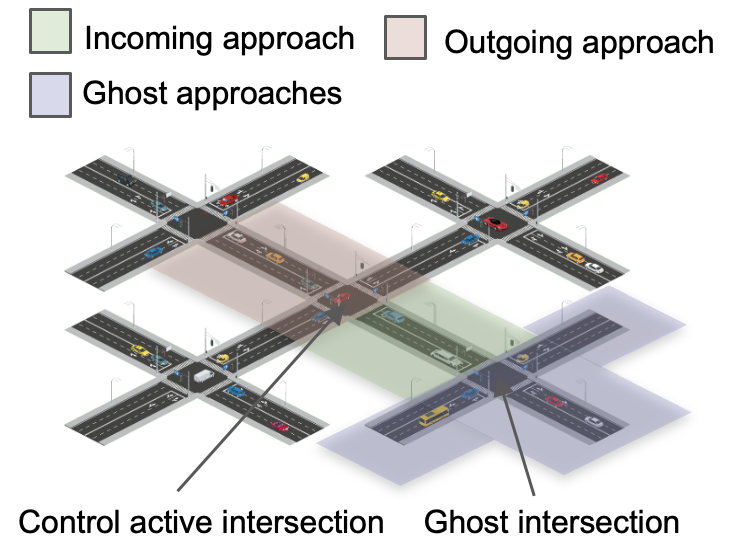}
  \caption{The control active intersection (where the vehicles are controlled for eco-driving), and ghost intersections (where the vehicles are not controlled for eco-driving). Each incoming approach is preceded by a ghost approach.} 
  \label{fig:ghost_intersection}
\end{wrapfigure}

Traffic-level modeling involves factors owing to the traffic scenario. These factors are modeled using traffic microsimulation. This includes using real-world intersection data to model intersection layouts, speed limits and vehicle inflows, vehicle age, vehicle type, fuel type distributions, etc. For human driver behaviors, we calibrate car-following models using real-world driving data. Since traffic flow at intersections is interconnected, with outflows from one affecting another, we model neighboring intersections where eco-driving is not active as \textit{ghost intersections}, while the intersection where eco-driving is active is termed the \textit{control active intersection} as illustrated in Figure~\ref{fig:ghost_intersection}. We consider optimal fixed-time traffic signal plans and assume pedestrian and cyclist compliance with road rules.

In vehicle-level modeling, we aim to capture the factors owing to the operating characteristics of individual vehicle engines. These factors are modeled using vehicle emission models. This approach allows us to account for factors such as temperature, humidity, seasonality, and weather conditions.
The industry-standard Motor Vehicle Emission Simulator (MOVES) by the US Environmental Protection Agency~\citep{epa_moves} provides accurate exhaust emissions for this purpose. Yet, it is impractical for our large-scale analysis, where we must assess millions of scenarios, given the MOVES takes about three minutes per scenario. To overcome this, we create a suite of efficient surrogate neural network emission models using MOVES as the data source~\citep{sanchez2022learning, sanchez2022learningJMLR}. Leveraging MOVES also allows us to capture the cumulative average effects of vehicle maintenance, vehicle model, vehicle characteristics, road types, aerodynamic drag, air conditioning, and vehicle load. 

Last, we focus on behavior-level modeling, which accounts for the factors owing to altered vehicle behaviors due to eco-driving optimization. These factors are modeled using deep reinforcement learning and rule-based control modeling. We consider behaviors with a substantial impact on emissions, such as idling, speeding, acceleration, deceleration, pulse-and-glide, and lane changes~\citep{xu2021overview}. Behavior-level modeling across diverse traffic scenarios is a core challenge addressed in this work, as discussed next.

\subsection{Eco-driving Control}
\label{mulit_scenario_learning}

Eco-driving control involves devising a set of eco-driving strategies for the representative scenario set $\Phi^{'}$. Eco-driving control for one million scenarios is a challenging task. Traditional methods like dynamic programming are computationally limiting~\citep{doan2018iterative}, while heuristics are often suboptimal~\citep{katsaros2011performance}. Trajectory optimizations are limited by the complexity of required vehicle dynamics models, particularly due to inter-vehicle dynamics and environmental variability introduced by many factors we consider~\citep{yang2016eco}.

Meanwhile, recent studies have shown the effectiveness of deep reinforcement learning in complex control tasks~\citep{wurman2022outracing, silver2017mastering, bellemare2020autonomous}. Neural vehicle control using deep reinforcement learning~\citep{flow, yan2022unified, jayawardana2022learning}, which can handle diverse traffic scenarios without manual fine-tuning, is also gaining popularity. However, training such a policy for each scenario is impractical due to the number of scenarios. Training a unified policy is complex due to multifaceted factors and the induced complex expectation~\citep{zhang2021survey}. Learning both discrete lane changes and continuous longitudinal accelerations presents optimization challenges, demanding diverse exploration strategies and distinct neural network representations. 

Thus, we introduce a custom training method involving three key steps: 1) similar scenario partitioning, 2) within partition multi-task deep reinforcement learning (MT-DRL), and 3) rule-based lane changes and a hierarchical policy selection strategy to accommodate lane changes. Below, we detail each of these steps in detail. 

\textbf{Similar scenario partitioning:} In this step, our aim is to partition scenarios by their similarities, allowing us to train separate MT-DRL policies for each partition. This strategy is inspired by the mixture of expert architecture widely used in real-world systems with generalization requirements~\citep{masoudnia2014mixture}. It can reduce scenario diversity within a partition and help reduce training complexity. While each scenario is defined by many factors, we only utilize intersection geometry and traffic flow-specific features as partitioning features because they strongly influence policy behaviors. 

\begin{figure*}[!h]
\centering
  \includegraphics[width=0.99\textwidth]{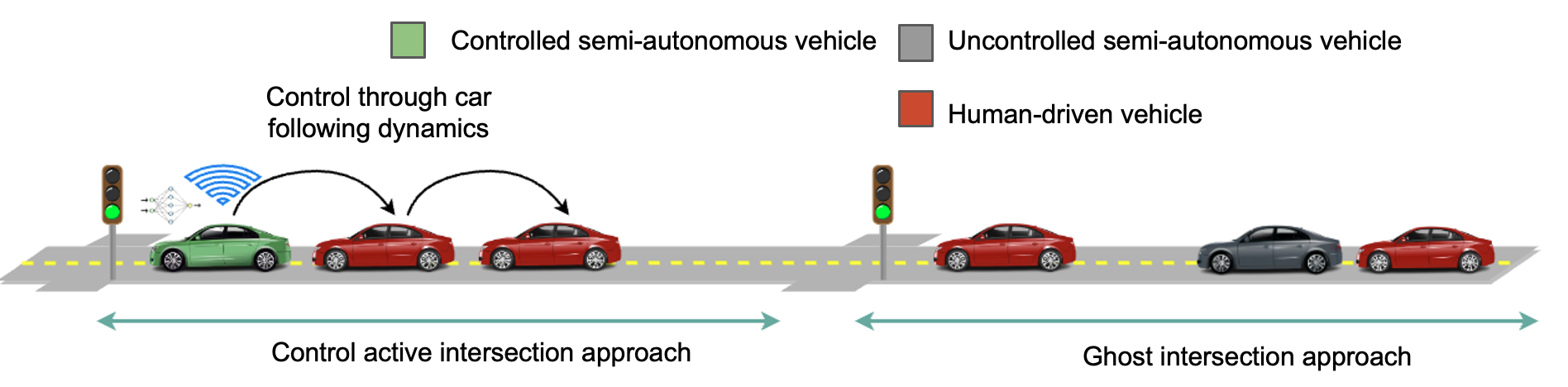}
  \caption{ A training environment is defined by a pair of incoming and outgoing approaches with the corresponding ghost approaches. In each training environment, a realistic traffic inflow to the control active intersection is generated using its incoming approaches and ghost approaches. Only semi-autonomous vehicles in the control active intersection approaches are controlled for emission reduction.} 
  \label{fig:training_environments}
\end{figure*}

However, since signalized intersections lack rotational invariance, comparing them for similarity is challenging. Therefore, we break down the intersection environment into non-conflicting pairs of incoming and outgoing approaches. This has added advantages. It accelerates training by simulating fewer vehicles and improves training stability by alleviating credit assignment challenges~\citep{zhou2020learning}. Further, it has no impact on modeling realism due to the decoupling nature of traffic signals (e.g., northbound traffic and eastbound traffic do not collide). These selected pairs, along with the ghost approaches connected to them, thus become our training environments and are illustrated in Figure~\ref{fig:training_environments}. 

The training environments are then partitioned into 28 partitions, considering the semi-autonomous vehicle penetration level, lane count, and traffic signal phase count as partition features. However, the environments in each partition may not form a uniform distribution. This may make environments sampled in training biased towards certain types of intersections for the most part (implicit imbalance problem). To overcome this, we adopt a synthetic environment distribution $\Phi_T$ derived from the full scenario set $\Phi$ for each partition with a uniform distribution for policy training as illustrated in Figure~\ref{fig:overall_method}. Additional details about the training environment distributions can be found in Appendix~\ref{appendix_training_environments}.

\textbf{Within partition MT-DRL:} Training an MT-DRL policy per partition involves solving an eco-driving contextual Markov Decision Process (CMDP) where each scenario is an MDP within the CMDP. Furthermore, MDPs within a CMDP can take on either single-agent or multi-agent forms. We adopt multi-agent control, necessitating implicit or explicit coordination among controlled vehicles for better emission reductions. Consequently, each MDP in the CMDP is modeled as a Partially Observable Markov Decision Process (POMDP).

\textit{Eco-POMDP}: A POMDP can be defined by extending an MDP when the agent's knowledge is limited to observing a partial view of the actual state rather than directly perceiving the complete state. POMDP is characterized by a seven-tuple that expands the five-tuple MDP by incorporating the observation space and conditional observation probabilities. The definitions of observations, rewards, and actions of our POMDP formulation for each vehicle (agent) are as follows. 

\textit{Observation}: Vehicle observation design is mainly influenced by the currently available technology, such as the requisite vehicle sensors (e.g., vehicle-to-infrastructure communication range), and the significance attributed by existing literature to specific information required for eco-driving. Thus, we define the observation for a vehicle (agent) $i$ as follows.

\begin{itemize}[topsep=0pt, itemsep=0pt]
    \item speed of the ego-vehicle
    \item relative distance to the traffic signal
    \item traffic signal state (red, green, or yellow) for the current phase
    \item time remaining in the current phase 
    \item time remaining until the traffic signal turns green for the second and third cycle 
    \item vehicle location (i.e., a flag indicating whether the vehicle is approaching the intersection, at the intersection, or exiting the intersection) 
    \item index of the vehicle's current lane 
    \item vehicle's intention to turn right, left, or go straight at the upcoming intersection
    \item for both the follower and the leader vehicles on the same lane, adjacent right lane, and left lane of the ego-vehicle:
    \begin{itemize}
        \item leader/follower speed
        \item relative distance to the leader/follower
        \item leader/follower turn signals status (turning right, left, or none)
    \end{itemize}
\end{itemize}

Furthermore, we also append a context vector describing the underlying intersection to the observation defined above.

\begin{itemize}[topsep=0pt, itemsep=0pt]
    \item number of lanes 
    \item length of the ego-lane
    \item signal timing plan for the traffic signal phase corresponding to the vehicle
    \item speed limit of the ego-lane
    \item penetration rate of semi-autonomous vehicles
\end{itemize}

\textit{Action}: The longitudinal acceleration of the vehicle is controlled by the policy.

\textit{Reward}: The reward function must encourage algorithmic generalization within the CMDP while considering various competing objectives like emission reduction and travel time efficiency, as we introduced in Section~\ref{optimal-control-problem}. Thus, the reward function for each semi-autonomous vehicle is expressed as $r_t^i = v_t^i + \alpha {1}_{v^i_t < \upsilon} + \beta e_t^i$, where $v_t^i$ represents the velocity of vehicle $i$ at time $t$, and $e_t^i$ represents its CO$_2$ emissions. In this equation, $\alpha$, $\beta$, and $\upsilon$ are hyperparameters. The indicator function ${1}_{v^i_t < \upsilon}$ indicates whether the vehicle is stopped, while the term $e_t^i$ encourages low emissions. The velocity term captures the effect on travel time. We note that we do not perform any explicit reward shaping to ensure the constraints of the regional eco-driving problem (Eq~\ref{regional-policy-throughput}, \ref{regional-policy-next-int-throughput}, and \ref{regional-policy-gliding}) are satisfied.

It’s generally accepted that limiting accelerations and idling reduces emissions~\citep{Mintsis2020survey}. But this may create traffic overflows or reduce intersection throughput. To account for these observations, we devise two policies per partition ${\psi_i}$: one, denoted as $\pi_{\psi_i}^1$, penalizes idling, while the other, $\pi_{\psi_i}^2$, places less weight on it. To achieve this we set $\alpha=35$, $\beta=3$ and $\upsilon=1$ for $\pi_{\psi_i}^1$ and $\alpha=10$, $\beta=3$ and $\upsilon=1$ for $\pi_{\psi_i}^2$.

\textbf{Hierarchical controller architecture.} While our MT-DRL policy controls vehicle accelerations, we leverage a rule-based default SUMO lane-changing controller for lane changes. It captures 1) strategic lane changes, 2) cooperative lane changes, 3) tactical lane changes, and 4) regulatory lane changes~\citep{erdmann2014lane}. At every time step, we check if the lane-changing controller would predict a lane-changing for the vehicle. If a lane-changing maneuver is warranted, precedence is given to the lane-changing maneuvers, and the lane-changing controller takes over the control of the vehicle to control both longitudinal and lateral movements of the vehicle for that time step. Otherwise, we use the MT-DRL policy by default to control the longitudinal movements of the vehicle. 

\subsection{Impact Assessment}
\label{impact_asssesment}

Impact assessment involves leveraging the trained eco-driving policies to assess the benefits of eco-driving in the representative scenario set $\Phi'$. Let $\Pi$ be the trained policy set for all 28 partitions. We further employ the zero-shot generalizability of these policies to conduct the final impact assessment. In particular, for a given semi-autonomous vehicle penetration level, we have seven partitions by varying the lane count and phase count. Recall that we denote each partition as $\psi_i$ where $i \in \{1,\cdots,7\}$ when the semi-autonomous vehicle penetration level is given. Let $\Pi^{'} = [(\pi_{\psi_i}^1, \pi_{\psi_i}^2)_{i \in \{1,\cdots,7\}} \cup \pi^b$] be the subset of trained policies for all seven partitions under a given semi-autonomous vehicle penetration level and the status quo driving baseline control policies $\pi^b \coloneqq (\pi^b_{\phi})_{\phi \in \Phi'}$. 

For a given scenario $\phi$ belonging to one of the seven clusters, we seek the best policy $\pi^*_{\phi}$, 

\begin{equation}
    \pi^*_{\phi} = \operatorname*{argmin}_{ \pi^k \in \Pi^{'}} (g(\pi^k, \pi^b_{\phi}, \phi))
\end{equation}

The function $g(\cdot)$ is defined as,

\begin{equation}
    g(\pi^k, \pi^b_{\phi}, \phi) =
    \begin{cases}
    f(\pi^k, \phi) & \text{if network decomposition constraints are satisfied} \\
    M & \text{otherwise} 
    \end{cases}
\end{equation}

such that $f(\pi^{k}, \phi)$ denoting the function that captures the CO$_2$ emission for the scenario $\phi$ under the control law $\pi^{k}$ and $M$ is large enough constant to indicate the policy $\pi^k$ is invalid due to any of the network decomposition constraint violations as listed in Equations~\ref{regional-policy-throughput}, \ref{regional-policy-next-int-throughput}, and \ref{regional-policy-gliding}. 
If any of the constraints are not satisfied, we default to the status quo driving baseline control policy $\pi^b$. Below, we describe how we check for these constraint violations. 

\textbf{Control active intersection throughput constraint}: To check if the control policy $\pi^k$ satisfies the throughput requirement, we compare its throughput with the status quo driving baseline $\pi^b$ throughput. If $n_{\pi^k}$ is the throughput of $\pi^k$ and $n_{\pi^b}$ is the throughput of $\pi^b$, then the constraint is satisfied if $n_{\pi^k} \geq n_{\pi^b}$. 

\textbf{Following intersection queue length to lane length ratio constraint}: If the eco-driving increases the throughput of the intersections by smoothing traffic flow, it is imperative to ensure that additional throughput is bearable by the intersection immediately ahead of the control active intersection. To ensure this, we leverage the average vehicle queue length to lane length ratio of each approach as it indicates how many additional vehicles the approach can accommodate without congestion in the equilibrium state. In particular, we check if the vehicle inflow to all the control active intersections is increased by $r$ percent (mimicking increased throughput from the preceding intersection), can we ensure that all control active intersections under policy $\pi^k$ have vehicle queue length to lane length ratio  $m_{\pi^k} \leq \tau$ where $\tau$ is a hyperparameter. When $\tau < 1$, the lane can accommodate more vehicles. In practice, we find the maximum $r$ with $\tau = 0.3$ such that at least 99\% of the intersections can accommodate the additional throughput. Once such $r$ is found, all presented analyses will ensure that the following intersection queue length to lane length ratio constraint is satisfied.

\textbf{Preceding intersection vehicle waiting time constraint}: When eco-driving is performed at the control active intersection, the changed vehicle flow may affect the outflow of the intersection immediately precedes it (e.g., when the vehicles glide at the control active intersection, slowing the rate other vehicles leave the preceding intersection). To ensure this does not happen, we check if $w_{\pi^k} \leq w_{\pi^b}$ where $w_{\pi^k}$ and $w_{\pi^b}$ are the average waiting time per vehicle in the preceding intersection under eco-driving policy $\pi^k$ and status quo driving baseline $\pi^b$, respectively. In simulations, this also includes any vehicle that has been scheduled to be spawned in the road network but yet to be spawned due to congested traffic settings.

Finally, by performing the above zero-shot transfer scheme to all scenarios in $\Phi^{'}$, we obtain the best policy $\pi^*_{\phi}$ for each scenario $\phi$. We then use the best policy of each scenario for subsequent impact assessments. In particular, \textit{regional eco-driving effectiveness} $E(\cdot)$ is used to assess the overall benefit level as per Equation~\ref{regional_assemenent-eq}. Each scenario is simulated five times for statistical significance, and results are averaged following the recommended practice~\citep{agarwal2021deep}.   

\textbf{Remark:} As we listed in each of the main sections, and more broadly in section~\ref{assumptions_limitations} in the Appendix, we make certain assumptions in our modeling efforts. Hence, the analyses presented here should be interpreted within the framework of these assumptions.

\section{Simulation-based Analysis}

\subsection{Experimental setup}

We use Proximal Policy Optimization (PPO)~\citep{schulman2017proximal} algorithm in training our MT-DRL policies with actor-critic architecture. The actor and critic do not share parameters and are separate neural networks. Centralized training and a decentralized execution paradigm are used for training. Section~\ref{hyperparameters} in the Appendix describes the architecture details and all the hyperparameter configurations used in training. 

\subsection{Impact of eco-driving}

Our eco-driving assessment encompasses diverse perspectives. First, we analyze the overall regional eco-driving effectiveness and the resultant variations in the intersection throughput in the three cities. Second, we analyze which intersection-related factors have a greater influence at different levels of eco-driving adoption. Third, we analyze the level of intersection compatibility needed to achieve the benefits, informing implementation plans. Fourth, we analyze eco-driving behaviors and examine the safety implications of eco-driving, an aspect that has not received adequate attention in previous work~\citep{carslaw2010comprehensive}. Last, we contextualize eco-driving by analyzing its role in joint consideration with travel growth projections and decarbonization technologies such as vehicle electrification and the adoption of hybrid vehicles. 

\subsubsection{Overall effectiveness}

\begin{figure}[H]
    \centering
    \subfloat[Emission benefits]{\includegraphics[width=0.4\textwidth]{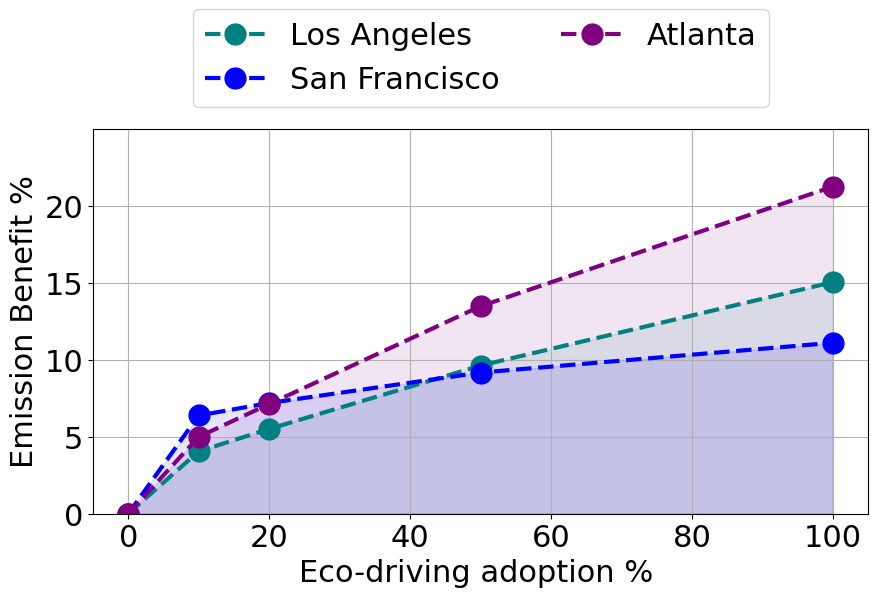}}
    \hspace{0.04\textwidth}
    \subfloat[Throughput benefits]{\includegraphics[width=0.4\textwidth]{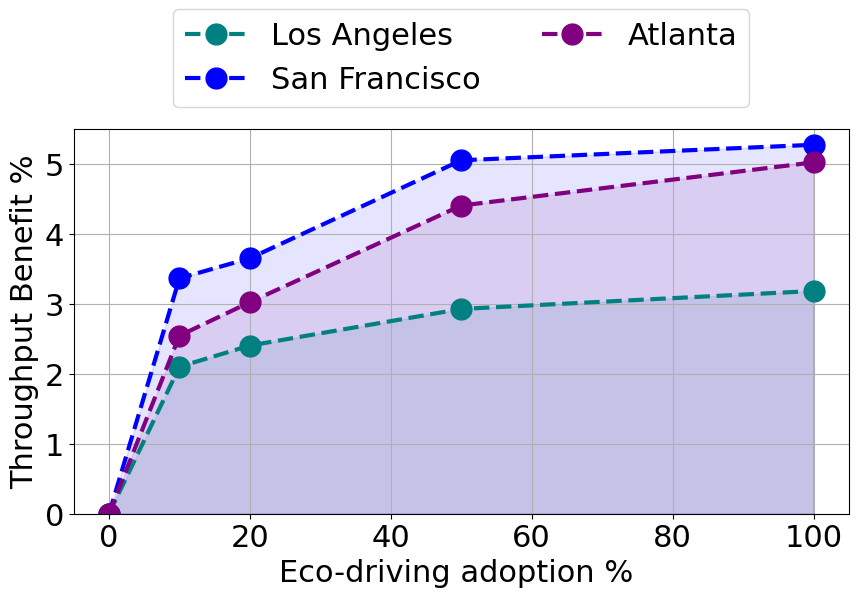}}

    \caption{ \textbf{Left} Annual average emission reduction vs. eco-driving adoption (by factoring all seasons and peak and off-peak hours). \textbf{Right} Throughput improvement with varying adoption levels. By design, the throughput improvement is at least 0\% when evaluating the emission reduction. Also, we ensure these throughput increases do not negatively affect the network-level traffic.}
    \label{fig:eco-driving-effectivness}
\end{figure}

\begin{figure}[H]
\centering
  \includegraphics[width=1.0\textwidth]{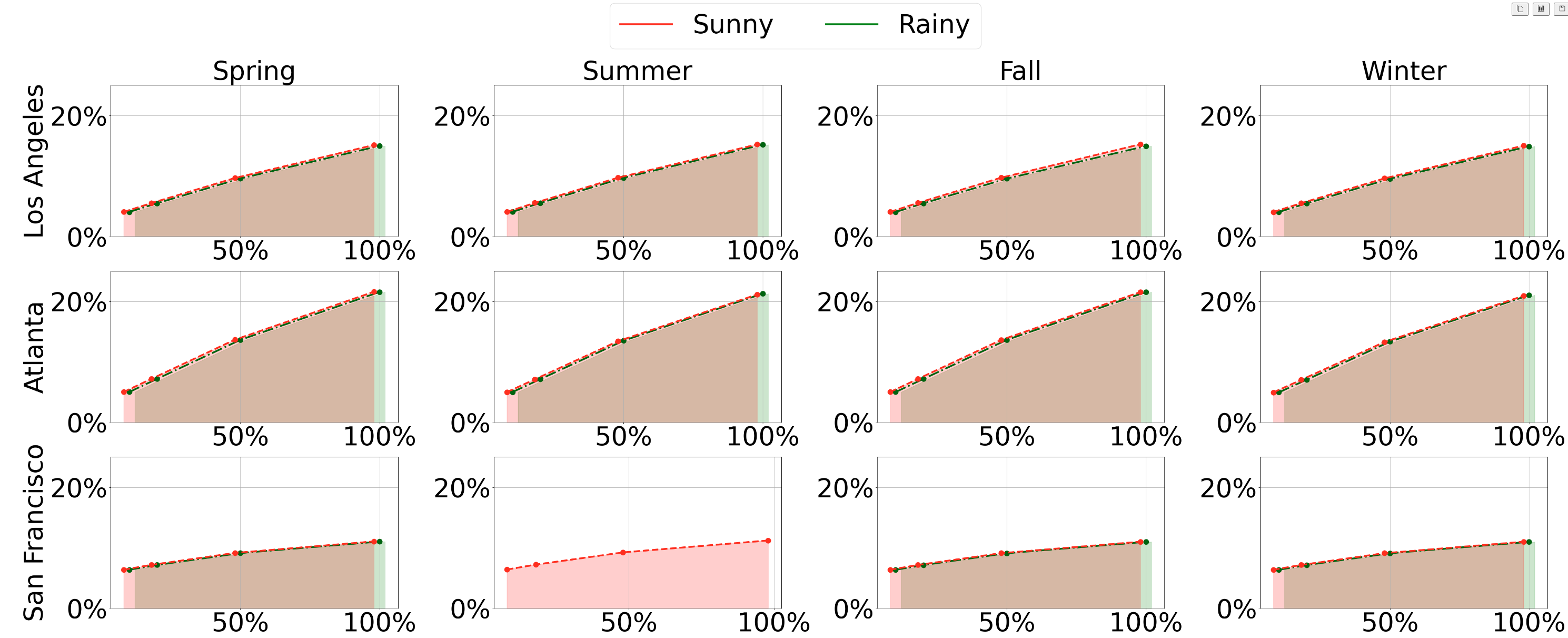}
\caption{Impact of weather on eco-driving benefits. For each city and season, we show emission benefits under sunny and rainy conditions. Some curves are offset by ±2\% on the x-axis for clarity. Not all weather types occur in every city-season pair. }
\label{fig:weather_eco_driving_allcities}
\end{figure}

\vspace{-0.2cm}
Figures~\ref{fig:eco-driving-effectivness} show the annual average effectiveness of eco-driving, assessing the percentage emission reduction and average throughput improvement. On the policy side, throughput gives an alternative argument supporting eco-driving. Eco-driving achieves a notable 11-22\% emission reduction, equivalent to the national carbon emissions of Israel and Nigeria, respectively (refer to Appendix~\ref{equal-benefit-estimation} for the estimation). Notably, a 10\% eco-driving adoption can yield 25-50\% of the total emission reduction, showcasing a non-linear benefit scaling and greater potential for short-term deployments to gain a majority of total benefits. Importantly, these gains incur no costs to intersection throughput. Additionally, Figure~\ref{fig:weather_eco_driving_allcities} demonstrates consistent emission benefits under different weather conditions across seasons, with minor fluctuations. This emphasizes that eco-driving can yield benefits throughout the year. In the Appendix~\ref{peak-off-peak-analysis}, we also show the peak vs off-peak emission benefit percentages. 

\vspace{-0.2cm}

\begin{figure}[H]
    \centering
    \subfloat[Atlanta - 10\% adoption]{\includegraphics[width=0.33\textwidth]{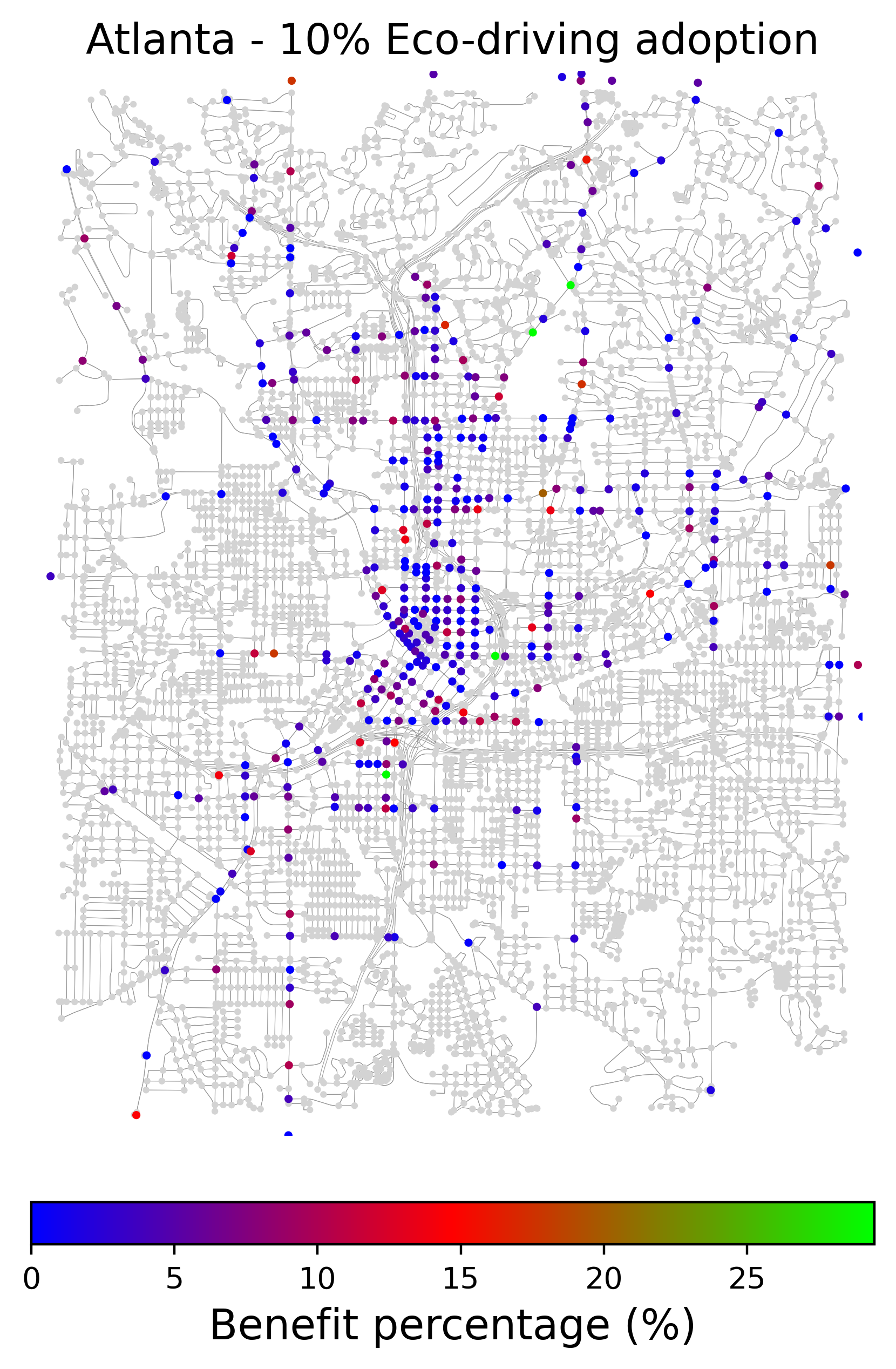}}
    \hfill
    \subfloat[Atlanta - 50\% adoption]{\includegraphics[width=0.33\textwidth]{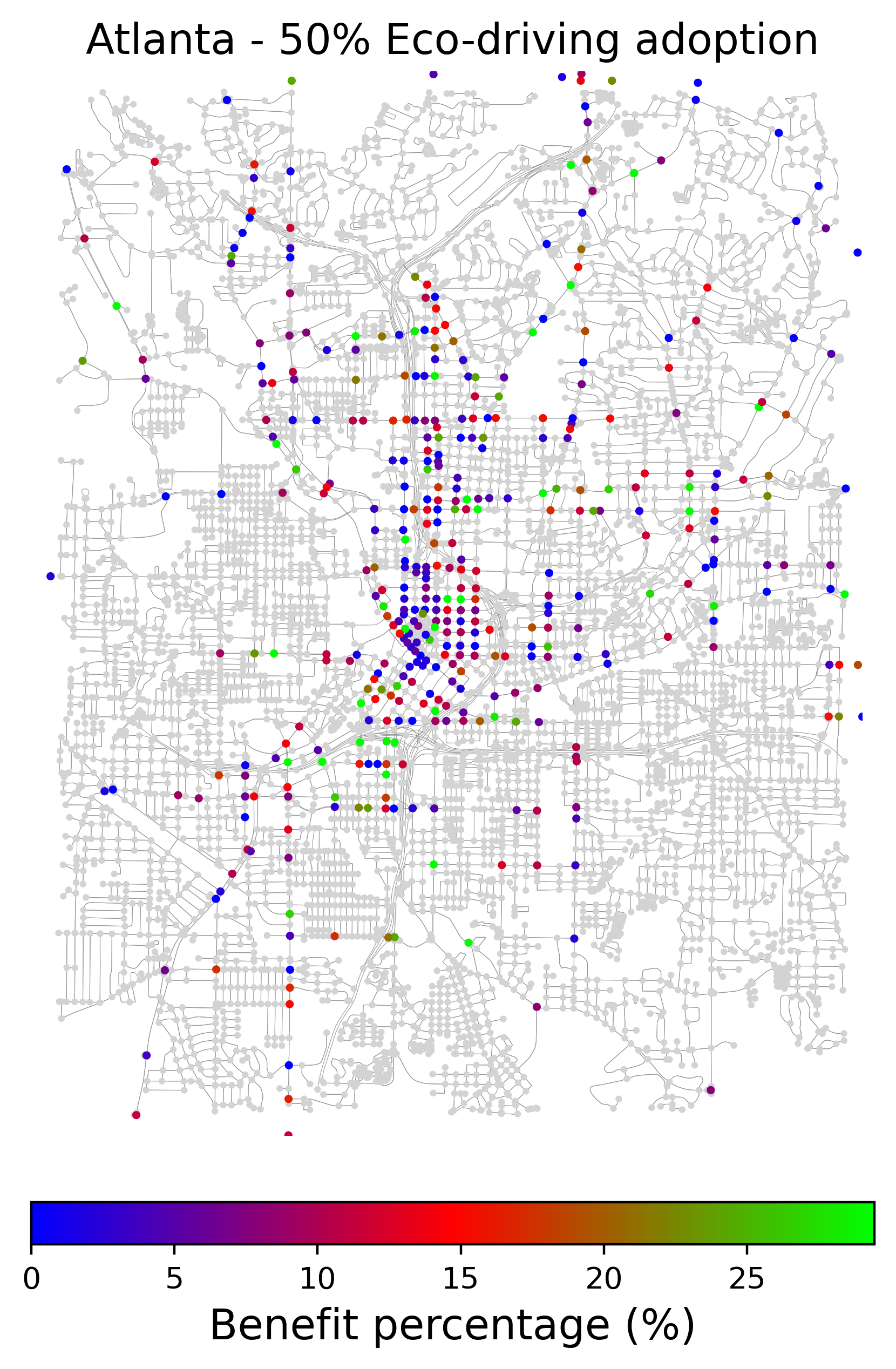}}
    \hfill
    \subfloat[Atlanta - 100\% adoption]{\includegraphics[width=0.33\textwidth]{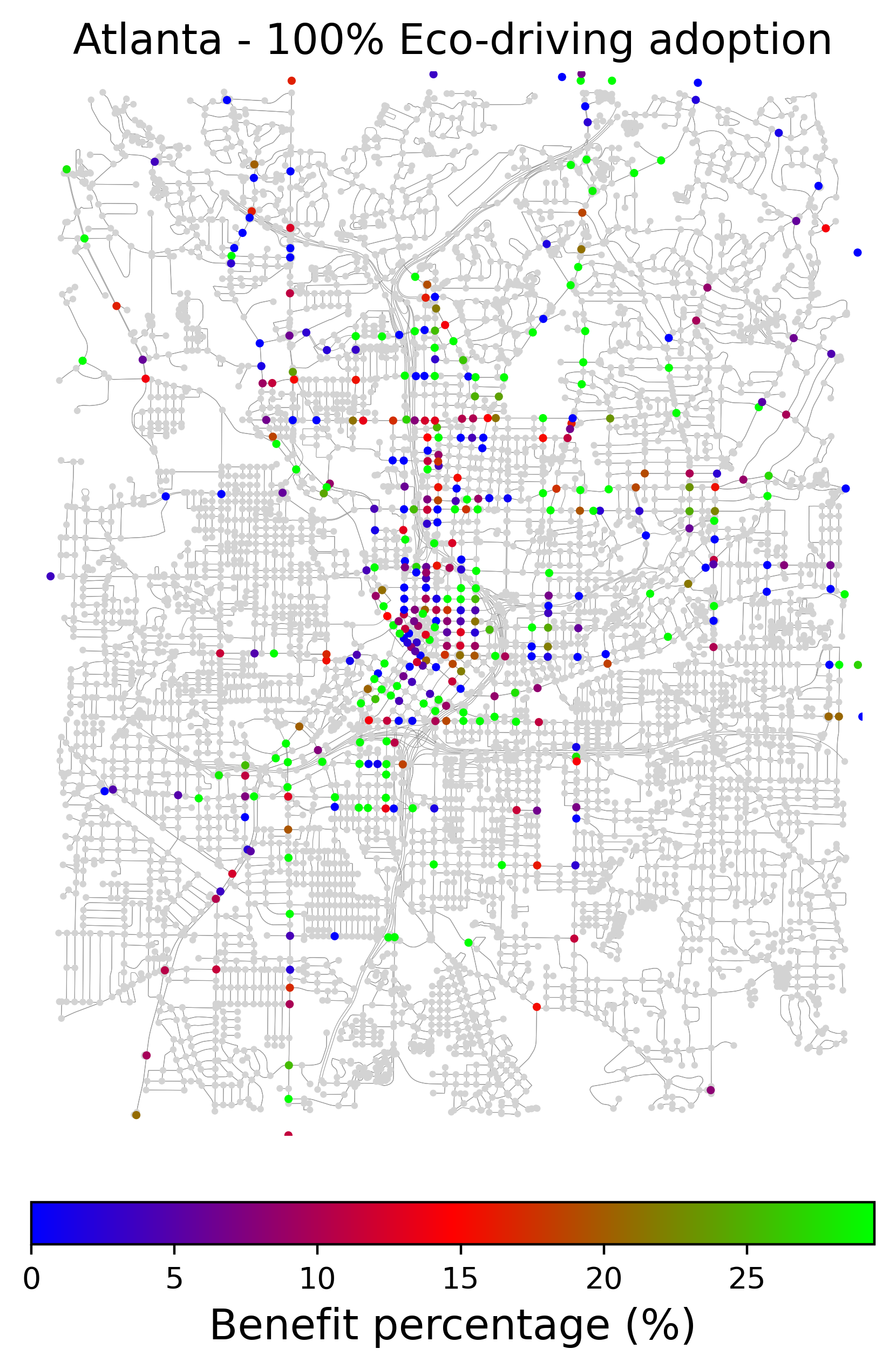}}
    \vspace{0.5cm}
    \subfloat[San Francisco - 10\% adoption]{\includegraphics[width=0.33\textwidth]{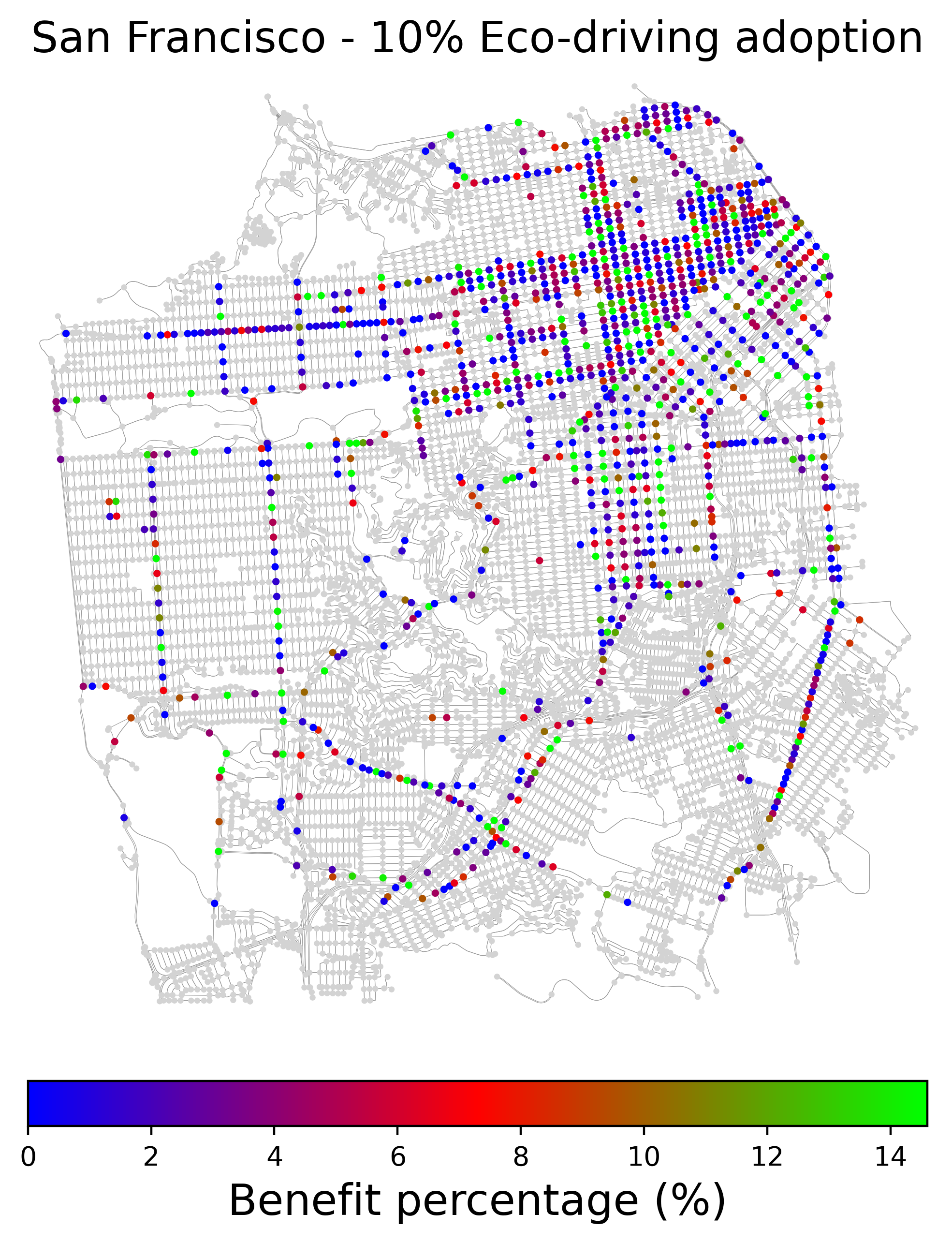}}
    \hfill
    \subfloat[San Francisco - 50\% adoption]{\includegraphics[width=0.33\textwidth]{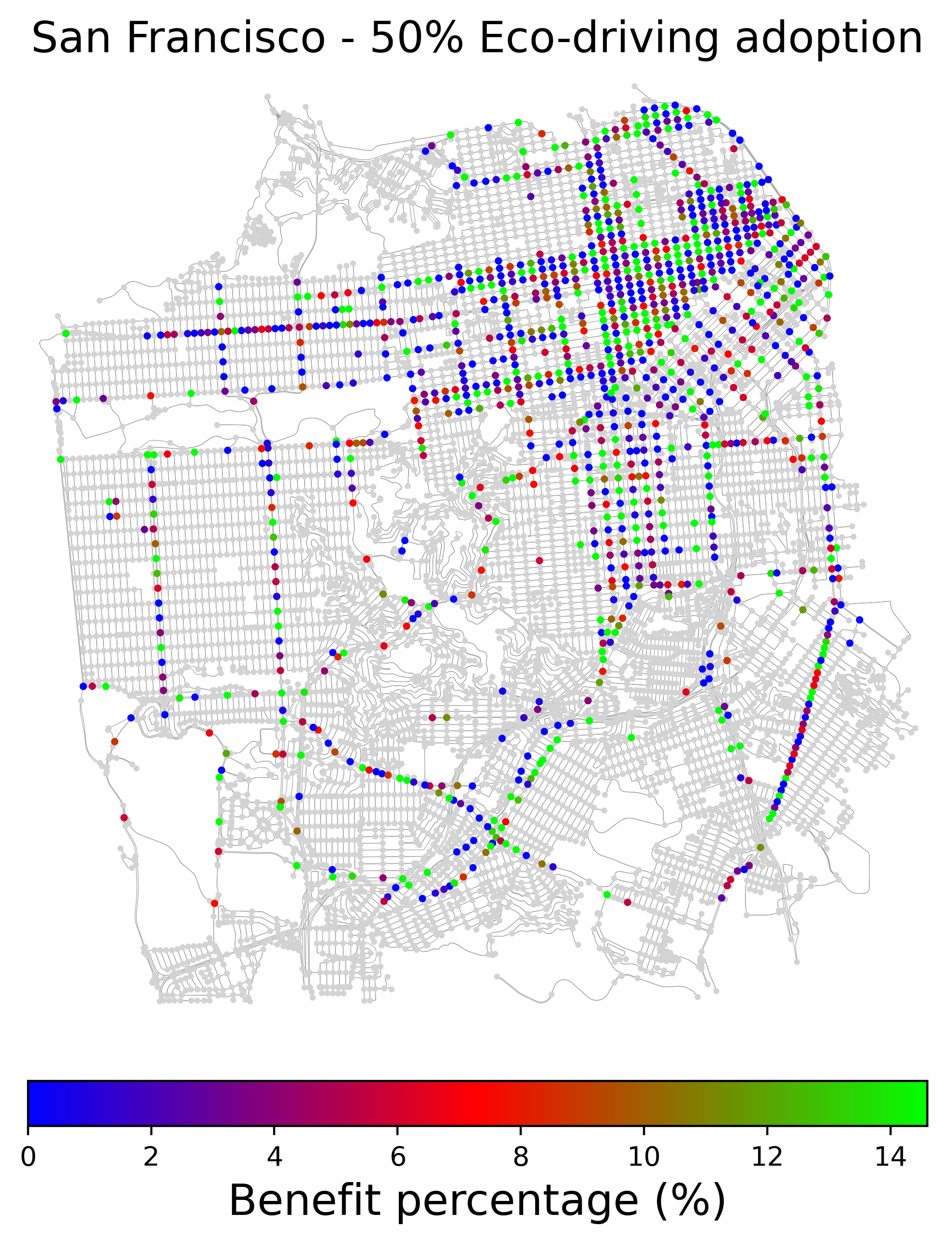}}
    \hfill
    \subfloat[San Francisco - 100\% adoption]{\includegraphics[width=0.33\textwidth]{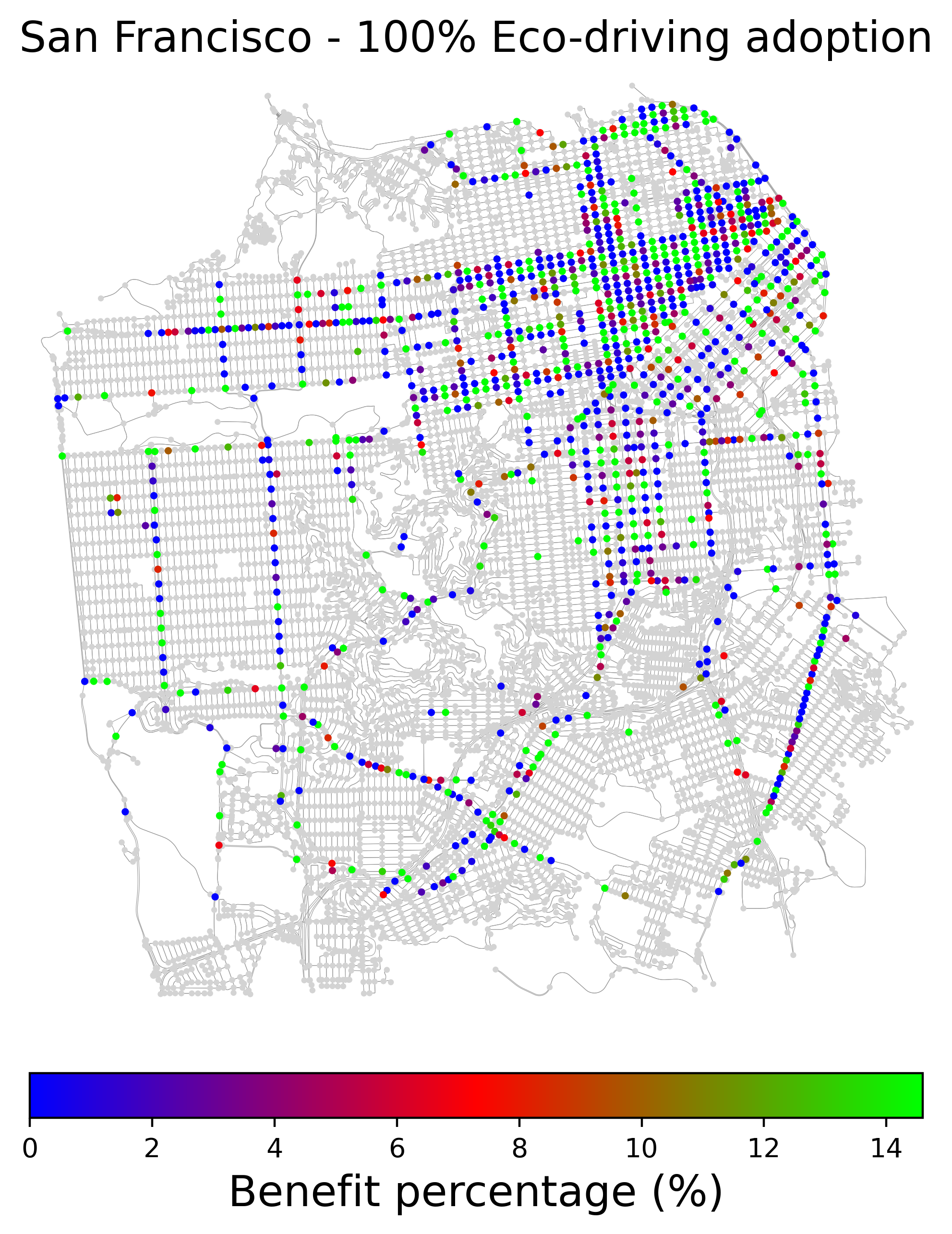}}
    \vspace{0.5cm}
    \subfloat[Los Angeles - 10\% adoption]{\includegraphics[width=0.33\textwidth]{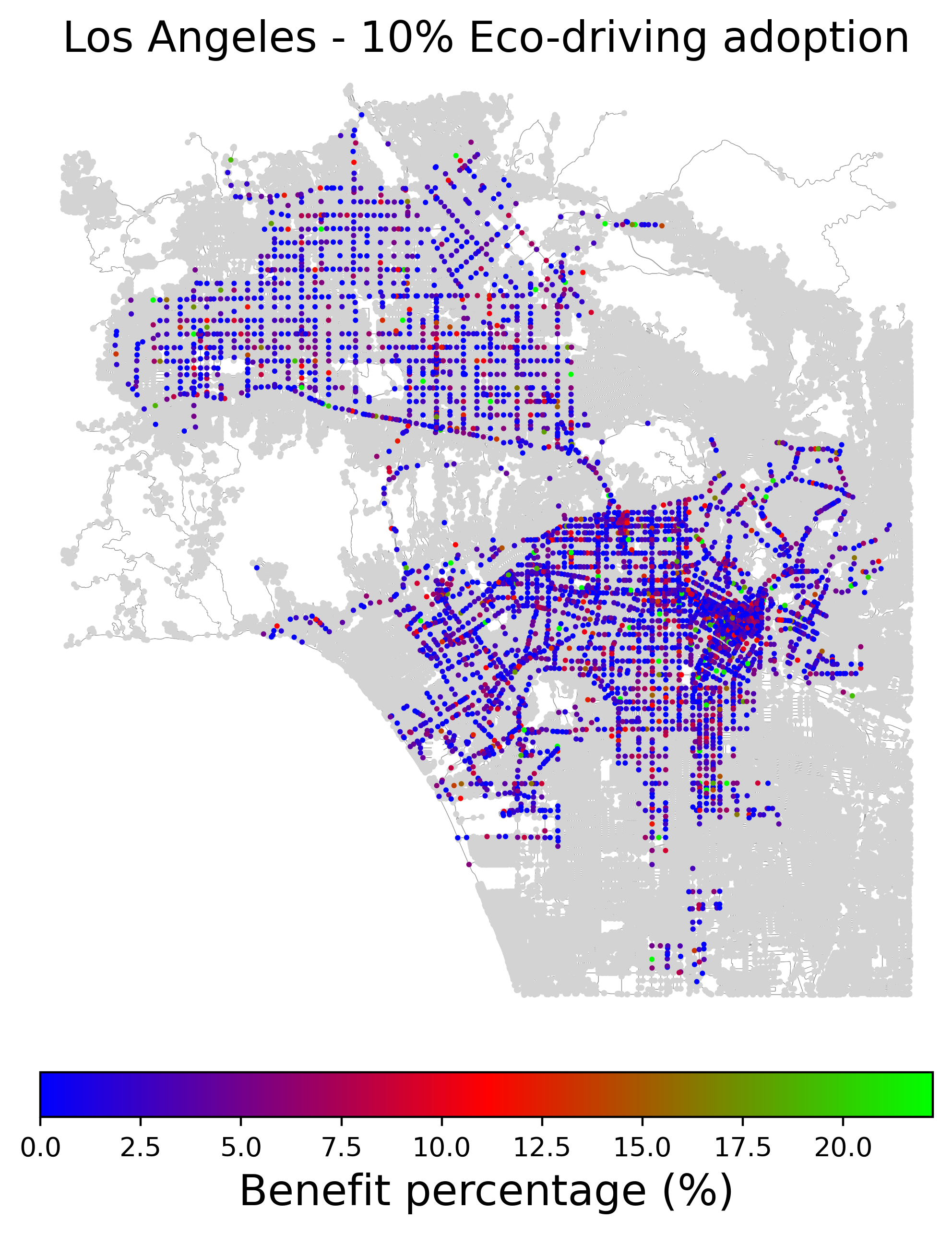}}
    \hfill
    \subfloat[Los Angeles - 50\% adoption]{\includegraphics[width=0.33\textwidth]{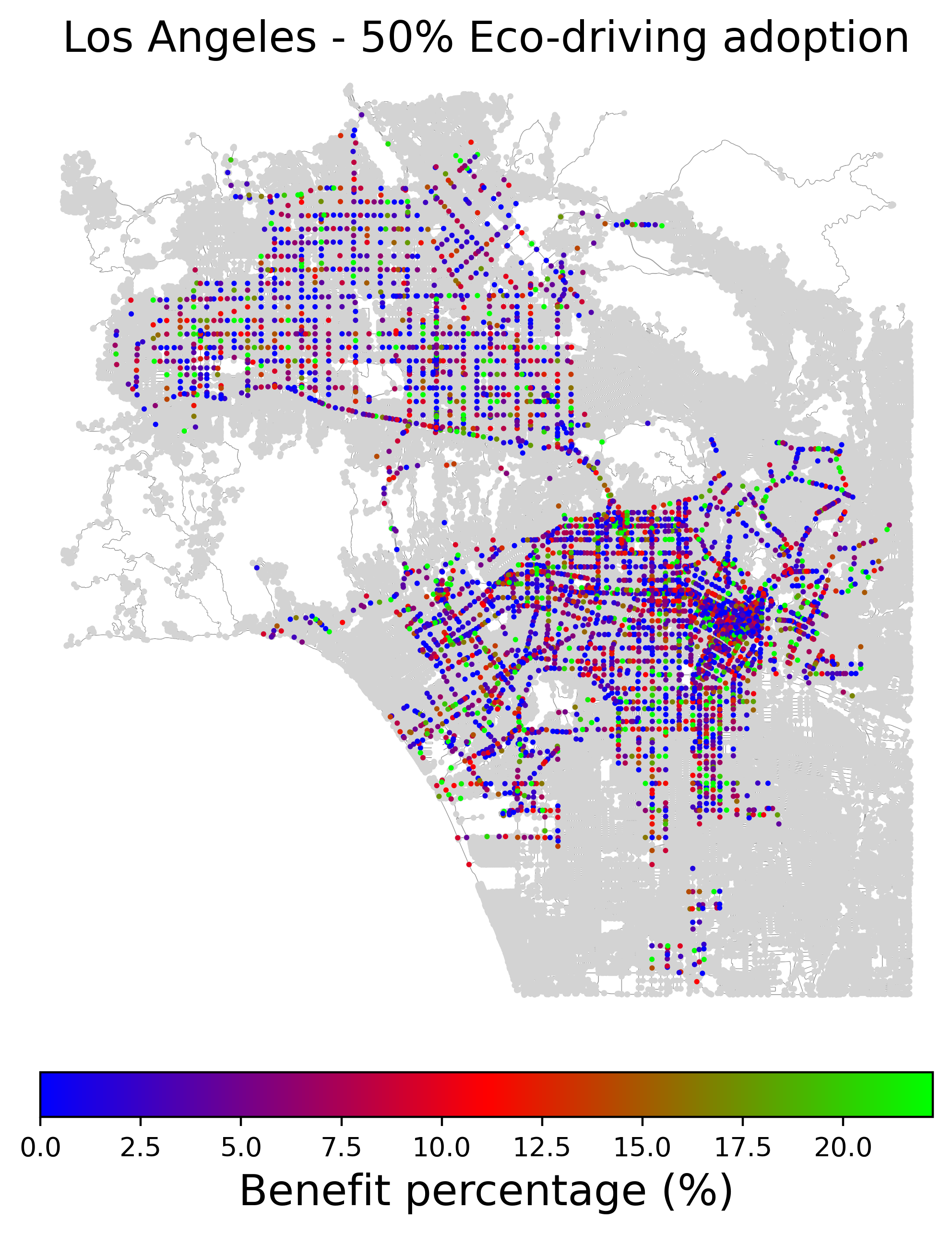}}
    \hfill
    \subfloat[Los Angeles - 100\% adoption]{\includegraphics[width=0.33\textwidth]{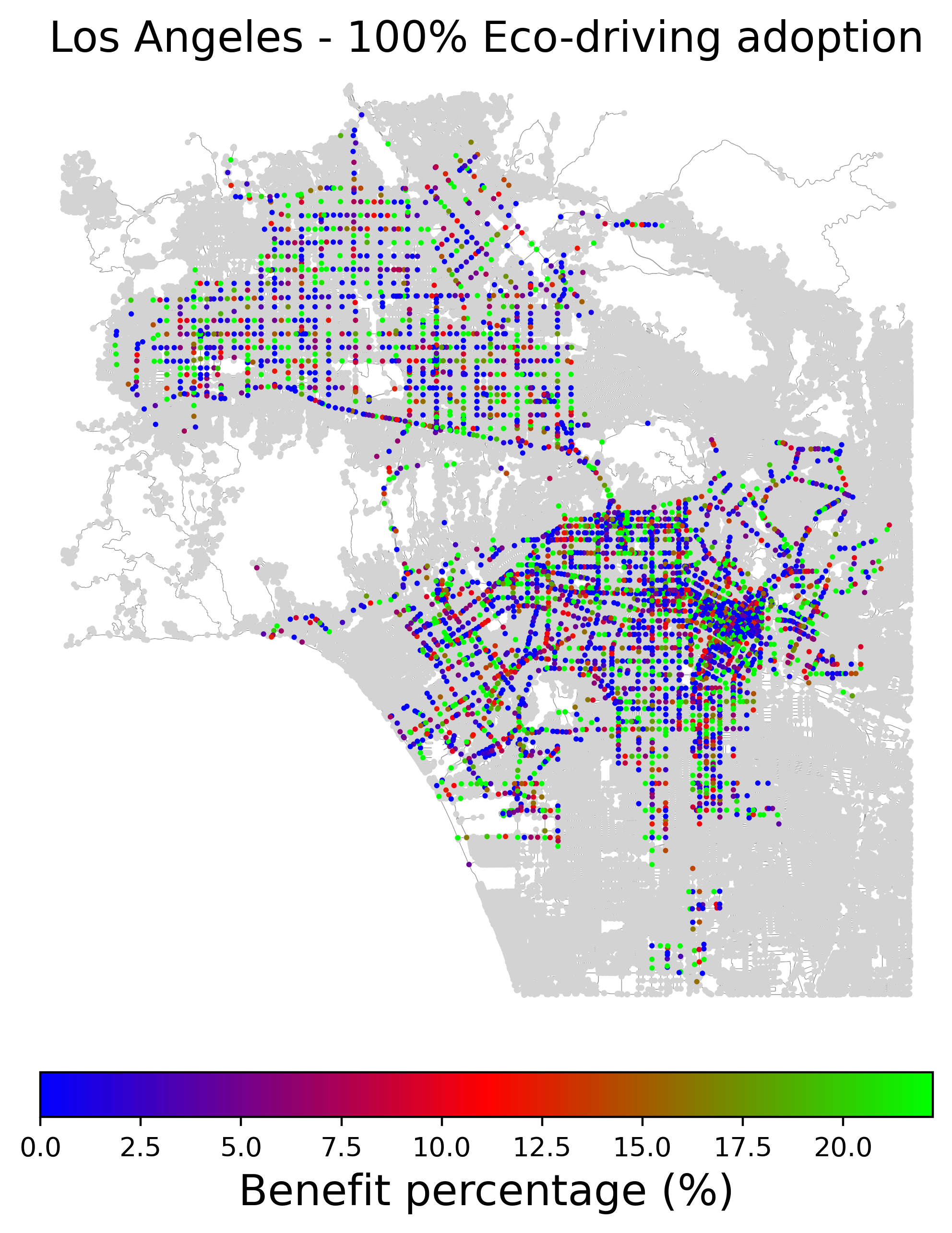}}
    
    \caption{Spatial distribution of eco-driving benefit at each of the three cities with varying eco-driving adoption levels.}
    \label{fig:emission_benefits_spatial}
\end{figure}

Figure~\ref{fig:emission_benefits_spatial} shows the spatial distributions of these emission benefits under different eco-driving adoption levels. The low benefits intersections in Figure~\ref{fig:emission_benefits_spatial} can happen due to three reasons. First, eco-driving at those intersections would not yield benefits due to their characteristics (e.g., not enough lane lengths for eco-driving). Second, the benefits of those intersections are defaulted to zero due to network decomposition constraint violation. Third, our MT-DRL policies are suboptimal in these intersections. Therefore, the benefits we observe in our analysis are a lower bound for the potential of eco-driving, and future work can address these challenges and increase the benefit levels. 

\subsubsection{Influential factors in eco-driving}

\begin{figure}[H]
    \centering
    \subfloat[Atlanta]{\includegraphics[width=1\textwidth]{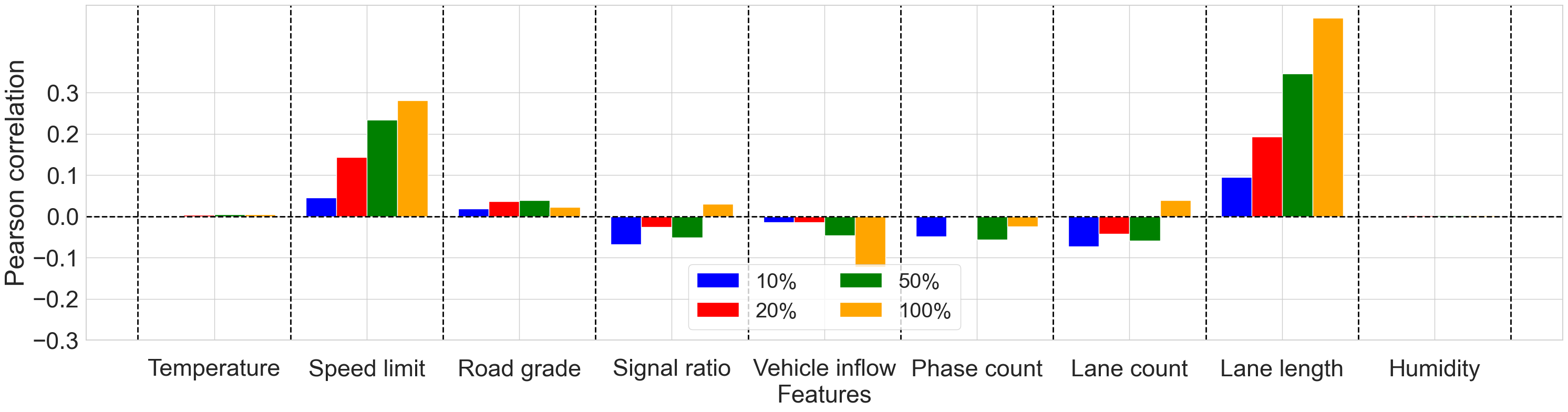}}
    \vspace{0.5cm}
    \subfloat[San Francisco]{\includegraphics[width=1\textwidth]{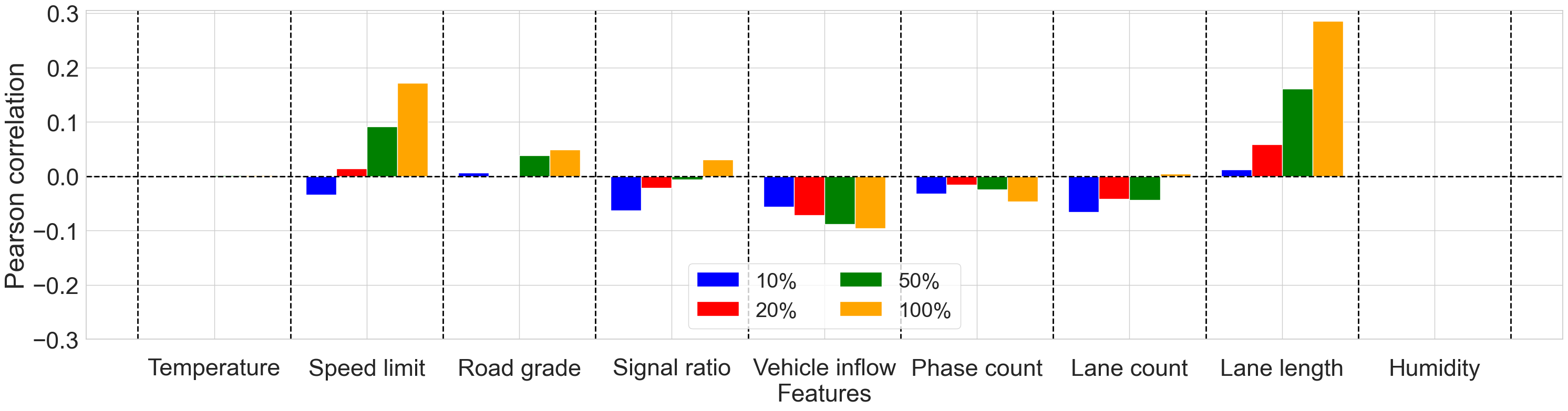}}
    \vspace{0.5cm}
    \subfloat[Los Angeles]{\includegraphics[width=1\textwidth]{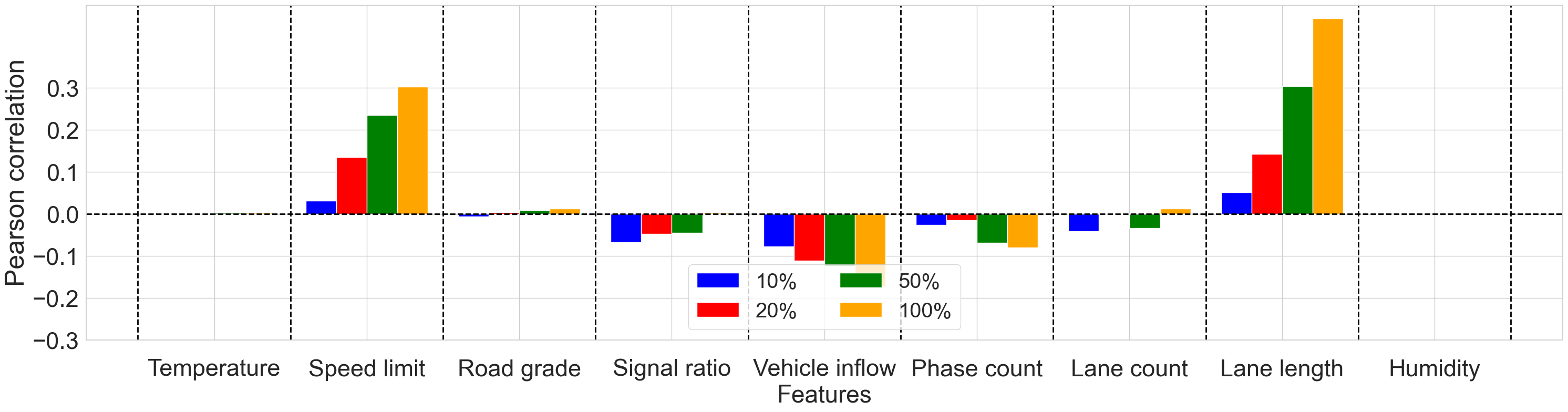}}
    \caption{Correlation analysis of eco-driving factors to the emission reduction benefits in all three cities under different eco-driving adoption levels.}
    \label{fig:pearson-analysis}
\end{figure}

Through a Pearson correlation analysis in Figure~\ref{fig:pearson-analysis}, we underscore key factors influencing emission benefits for informed intersection design and city planning. We consider nine key eco-driving factors that are known to have a significant impact on emission benefits and are directly observable, including temperature, humidity, speed limit, road grade, signal time ratio, vehicle inflow, phase count, lane count, and lane length. The signal ratio denotes the ratio between approach-related phase times and total cycle time. Notably, speed limits, vehicle inflow, and lane lengths significantly impact emission benefits. In particular, when the speed limit and lane length increase, the benefit levels increase. On the other hand, when the inflow increases, the benefit decreases. This explains why the three cities we consider have different benefit levels. As discussed in Appendix Section~\ref{int-count}, different cities have different intersection feature distributions (refer to Figure~\ref{fig:city_feature_distributions} in the Appendix). Thus, one plausible explanation is that cities like Atlanta demonstrate higher benefits, given their higher speed limits and longer incoming approaches, compared to relatively low-benefit cities like San Francisco. When the cities are denser (short incoming approaches), there is less opportunity for eco-driving, and when the speed limits are high, there is more room for vehicle speed changes. However, we note that while this is a plausible explanation, the final benefit levels are determined by the combined influence of multiple interacting factors that we consider. We also note that Atlanta, San Francisco, and Los Angeles typically experience relatively mild variations in average temperature and humidity across the four seasons. Since our analysis is based solely on seasonal average temperatures, this limited variability may explain why these factors do not show a significant influence. We also provide a statistical significance test for the analysis performed in this section in Appendix~\ref{stat-test-factor-correlation}.

Moreover, a noteworthy shift in influential factors is observed depending on the level of eco-driving adoption. In scenarios with lower adoption (10\%), factors such as signal ratio and lane count carry more weight, whereas, at higher adoption levels (100\%), speed limits, vehicle inflow, and lane length become more significant. This observation suggests that optimizing intersections for eco-driving is non-trivial, as benefits may either rise or decline with changing adoption rates over time. Furthermore, the impact of the signal ratio goes from a negative correlation to a positive correlation with the increasing adoption level. This highlights that traffic signal timing guidelines will need to be updated frequently over time with the adoption of eco-driving in order to retain the benefit levels. 

\subsubsection{Where to implement eco-driving}

\begin{figure}[H]
    \centering
    \subfloat[Pareto charts of Atlanta (From left to right: plots under 10\%, 20\%, 50\% and 100\% eco-driving adoption)]{\includegraphics[width=1\textwidth]{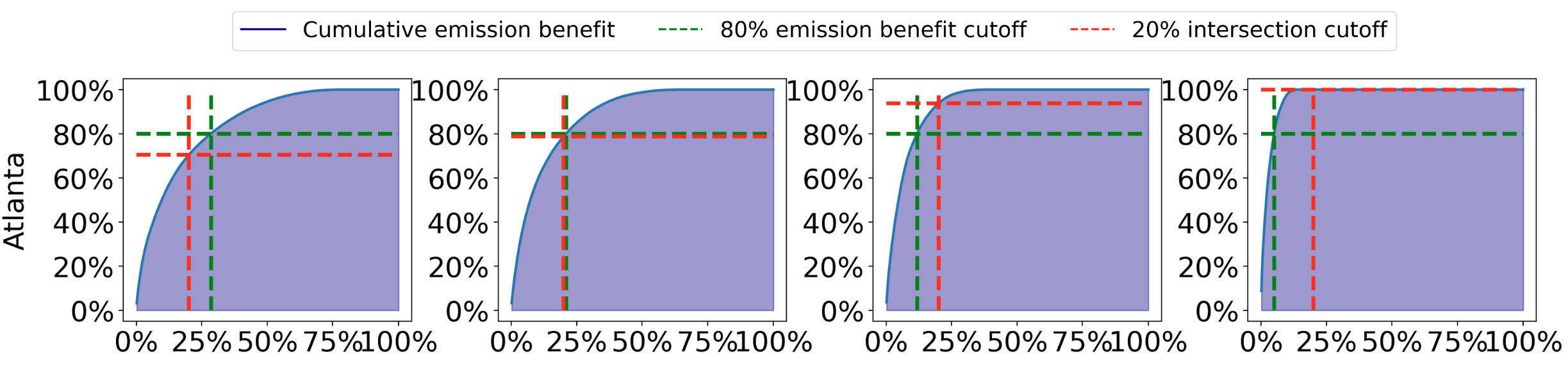}}
    \vspace{0.5cm}
    \subfloat[Pareto charts of San Francisco (From left to right: plots under 10\%, 20\%, 50\% and 100\% eco-driving adoption)]{\includegraphics[width=1\textwidth]{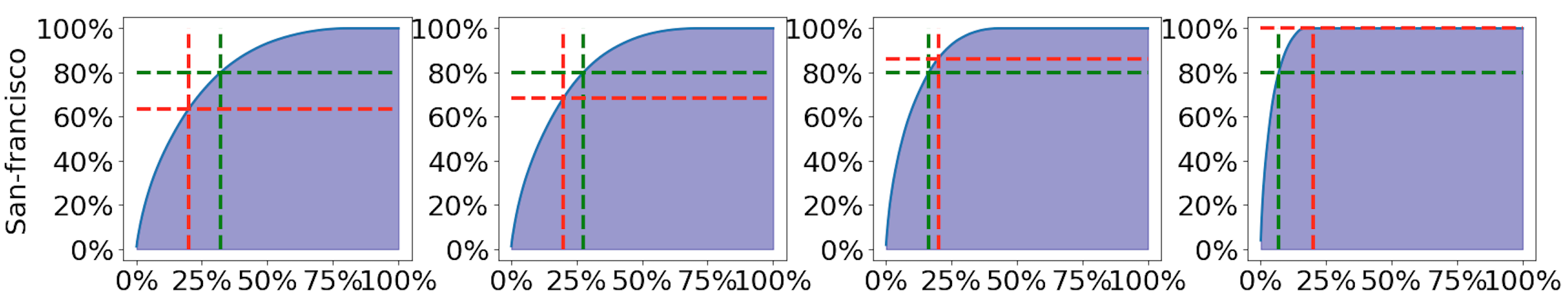}}
    \vspace{0.5cm}
    \subfloat[Pareto charts of Los Angeles (From left to right: plots under 10\%, 20\%, 50\% and 100\% eco-driving adoption)]{\includegraphics[width=1\textwidth]{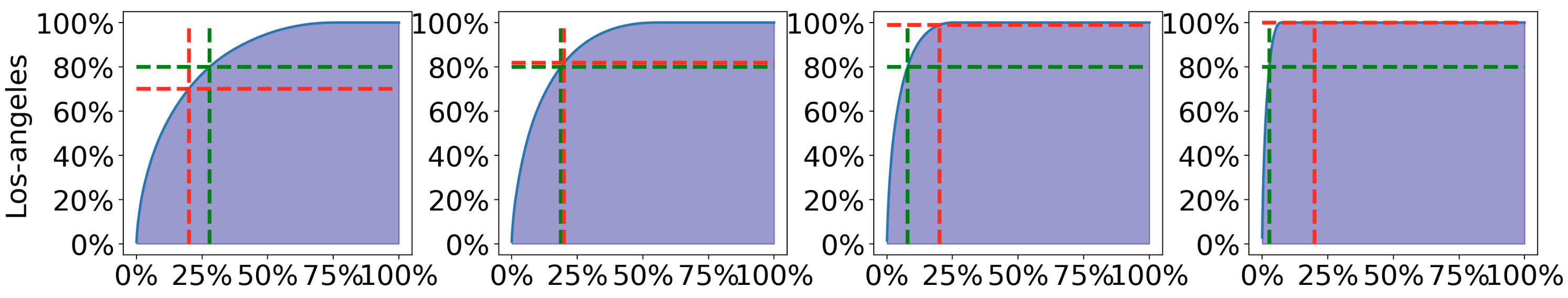}}
    \caption{For each city, the four plots present Pareto charts displaying emission benefits vs intersections in each city at different eco-driving adoption levels. For each plot, the y-axis denotes the cumulative emission benefit percentage, and the x-axis denotes the percentage of intersections that produce a given cumulative benefit level. In plotting the x-axis, the intersections are ordered based on their emission reduction contributions. }
    \label{fig:pareto-charts}
\end{figure}

\begin{figure}[]
    \centering
    \subfloat[Atlanta]{\includegraphics[width=0.33\textwidth]{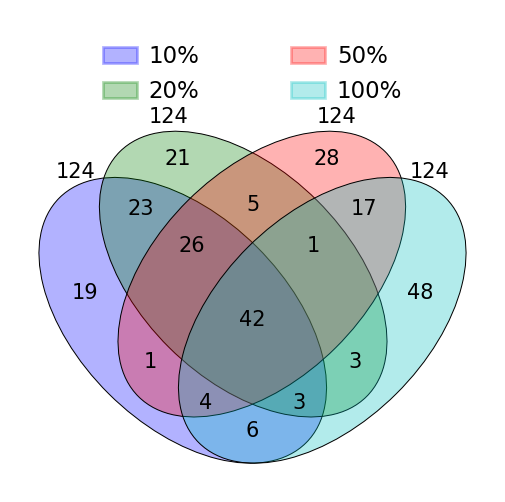}}
    \hfill
    \subfloat[San Francisco]{\includegraphics[width=0.33\textwidth]{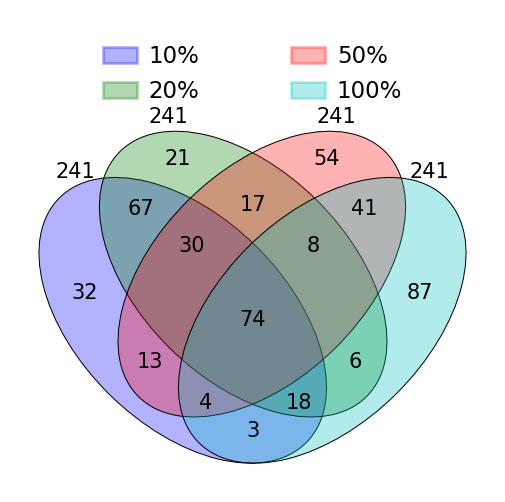}}
    \hfill
    \subfloat[Los Angeles]{\includegraphics[width=0.33\textwidth]{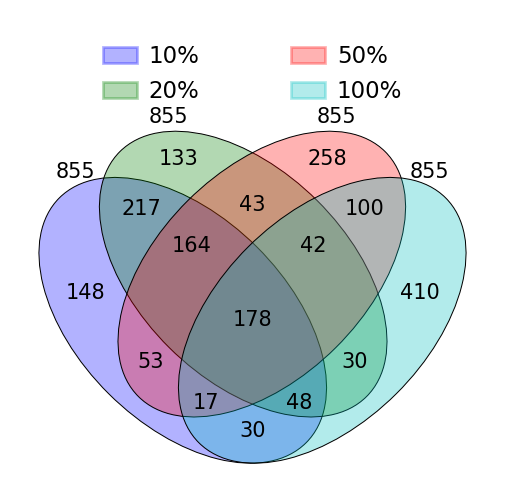}}
    \vspace{0.5cm}
    \subfloat[Atlanta]{\includegraphics[width=0.27\textwidth]{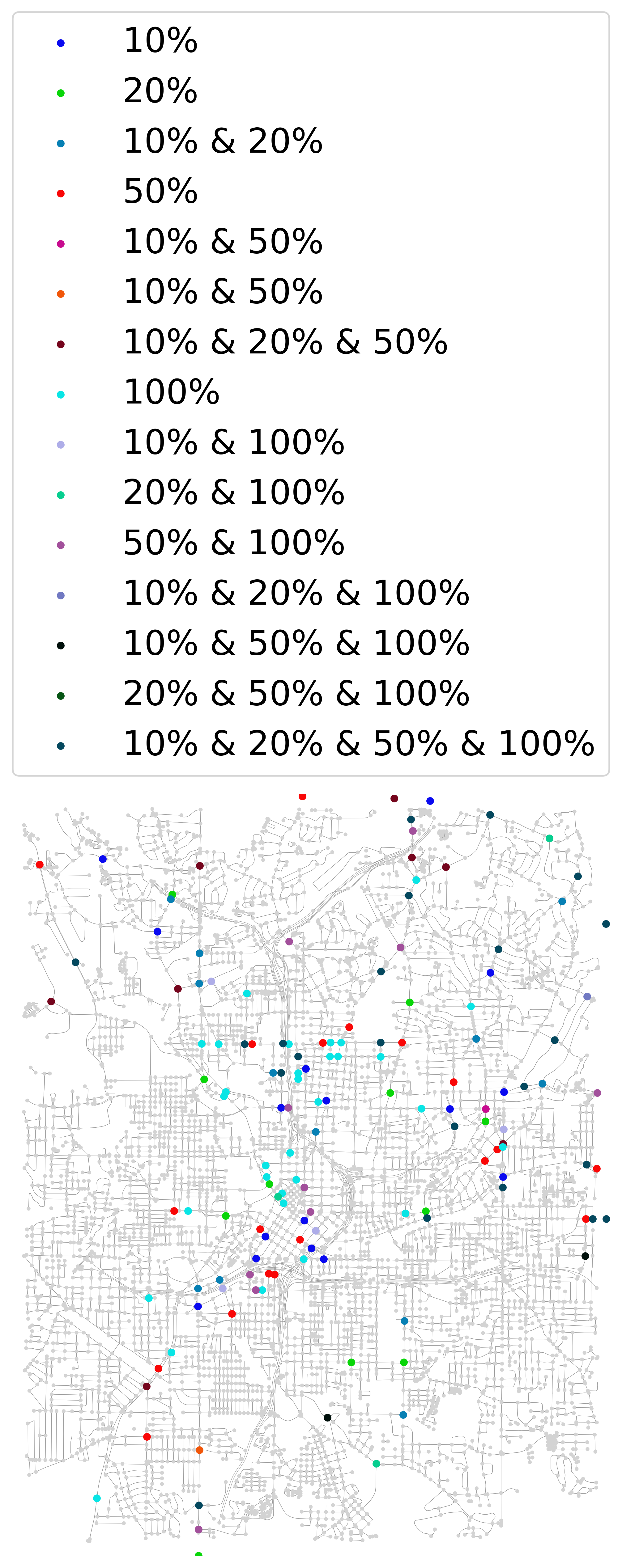}}
    \hfill
    \subfloat[San Francisco]{\includegraphics[width=0.33\textwidth]{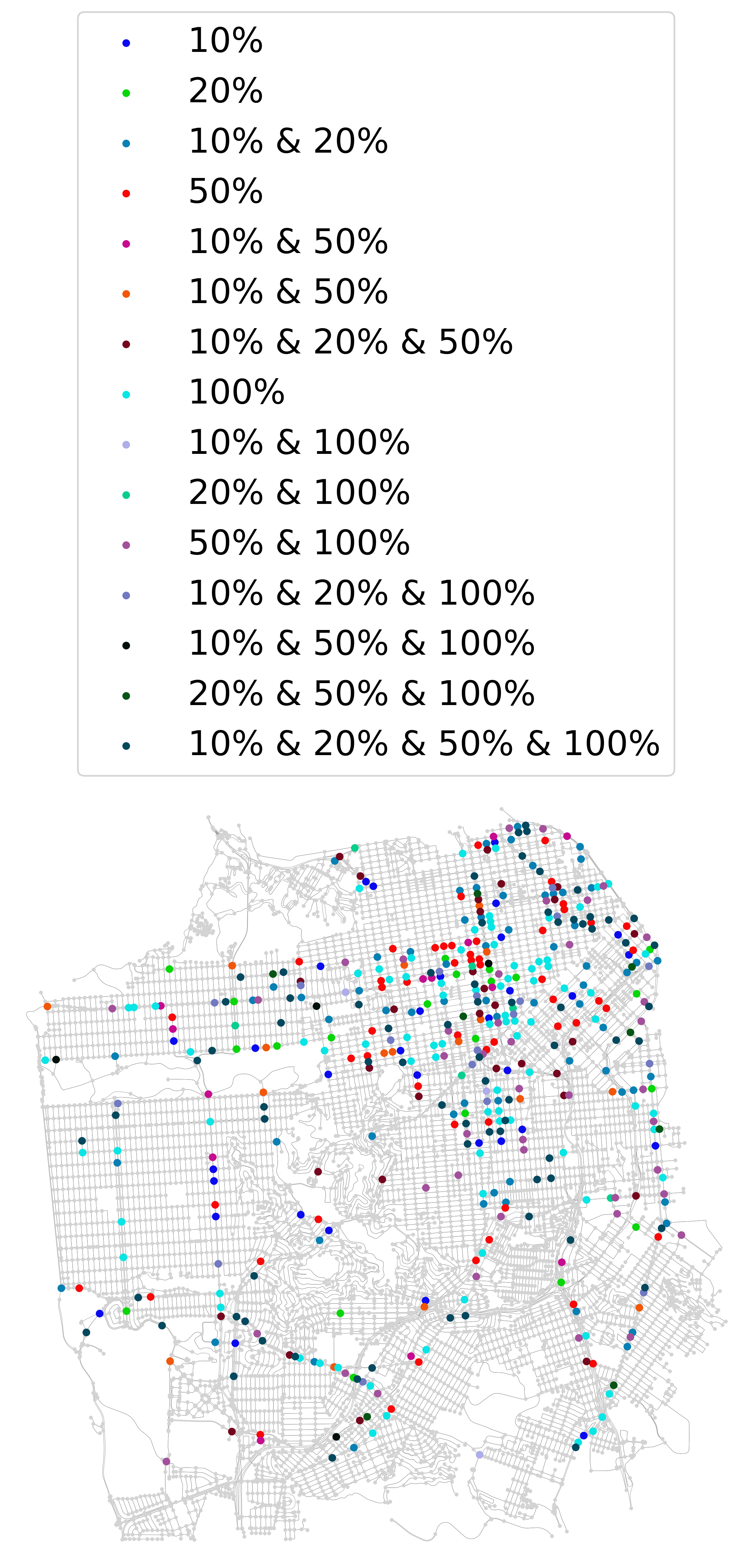}}
    \hfill
    \subfloat[Los Angeles]{\includegraphics[width=0.32\textwidth]{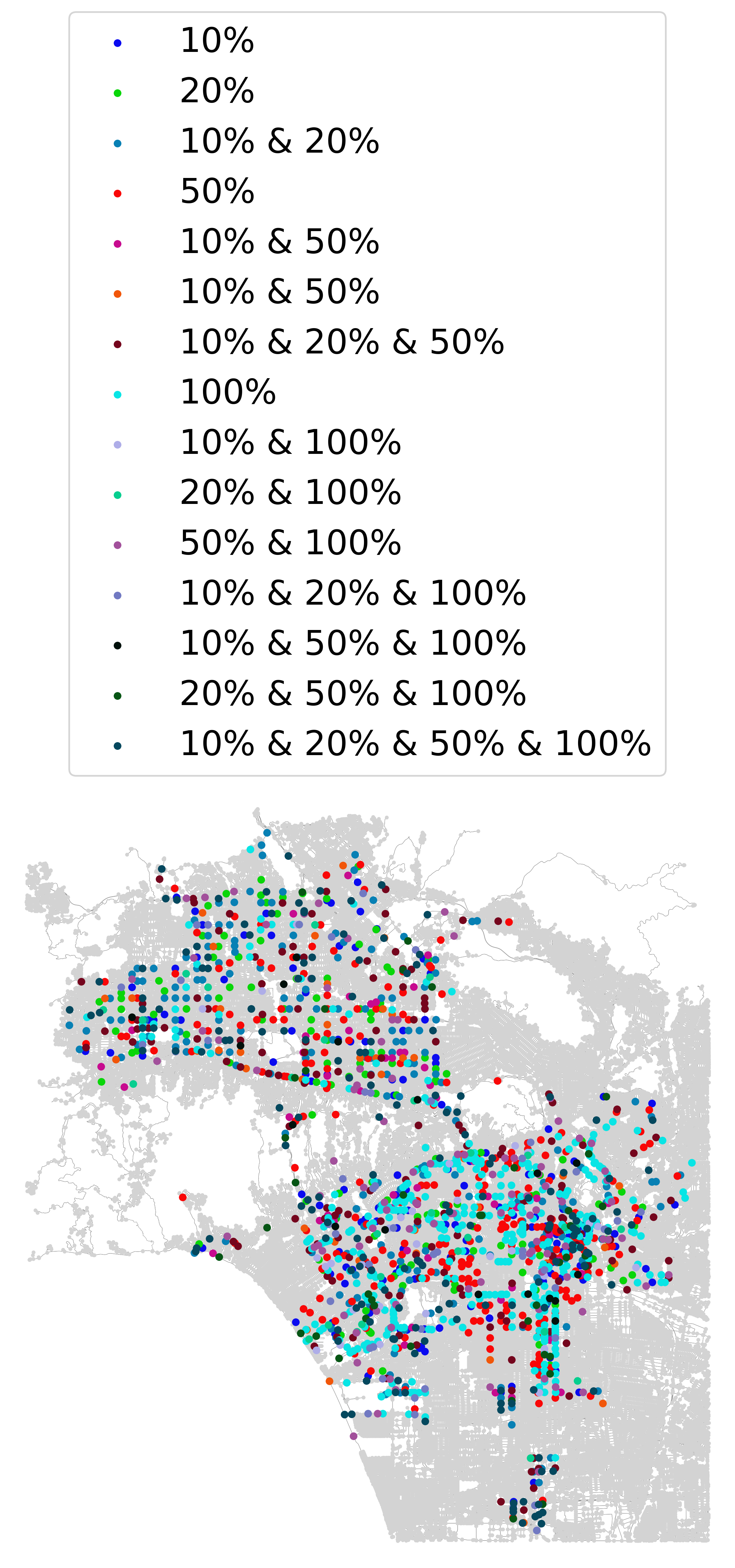}}
    \caption{The a,b, and c plots provide Venn diagrams for the three cities depicting the 20\% of intersections that yield the most significant emission benefits under each eco-driving adoption level. These figures show how intersections evolve with changing adoption levels (the numbers on the figure are the intersection counts). A closer look at the figure underscores the need for meticulous planning when deploying eco-driving, as certain intersections that prove advantageous at lower eco-driving adoption rates may not exhibit the same level of effectiveness when adoption rates are high. Figures d,e, and f show the spatial distribution of the most effective 20\% of intersections and how the overlaps between eco-driving adoption levels vary.}
    \label{fig:venn_diagrams}
\end{figure}

In this analysis, we identify which intersections hold the most promise for eco-driving. This is a key question, as enabling eco-driving may require the installation of roadside units at intersections (for V2I communication). Prioritizing intersections that offer the greatest potential benefits can support a more sustainable and strategic deployment plan.

Figure~\ref{fig:pareto-charts} shows Pareto charts for all three cities, revealing that at least 70\% of emission benefits can be achieved by implementing eco-driving in just 20\% of intersections at each adoption level. This emphasizes the potential for an efficient phased deployment, focusing on a specific subset of intersections. However, the Venn diagram in Figure~\ref{fig:venn_diagrams} illustrates the specific 20\% of intersections delivering the most significant emission benefits at every adoption level may have little overlap between eco-driving adoption levels. It highlights that intersections effective at lower eco-driving adoption levels may not maintain the same efficacy as eco-driving adoption increases, emphasizing the need for careful deployment planning, such as which intersections to be embedded with roadside units (RSU) for Signal and Phase Timing (SPaT) communication. In broad terms, this would question what infrastructure is considered long-term value for eco-driving, as transportation agencies are often constrained by monetary budgets for infrastructure.

\subsubsection{Eco-driving behaviors}

\begin{figure}[!ht]
    \centering
    \begin{subfigure}[b]{0.32\textwidth}
        \centering
        \includegraphics[width=\textwidth]{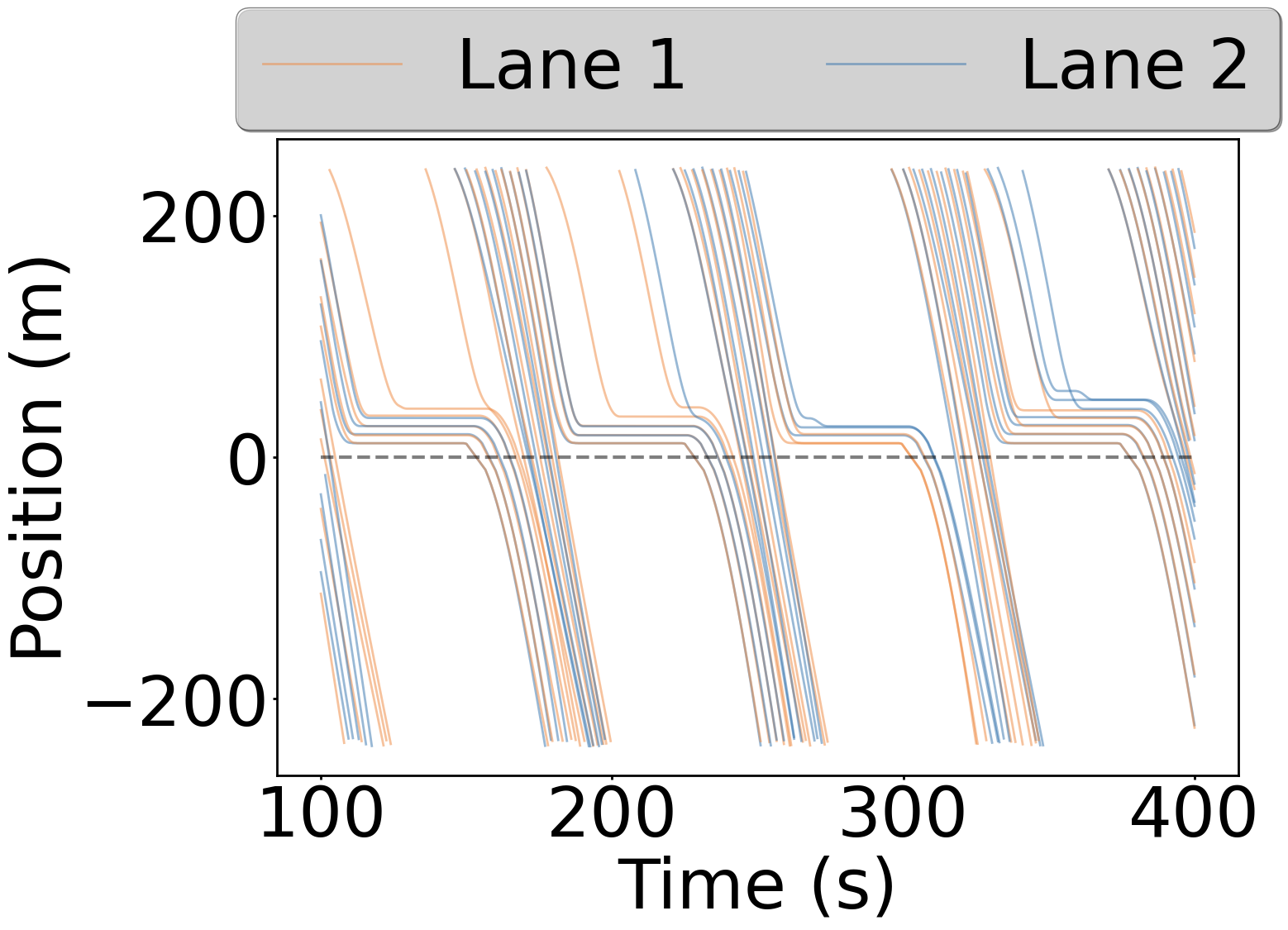}
        \caption{100\% human-driven vehicles}
        \label{R1-1}
    \end{subfigure}
    \hfill
    \begin{subfigure}[b]{0.32\textwidth}
        \centering
        \includegraphics[width=\textwidth]{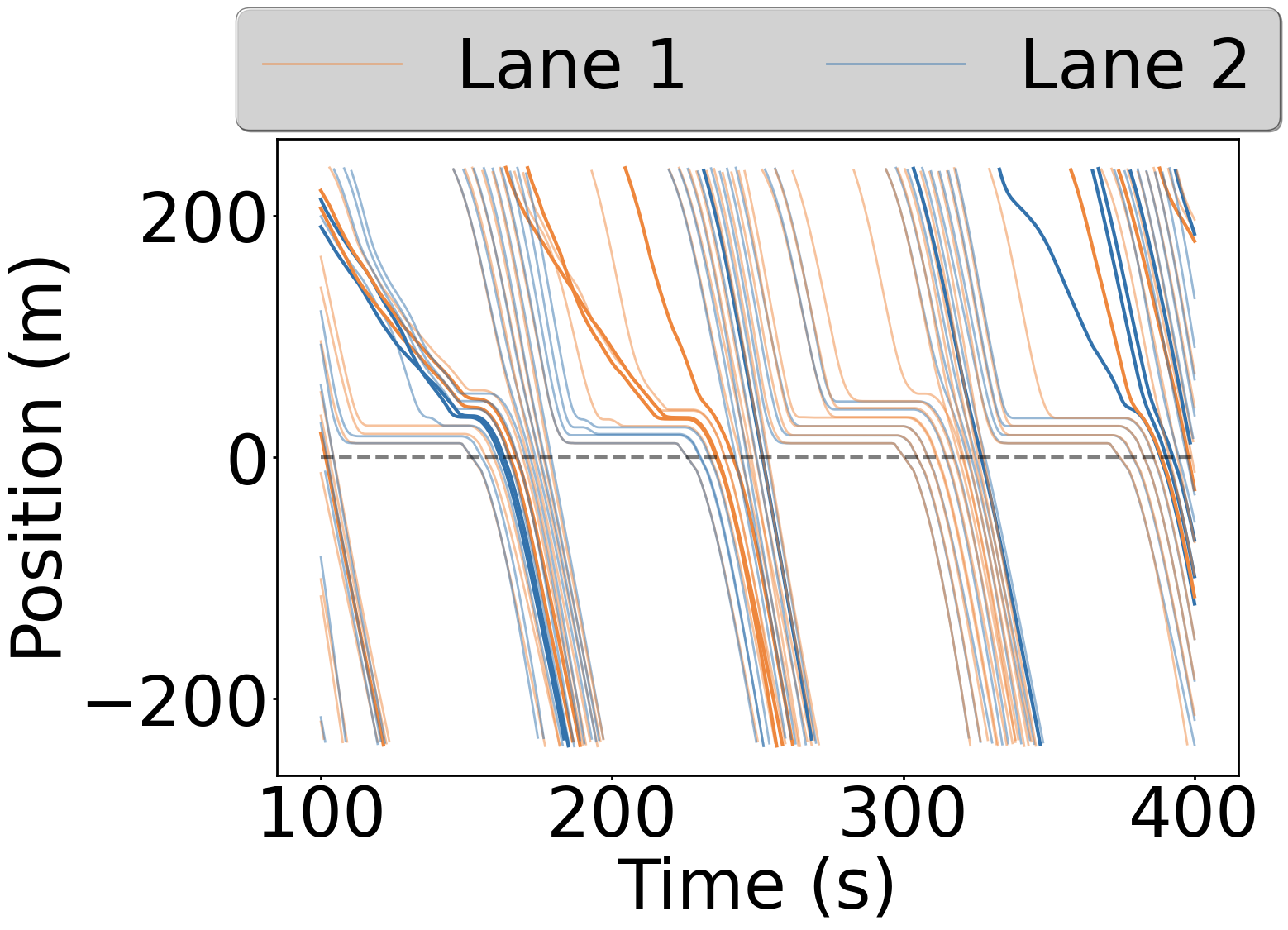}
        \caption{10\% eco-driving with $\pi^1$ policy}
        \label{R1-2}
    \end{subfigure}
    \hfill
    \begin{subfigure}[b]{0.32\textwidth}
        \centering
        \includegraphics[width=\textwidth]{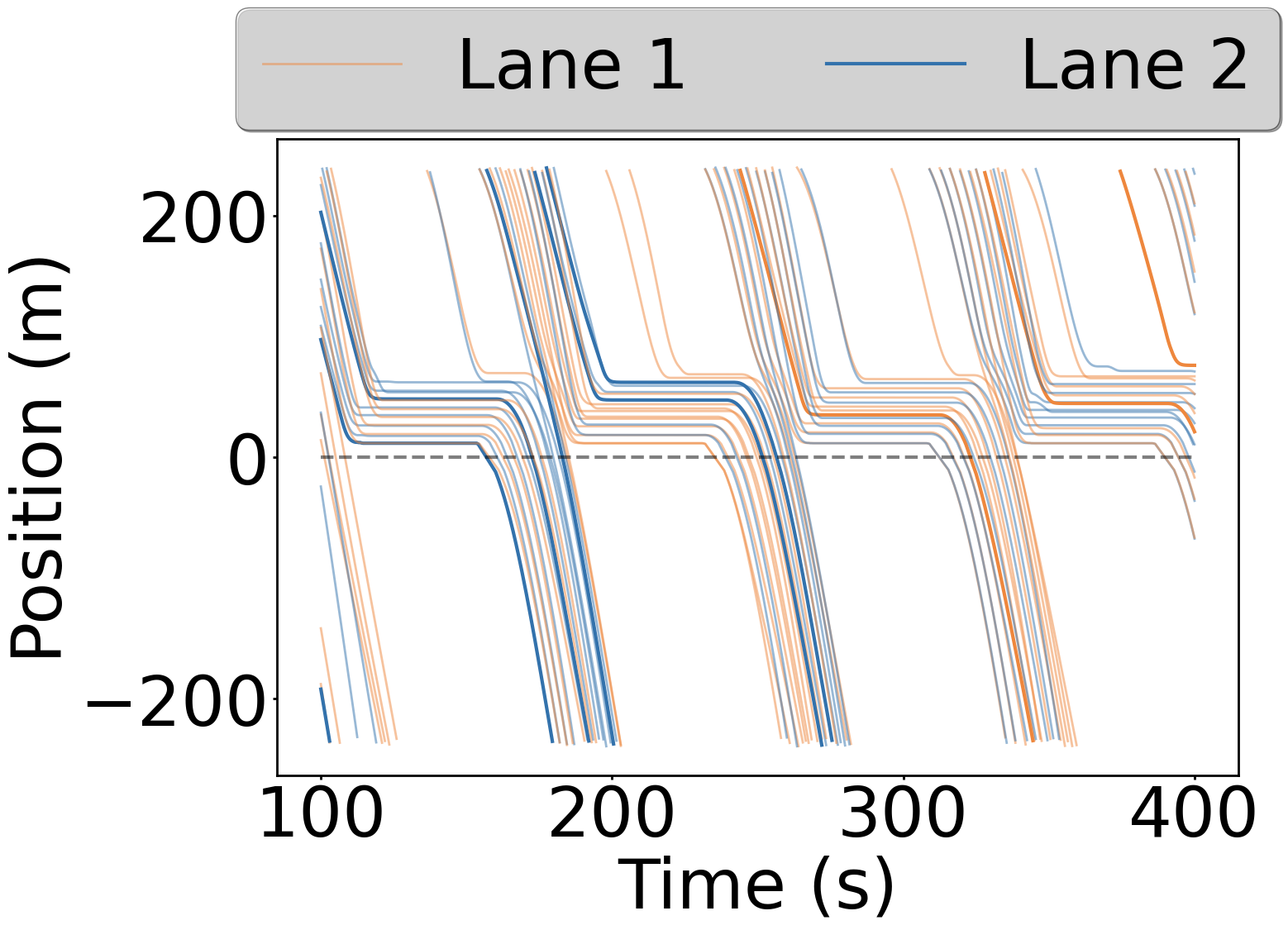}
        \caption{10\% eco-driving with $\pi^2$ policy}
        \label{R1-3}
    \end{subfigure}
    \vfill
    \hspace*{\fill} % Add horizontal space to center the next row
    \begin{subfigure}[b]{0.32\textwidth}
        \centering
        \includegraphics[width=\textwidth]{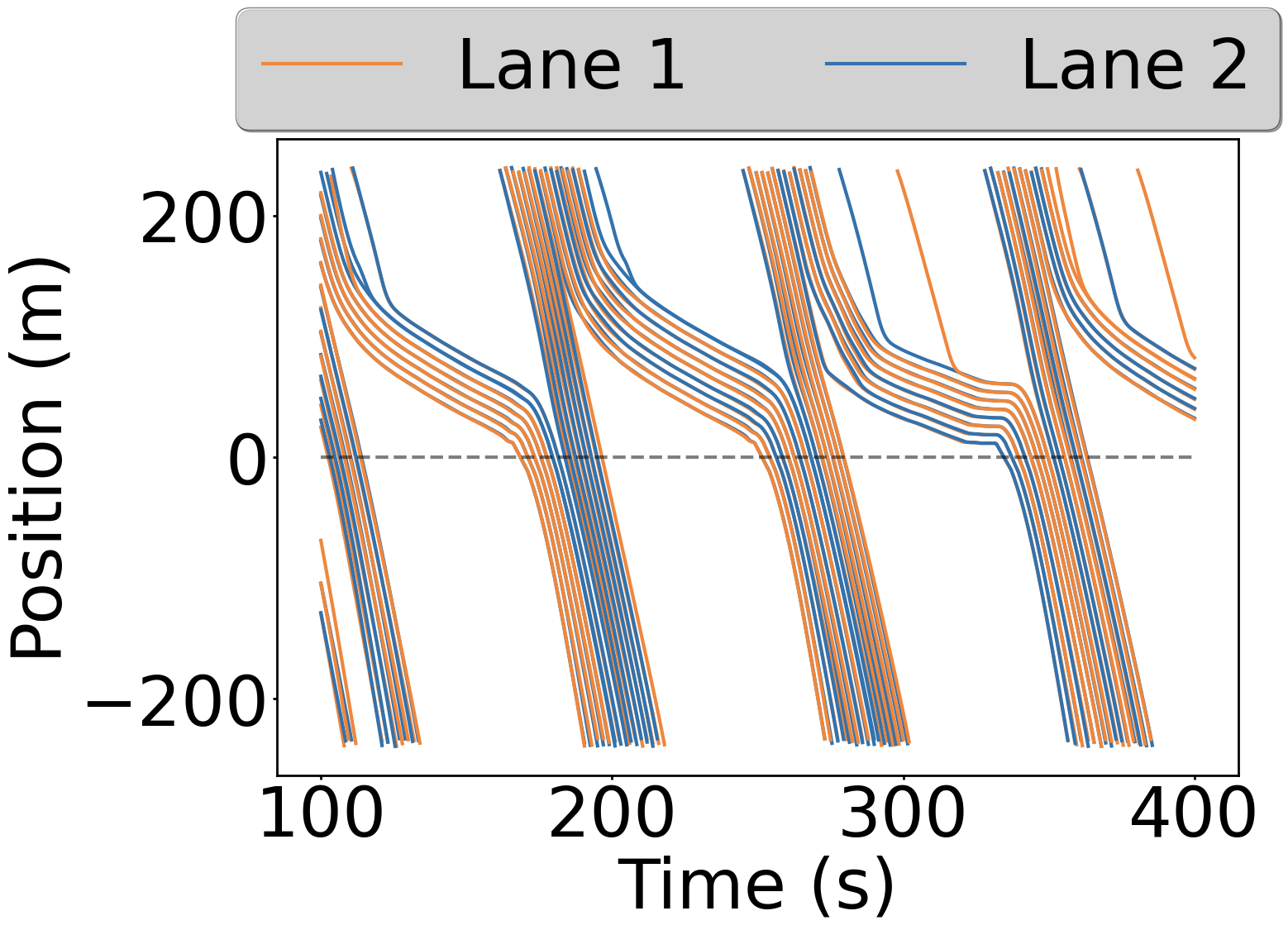}
        \caption{100\% eco-driving with $\pi^1$ policy}
        \label{R2}
    \end{subfigure}
    \hspace{0.05\textwidth} % Add a small horizontal space between the images
    \begin{subfigure}[b]{0.32\textwidth}
        \centering
        \includegraphics[width=\textwidth]{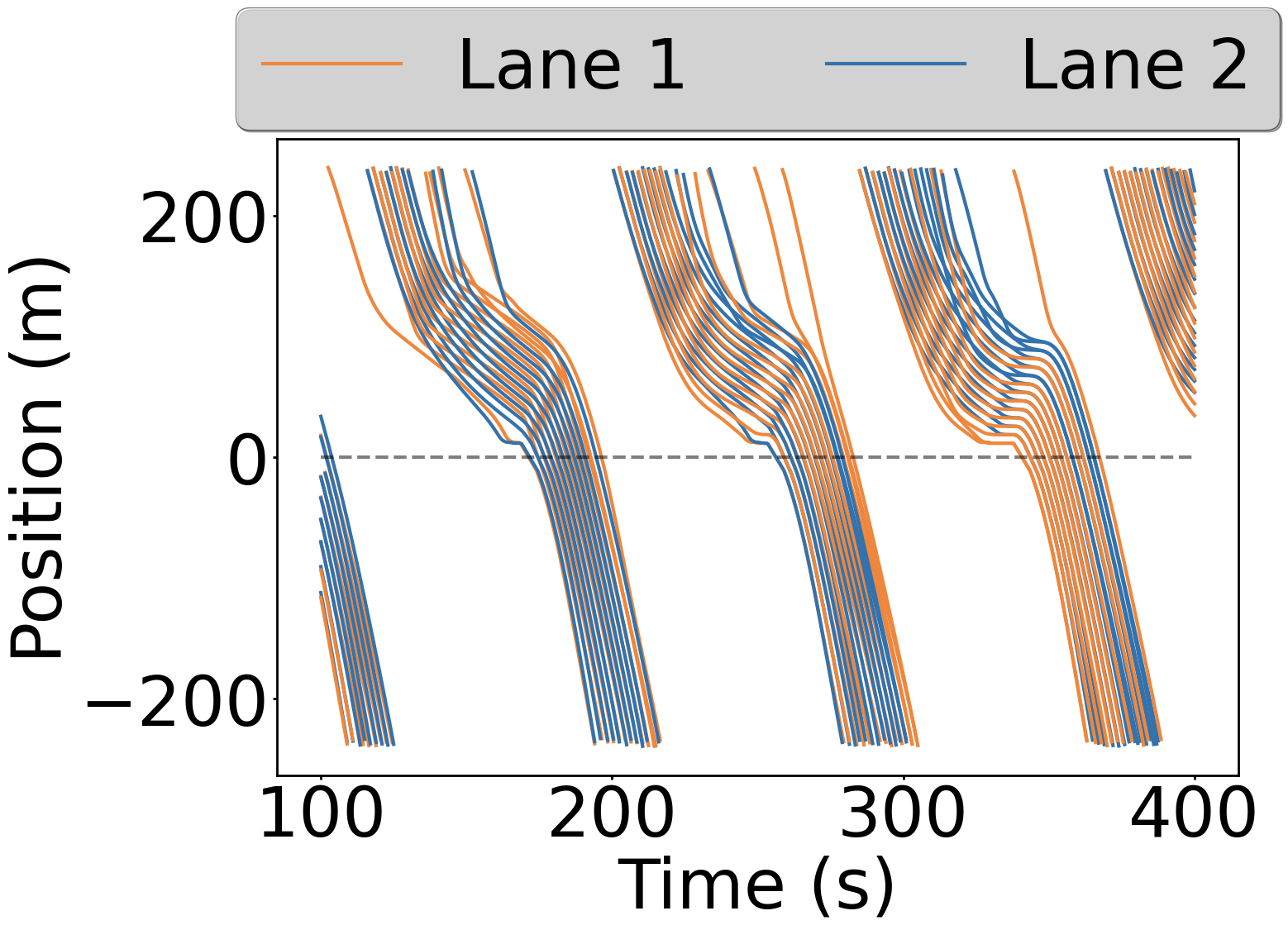}
        \caption{100\% eco-driving with $\pi^2$ policy}
        \label{R3}
    \end{subfigure}
    \hspace*{\fill} % Add horizontal space to center the next row
    \caption{Time-space diagrams of semi-autonomous vehicles (highlighted) and human-driven vehicles (muted) for a selected representative two-lane intersection. The vehicles are driven from 200m relative position ot -200m relative position.}
    \label{fig:time-space}
\end{figure}

In this section, we examine the behavioral changes in vehicles via learned policies compared to human-like driving. In Figure~\ref{fig:time-space}a, a two-lane representative intersection scenario with human-driven vehicles shows typical stopping at a traffic signal, resulting in unwanted idling. In contrast, policy $\pi^1$ in Figures~\ref{fig:time-space}b and \ref{fig:time-space}d exhibits a non-stopping behavior, with most vehicles smoothly passing through the intersection without coming to a halt. Policy $\pi^2$ in Figure~\ref{fig:time-space}e mirrors this behavior but adopts a more human-like approach at lower adoption due to less weight on penalizing stopping, as seen in Figure~\ref{fig:time-space}c.

\begin{figure}[!h]
\centering
  \includegraphics[width=1.0\textwidth]{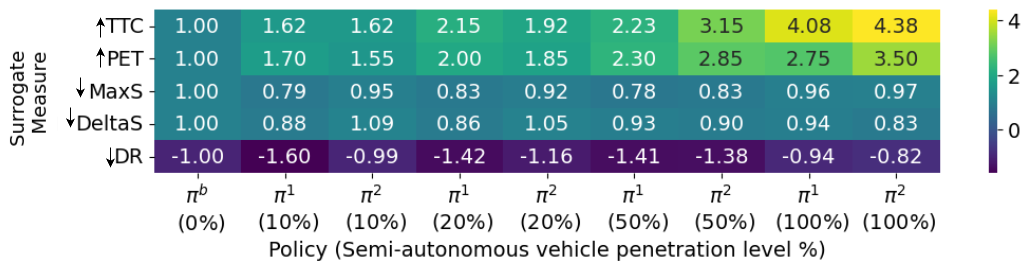}
\caption{The safety of vehicle flow behavior measured using safety surrogate measures (SSM). The score for safety surrogate measures of learned policies in comparison to human-like driving. Each row is normalized by the human driving baseline $\pi^b$. Arrows indicate whether higher or lower is preferred. }
\label{fig:ssm}
\end{figure}

We further analyze the induced road safety metrics of these learned behaviors compared to human-like driving. Figure~\ref{fig:ssm} presents a roadway safety analysis using widely used surrogate safety measures (SSMs) commonly employed in assessing roadway safety~\citep{wang2021review}. We analyze five SSMs, namely: time to collision (TTC), post-encroachment time (PET), maximum speed (MaxS), change in speed (DeltaS), and deceleration rate (DR). Further details on the analysis and used surrogate measures are in Appendix Section~\ref{safety_assessment}. Results in Figure~\ref{fig:ssm} suggest that the learned behaviors are as safe as human-like driving, resulting from smoother vehicle behaviors. However, given that eco-driving behaviors of controlled vehicles are emerging behaviors that may come as counterintuitive for regular human drivers, it may trigger unforeseen human-driver responses, requiring additional human-factor studies for a better understanding of such human factors at play~\citep{10422120}.  

\begin{figure*}[!h]
\centering
  \includegraphics[width=1.0\textwidth]{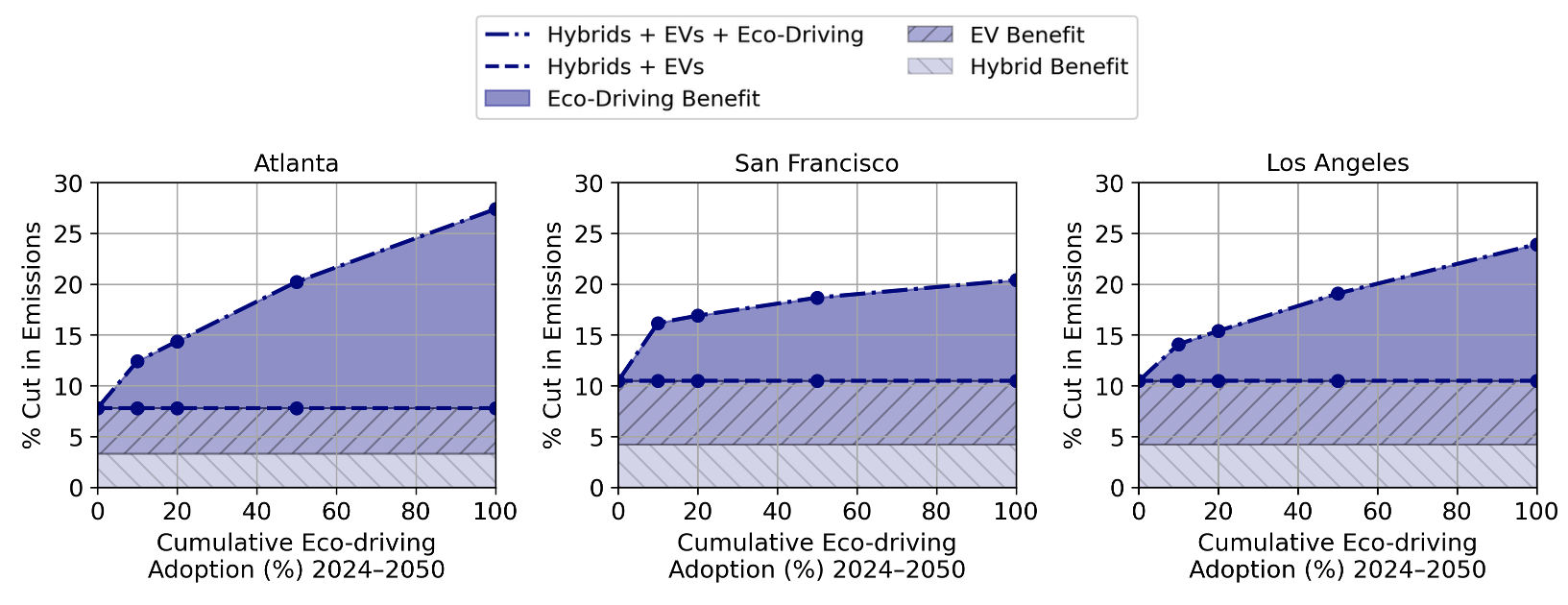}
\caption{Eco-driving benefits in the presence of vehicle electrification and hybrid engines as alternative decarbonization strategies. Figure shows total reductions over 2024-2050 in carbon emissions at various cumulative average eco-driving adoption levels amidst projected electric vehicle and hybrid adoption in each city. Eco-driving yields additional benefits relative to a future with hybridization or electrification alone. For example, with a cumulative average total adoption rate of 20\% from 2024 to 2050, eco-driving alone would cut San Francisco emissions in that period by about 7\%, and when combined with projected electric vehicle and hybrid adoption, would cut emissions by about 17\%. Results are relative to an all internal combustion engine vehicles baseline (i.e., without electrification or hybrid adoption.)}
\label{fig:electric_hybrid}
\end{figure*}

\subsubsection{Role of eco-driving in the presence of alternatives}

Next, we analyze the potential role of eco-driving in the presence of other popular transportation decarbonization strategies. Our analysis reveals a complementary relationship between eco-driving and the growing prevalence of hybrid and electric vehicle adoption. Figure~\ref{fig:electric_hybrid} demonstrates the concurrent reduction in carbon emissions from increased eco-driving and the projected adoption of electric and hybrid vehicles in the US fleet up to the year 2050 for the three cities. Because the electric grid is not carbon-free, energy savings from eco-driving behavior continue to result in carbon emissions reductions even as more electric vehicles are adopted. Further details of the analysis can be found in Appendix Section~\ref{alternatives_and_eco_driving}.

\section{Conclusion}

Eco-driving has long captured the attention of the automotive industry, academics, city planners, and policymakers, yet a comprehensive understanding of its impact has remained elusive for decades, resulting in no large-scale deployments. The multitude of traffic scenario variations at scale has been a major obstacle, impeding progress in the field. In response to this longstanding call for a thorough and multi-factorial study that goes beyond existing efforts, we assess the prospective impact of eco-driving on a scale nearly four orders of magnitude larger than any previous study. Through our comprehensive analysis, we provide insightful perspectives on eco-driving, demonstrating its ability to yield substantial benefits in the pursuit of national climate change mitigation goals. We find the operational plans for eco-driving deployments require careful planning, accommodating the changing benefit dynamics with adoption rates. City planning and policy-making to support eco-driving are also informed through the analysis, especially in concert with other decarbonization strategies. 

\section{Acknowledgments}

This research was partially supported by the Utah Department of Transportation (Project F-ST99(783)), the National Science Foundation (NSF) CAREER award (\#2239566), and an Amazon AWS Machine Learning Research Award. The authors extend their appreciation to Michael Sheffield, Christopher Siavrakas, and Kelly Njord at the Utah Department of Transportation for constructive discussions and providing their expert opinions on the work. Authors also extend their appreciation to Steven Barrett, Samitha Samaranayake, Zuduo Zheng, Yashar Farid, Jinhua Zhao, and Felice Frankel for their constructive comments. The authors also thank Sunera Avinash, Anirudh Valiveru, Jiaxin He, and Dajiang Suo for their help in numerous stages of the project. The authors acknowledge MIT SuperCloud and the Lincoln Laboratory Supercomputing Center for providing computational resources supporting the research results in this paper.

\putbib[main] 
\end{bibunit}

% appendix
\newpage
\begin{bibunit}[plainnat]
\section{Appendix}

\subsection{Back-of-the-envelope calculation}
\label{back-of-the-envelop}

Our initial estimation of emissions resulting from unproductive activities at intersections in the US, derived from preliminary analysis, is supported by data from various sources. To begin, we calculate the proportion of vehicle miles traveled at signalized intersections in the US, using 2019 statistics reported by the Federal Highway Administration~\citep{back-of-the-envelop}. Refer to Table~\ref{miles-travelled} for the detailed calculation. Our analysis indicates that approximately 51.9\% of vehicle miles in the US occur at signalized intersections.

In addition, we refer to 21 previous studies on eco-driving as reported by~\cite{Mintsis2020survey}, which highlight the environmental benefits of adopting eco-friendly driving practices. These studies demonstrate an average eco-driving benefit of 17\%, with a standard deviation of 11\%. Using this information, we estimate that the upper bound for eco-driving benefits at intersections is $51.9\% \times (17\% + 11\%) = 14.6\%$. Consequently, this represents the current portion of CO$_2$ emissions wasted at intersections due to inefficient driving practices.

\vspace{-0.3cm}

\begin{table}[hbt!]
\centering
\begin{tabular}{ |c|c|c|c|c| } 
\hline
Category & Miles(mi) & & \\
\hline
\hline
\textbf{Rural driving} & & &\\ 
principal arterial & 197,630 & &\\ 
minor arterial & 147,457 & &\\ 
major collectors & 164,153 & &\\ 
minor collectors & 42,615 & &\\ 
\makecell{Total rural vehicle \\ miles at intersections} & & 551,855 &\\
\hline
\textbf{Urban driving} & & &\\ 
principal arterial & 483,557 &  &\\ 
minor arterial & 418,151 &  &\\ 
major collectors & 217,681 &  &\\ 
minor collectors & 22,355 &  &\\ 
\makecell{Total urban vehicle \\ miles at intersections} &  & 1,141,744 &\\
\hline
\makecell{Total vehicle-miles \\ in the US } & & 3,261,772 &\\ 
\hline
\hline
\makecell{Intersection driving \\ as a percentage} &  &  & 51.9\% \\ 
\hline
\end{tabular}
\caption{Vehicle miles driven at signalized intersections}
\label{miles-travelled}
\end{table}

\begin{table}[hbt!]
\centering
\small
\begin{tabular}{|p{7cm}|p{2.5cm}|p{2.5cm}|}
\hline
\textbf{Description} & \textbf{Percentage} & \textbf{Value/Note} \\
\hline
\multicolumn{3}{|l|}{\textbf{US Emissions}} \\
\hline
Transportation contribution to total GHG in US~\citep{us_green_house_gas} & 29\% & - \\
Land transportation-related emissions percentage~\citep{us_green_house_gas_fast_facts} & 81\% & - \\
Driving around intersection compared to total land transportation [Table ~\ref{miles-travelled}] & 51.9\% & - \\
Eco-driving emission benefit level from simulations & 11 - 22\% & - \\
Total US CO$_2$ reduction & - & 1.34 - 2.68\% \\
\hline
\multicolumn{3}{|l|}{\textbf{Global Context}} \\
\hline
US CO$_2$ emission in the global scale~\citep{world-co2-emissions} & 12.61\% & - \\
US CO$_2$ reduction as a global reduction & 0.17 - 0.34\% & - \\
A country that contributes 0.17\% to global CO$_2$~\citep{world-co2-emissions} & - & Israel \\
A country that contributes 0.34\% to global CO$_2$~\citep{world-co2-emissions} & - & Nigeria \\
\hline
\end{tabular}
\caption{Eco-driving benefit scaling estimation}
\label{country-level-emission-estimate}
\end{table}

\subsection{Alternative benefit estimation}
\label{equal-benefit-estimation}

Our finding indicates that the total benefits of eco-driving can range from 11\% to 22\% under 100\% eco-driving adoption. We estimate that these emission reduction benefits are equivalent to the total carbon emissions of Israel and Nigeria, respectively. In Table~\ref{country-level-emission-estimate}, we detail how we devise these estimations. We operate under the primary assumption that the eco-driving benefits range derived from evaluating the three metropolitan cities are representative of the entire United States.

\subsection{Eco-driving factors}
\label{eco_driving_factors_related_work}

Our literature review and expert consultation identified 33 factors that are presumed to have an impact on eco-driving. We list these features and their known impact levels in Table~\ref{tab:eco-driving-factors-table1}, \ref{tab:eco-driving-factors-table2}, \ref{tab:eco-driving-factors-table3} and \ref{tab:eco-driving-factors-table4}. In review, we list all factors that have demonstrated impact on eco-driving both at signalized intersections and highway settings. We also list impact levels on fuel consumption, as fuel consumption levels and emission levels are known to have a proportional relationship~\citep{pinto2008using}.

\subsection{Additional details on scenario modeling}
\label{extra_modeling}

For scenario modeling, three major US metropolitan cities, including San Francisco, Los Angeles, and Atlanta, are selected. These cities were chosen to represent both high-emission urban areas and diverse geographical regions~\citep{gately2015cities}. City boundary data was acquired from 2021 TIGER/LINE Shapefiles for zip code tabulation areas (ZCTAs) provided by the US Census Bureau~\citep{city-boundaries}. We locate all zip code regions linked to the city and then concatenate them by taking the union of these areas, resulting in a total of 6,110 identified signalized intersections across the three selected cities. In Tables~\ref{tab:factor_curation_1}, \ref{tab:factor_curation_2}, \ref{tab:factor_curation_3} and \ref{tab:factor_curation_4}, we present specific details on how we treat each of the eco-driving factors in the simulations and what values we consider for scenario pruning and also provide additional details below. 

\begin{table*}[t!]
\centering
\begin{tabular}{ |p{2cm}|p{2cm}|p{2cm}|p{8.5cm}| } 
\hline
\textbf{Factor} & \textbf{Potential values} & \textbf{Modeled values} & \textbf{Explanation} \\
\hline
Vehicle model & Many & Not explicitly modeled & 
While each vehicle model features its own distinct emission model, these models are typically proprietary and not publicly accessible. Consequently, our approach does not involve modeling specific individual vehicle models. Instead, we employ general emission models categorized by vehicle types, like passenger cars, passenger trucks, buses, and trucks. These higher-level emission models are formulated to encompass the collective emission characteristics of various models within each vehicle type.   \\ 
\hline
Vehicle maintenance & Many & Not explicitly modeled & Vehicle maintenance is complex and lacks precise definitions. Our approach involves assuming that vehicles are adequately maintained. To account for wear and tear as well as other effects associated with vehicle aging, we define emission models based on the age of the vehicle. \\ 
\hline
Vehicle characteristics & Many & Engine types: Internal combustion, hybrid, and electric engines & Engine technology can have a significant impact on the emission levels. Therefore we consider engine technologies in our analysis. We do not model other vehicle characteristics as they are often manufacturer or vehicle model-specific.\\ 
\hline
Air conditioning & Many & Not explicitly modeled & Air conditioning usage is at the discretion of passengers and is considered a personal choice. As a result, we do not explicitly model it in our analysis. However, our emission modeling encompasses the average air conditioning usage.  \\ 
\hline
Vehicle load & Many & Not explicitly modeled & Vehicle load is determined by both passengers and the driver based on their preferences and needs. Consequently, we do not explicitly model specific load configurations. However, our emission modeling represents the average emissions associated with varying vehicle loads.  \\ 
\hline
Routing & Many & Intersection-level turning-related routing & Network level routing are not applicable as we model intersections. But it's essential to consider turning (left or right). However, specific turn data for intersections are not publicly available. Therefore, we implement a simple yet reasonable heuristic: we assume that 5\% of inflow vehicles will turn for each available turn lane (left or right). \\
\hline
Traffic demand & Many & Peak and off-peak hour traffic inflows for every intersection in every city & Utilizing realistic traffic flow rates is crucial, considering that intersections are engineered to handle specific traffic volumes. We model traffic demands for representative peak and off-peak hours by leveraging the Annual Average Daily Traffic (AADT) data for each city. Our approach involves applying standardized conversion rates to AADT data to effectively model both peak-hour and off-peak-hour traffic demands.\\
\hline
Aerodynamical drag & Many & Not explicitly modeled  & Aerodynamic drag, in comparison to other influential factors, has a relatively minor effect. Moreover, its significance becomes prominent mainly in highly automated platooning configurations. Since our current setup doesn't specifically replicate precisely defined platooning scenarios, aerodynamic drag is not a primary focus in our analysis.\\
\hline
\end{tabular}
\caption{\textbf{Summary of how each eco-driving factor is modeled to build the representative scenario tree (1/4 tables).} Factor defines the eco-driving factor that is presumed to have an impact on the overall emission benefits. Potential real-world values indicate what range of values can the given factor take in the real-world. Modeled values indicate our curated value selection used to construct the scenario tree. Finally, we provide a detailed explanation for each factor curation based on related literature and expert opinions.  }
\label{tab:factor_curation_1}
\end{table*}

\begin{table*}
\centering
\begin{tabular}{ |p{2cm}|p{2cm}|p{2cm}|p{8.5cm}| } 
\hline
\textbf{Factor} & \textbf{Potential values} & \textbf{Modeled values} & \textbf{Explanation} \\
\hline
Road grade & Many & Same as real-world & 
Road grades play a crucial role in emissions due to their direct influence on the energy needed to propel a vehicle. To account for this impact, we leverage U.S. geological data to accurately model the road grade for each approach at every intersection.  \\
\hline
Road type & Mainly paved roads & Not explicitly modeled  & Our assumption considers that all roads in our analysis are paved, leading to reduced friction from road materials, especially at lower speeds typical of intersections as opposed to higher speeds experienced on highways.\\
\hline
Speed limit & Many (within a known range) & Same as real-world & We incorporate real-world speed limits obtained from Open Street Maps and employ established imputation methods for handling missing values. This allows us to model realistic speed limits for each approach at intersections. \\
\hline
Number of lanes & One to seven & Same as real-world & We model the lane count for each intersection by leveraging real-world data sourced from Open Street Maps. Employing proven imputation methods for handling any missing values, we ensure a reasonable representation of the actual number of lanes in our modeling. \\
\hline
Turn lane configuration & One to five turn lanes & Same as real world & We model each intersection with real-world turn lane configuration using data from Open Street Maps and known imputation methods for missing values. \\
\hline
Lane length & Many & Same as real-world with a maximum length of 750m & Real-world lane lengths are used based on data from Open Street Maps and known imputation methods for missing values. \\
\hline
Traffic signal timing & A few plans for a given intersection & Optimal plan for the controlled intersection & Adjusting traffic signal timings present an extensive range of possibilities, particularly when factoring in ghost intersections. This expansive combinatorial nature requires significant tuning time. To streamline the process and improve efficiency, we conduct an exhaustive search within fixed-time traffic signal control plans for each control active intersection, ensuring the ghost intersection plans are set to default viable programs.\\
\hline
Pedestrians & Many behaviors & Not explicitly modeled & We operate under the assumption that pedestrians strictly adhere to road rules, eliminating any potential interference with vehicles. This assumption allows us to model scenarios where pedestrians consistently follow traffic regulations and refrain from jaywalking. \\
\hline
Cyclists & Many behaviors & Not explicitly modeled & We operate under the assumption that cyclists consistently adhere to road rules and exclusively utilize designated bike lanes. \\
\hline
Nearby intersections & Intersections form a network & Immediate one-hop intersections connecting every incoming approach (ghost intersections) & While eco-driving analyses often rely on uniform inflow rates, real-world traffic flows at intersections rarely exhibit uniformity and can be influenced by nearby intersections. Thus, we incorporate a modeling approach that reflects the dynamic nature of traffic, particularly the one-hop neighbors of each intersection. This strategy allows for a more realistic representation of vehicle inflow at the control active intersection. \\
\hline
\end{tabular}
\caption{\textbf{Summary of how each eco-driving factor is modeled to build the representative scenario tree (2/4 tables).} Factor defines the eco-driving factor that is presumed to have an impact on the overall emission benefits. Potential real-world values indicate what range of values can the given factor take in the real-world. Modeled-values indicate our curated value selection used to construct the scenario tree. Finally, we provide a detailed explanation for each factor curation based on related literature and expert opinions.  }
\label{tab:factor_curation_2}
\end{table*}

\begin{table*}
\centering
\begin{tabular}{ |p{2cm}|p{2cm}|p{3cm}|p{7.5cm}| } 
\hline
\textbf{Factor} & \textbf{Potential values} & \textbf{Modeled values} & \textbf{Explanation} \\
\hline
Temperature & Many & Average temperatures under each season (summer, fall, spring, and fall) and under each weather condition (clear, rainy, snow) in each season under both peak and off-peak hours & Temperature values can vary across a wide spectrum, yet within a specific city and prevailing season or weather conditions, they typically fall within a defined range. \\
\hline
Humidity & Many & Average humidities under each season (summer, fall, spring, and fall) and under each weather condition (clear, rainy, snow) in each season under both peak and off-peak hours & Humidity values can vary across a wide spectrum, yet within a specific city and prevailing season or weather conditions, they typically fall within a defined range. \\
\hline
Weather conditions & Clear, snow, and rain & Clear, snow, and rain & In our modeling approach, we represent each weather condition by considering alterations in temperature and humidity levels. It's important to recognize that various weather conditions can influence driving behaviors, potentially due to slippery roads or reduced visibility. However, under low-speed limits typical at signalized intersections, we assume that these influences are not significant.\\
\hline
Seasonal effects & Summer, fall, winter, spring & Summer, fall, winter, spring & We model each season as a change in temperature and humidity levels. The season may also cause changes to vehicle inflows. We assume this impact is not significant at the level of intersections.\\
\hline
Vehicle age distribution & Up to 30+ years & 1-10 years in increments of 1 year, and all other vehicles as vehicles with age 10+ years & Vehicle age can span up to 30+ years; however, a significant portion typically falls within the range of 1-10 years in the U.S. Although age distribution may vary slightly across different cities, we streamline our modeling process by utilizing the MOVES database to extract the national-level vehicle age distribution. \\
\hline
Vehicle type distribution & Passenger cars, passenger trucks, buses, trucks, vans, pickup trucks, sports cars, and more & Passenger cars, passenger trucks, buses, trucks & While various vehicle sub-types exist, a significant majority in the US can be broadly categorized into passenger cars, passenger trucks, buses, and trucks. Although the distribution of vehicle types may exhibit slight variations across different cities, we streamline our modeling approach by employing the MOVES database to extract the vehicle type distribution at the national level. \\
\hline
\end{tabular}
\caption{\textbf{Summary of how each eco-driving factor is modeled to build the representative scenario tree (3/4 tables).} Factor defines the eco-driving factor that is presumed to have an impact on the overall emission benefits. Potential real-world values indicate what range of values can the given factor take in the real-world. Modeled-values indicate our curated value selection used to construct the scenario tree. Finally, we provide a detailed explanation for each factor curation based on related literature and expert opinions.  }
\label{tab:factor_curation_3}
\end{table*}

\begin{table*}
\centering
\begin{tabular}{ |p{2cm}|p{2cm}|p{2cm}|p{8.5cm}| } 
\hline
\textbf{Factor} & \textbf{Potential values} & \textbf{Modeled values} & \textbf{Explanation} \\
\hline
Driver mix distribution & Many distributions based on city & A single distribution based on real-world driving data at a signalized intersection in Florida. & 
Driver demographics can vary from one city to another, potentially affecting driving behaviors. However, at the intersection level, these variations may not exhibit significant differences compared to highways. To enhance the simplicity and accuracy of our modeling, we utilize real-world driving data obtained from a signalized intersection in Florida. We devise driver mix distributions separately for passenger cars, passenger trucks, busses and trucks. Additionally, we employ well-established methods to calibrate human driver behavior, ensuring a realistic representation within our models. \\
\hline
Fuel type distribution & Gasoline, diesel, ethanol, biodiesel, propane (liquefied petroleum), compressed natural gas, electricity, hydrogen & Gasoline, diesel & Among the various fuel types available, the predominant vehicles on the roads primarily run on gasoline, diesel, or electricity.\\
\hline
Semi-autonomous vehicle penetration & Any value between 0\% to 100\% & 10\%, 20\%, 50\% and 100\% & 
Our objective is to model different levels of penetration—ranging from lower levels (10\%, 20\%), a medium level (50\%), to a higher level (100\%)—to comprehensively grasp the overall impact at each penetration level. This approach allows us to identify patterns and trends in benefits associated with varying levels of penetration.\\
\hline
Idling & vehicles idle at intersections due to traffic signals & minimize idling & Idling is known to increase vehicle emissions, and therefore we aim to reduce idling.\\
\hline
Speed & Speeds under the speed limits & Speeds under the speed limits & We attempt to make the speed profiles smoother as that has been known to reduce emissions.  \\
\hline
Acceleration & Within the actuator limits of the vehicles & Within the actuator limits of the vehicle & Our goal is to minimize accelerations, given the established knowledge that it effectively leads to a reduction in emissions. \\
\hline
Deceleration & Within the actuator limits of the vehicles & Within the actuator limits of the vehicles & We aim to reduce deceleration as that can lead to stop and go waves.\\
\hline
Pulse and glide & Not a popular driving strategy & Pulse and glide when approaching the intersection & We aim to perform pulse and glide when approaching the intersection as it helps reduce emissions and enable platooning. \\
\hline
Lane changing & Many (overtaking, turning, speed gain, keep right, etc) & Same as real-world & We employ a popular rule-based SUMO lane changing model to simulate lane changes for both semi-autonomous and human-driven vehicles.\\
\hline
\end{tabular}
\caption{\textbf{Summary of how each eco-driving factor is modeled to build the representative scenario tree (4/4 tables).} Factor defines the eco-driving factor that is presumed to have an impact on the overall emission benefits. Potential real-world values indicate what range of values can the given factor take in the real-world. Modeled-values indicate our curated value selection used to construct the scenario tree. Finally, we provide a detailed explanation for each factor curation based on related literature and expert opinions.  }
\label{tab:factor_curation_4}
\end{table*}

\textbf{Intersection modeling.} 
To capture the effect of most of the intersection-related factors, we model them in micro-simulations. We use the well-known open-source SUMO microscopic traffic simulator~\citep{SUMO2018} for this purpose. Our primary data source for intersection-level details is the OpenStreetMap (OSM)~\citep{mordechai2008osm}, from which we extract real-world intersection data, including lane lengths, lane counts, turn lane configurations, and speed limits. Road grades are obtained from US geological surveys~\citep{USGS}. The OSM data is processed to create corresponding digital replicas of intersections within SUMO. 

However, while generally reliable, OSM does contain missing values in its data. \cite{Qu2022OSMint} address this limitation and provide guidelines to impute missing values and build realistic signalized intersections from OSM data. We follow the same pre-processing steps recommended. However,~\cite{Qu2022OSMint} caution that the pre-processing steps may not fully recover the real-world conditions for certain intersections. As such, we acknowledge that our modeling may inherit these limitations.

To model vehicle inflows, we leverage Annual Average Daily Traffic (AADT)~\citep{HUNTSINGER202217}, which is the most reliable openly available arterial traffic flow data for all three cities. These datasets are published by the Department of Transportation of each state/city. Since these AADT counts are link-based, we are able to directly assign the relevant count to each incoming approach at the intersection, eliminating the need for an OD calibration. The quality of AADT data varies across states. In the state of California, only a limited number of streets have recorded data, necessitating reliable estimation methods. For them, we formulate a flow decomposition problem isnpired by popular algorithms for origin-destination matrix and route estimation methods~\citep{}, treating the available flow information for each road as the expected flow rather than its capacity. Algorithm~\ref{graphsimp} describes this formulation. 

\begin{table*}[!ht]
\begin{tabular}{ |p{3cm}|p{12cm}| } 
\hline
\textbf{Factor} & \textbf{Description}  \\
\hline
Vehicle model & The model of the vehicle used for eco-driving greatly affect the benefits. \cite{sivak2012eco} demonstrate varying fuel economies among vehicle classes (e.g., cars, pickup trucks), with the best cars being nine times and the best pickup trucks being two times more fuel efficient than their worst-case models. \\ 
\hline
Vehicle maintenance & Vehicle maintenance mainly covers engine and tire maintenance apart from other maintenance required. Maintaining vehicles with tuned engines can improve their gas mileage by an average of 4\%~\citep{epa_website}. A 10\% change in the nominal rolling resistance of tires could cause a 1-2\% change in fuel economy~\citep{fueleconomy_website}. Also, a one psi drop in tire inflation can cause a 0.3\% reduction in fuel economy~\citep{fueleconomy_website}. \\ 
\hline
Vehicle characteristics & Vehicle characteristics encompass many factors from engine technology to cabin technologies to overall built. However, a main characteristic is the engine technology. Internal Combustion Engine (ICE) vehicles can achieve 15-25\% of fuel efficiency gains through driving style adjustments~\citep{fafoutellis2020eco}. Hybrid and electric vehicles, benefiting from kinetic energy regeneration during braking, are more responsive to driving style changes~\citep{xu2021overview}. Hybrid electric vehicles show a 12.6\% improvement, and electric vehicles show 11.7-18.4\% benefit from eco-driving, highlighting the considerable advantages of eco-driving across vehicle technologies~\citep{aims}. \\ 
\hline
Air conditioning & Air conditioning in vehicles is a major contributor to increased fuel consumption, being the largest auxiliary load~\citep{farrington2000impact}. When turned on, it uses engine energy to power the cooling compressor, leading to higher fuel usage. Studies suggest a 50\% reduction in air conditioning energy could save 3.6 billion gallons (13.5 billion liters) of fuel in the US, cutting oil imports by 5\%~\citep{johnson2002fuel}. The choice between air conditioning and windows-down for ventilation impacts fuel efficiency; windows-down is more efficient at low speeds, but at higher speeds, air conditioning is preferred due to aerodynamic drag and its effect on fuel consumption~\citep{huff2013effects}. \\ 
\hline
Vehicle load & The weight added to a vehicle, including passengers, cargo, and vehicle itself, impacts the power needed to overcome increased resistance. Just 45 kg of extra weight can increase fuel consumption by 1-2\%, especially pronounced in smaller vehicles~\citep{alam2014critical, fueleconomy_website}. Since 1988, annual fuel consumption has risen by at least 272 million gallons due to average passenger weight increases. When measured from 1960, this amounts to about 938 million gallons annually, approximately 0.7\% of the nation's yearly fuel consumption or nearly three days of fuel consumption by automobiles~\citep{jacobson2006economic}. \\ 
\hline
Routing & Route choice can impact the overall fuel saving or emission levels. Studies suggest that eco-routing enables 12-33\% fuel improvement~\citep{dhaou2011fuel, zhou2016review}. However, eco-routing has the risk of increased travel time due to taking alternative routes~\citep{zeng2017application, zeng2016prediction}.
 \\ 
\hline
Traffic demand & Fuel economy reduction in vehicles can range from 20-40\%  when transitioning from the free flow to forced or breakdown flow on highways, with a significant drop in fuel economy when flow becomes unstable~\citep{facanha2009effects}. \cite{barth2009energy} observe that there is little benefit under free-flow conditions, while the savings are considerable in severe congestion. In contrast, \cite{kobayashi2007eco} show that in a network of two signalized intersections, the emission levels drop significantly when the congestion is high as the eco-driving vehicle accelerates too slowly such that it would hold back the following vehicles, causing congestion and therefore emitting more CO$_{2}$.  \\ 
\hline
Aerodynamical drag & Air resistance increase engine workload and are influenced by air pressure, density, wind speed, and direction. Open windows and additional exterior vehicle parts can raise fuel consumption by 20\% at high speeds due to aerodynamic drag~\citep{huang2018eco}. Roof and rear cargo boxes can further decrease it by 2-25\%, depending on driving conditions~\citep{fueleconomy_website}.\\ 
\hline
\end{tabular}
\caption{Summary of the known impact of each factor on eco-driving emission benefits (1/4 tables).}
\label{tab:eco-driving-factors-table1}
\end{table*}
\clearpage

\begin{table*}[!ht]
\centering
\begin{tabular}{ |p{3cm}|p{12cm}| } 
\hline
\textbf{Factor} & \textbf{Description}  \\
\hline
Road grade & Road grade has a direct relationship with fuel consumption and emissions~\citep{kamal2011ecological}. Routes with varying grades (flat routes and hilly routes) can lead to a fuel difference of up to 15-20\% in favor of flatter routes~\citep{boriboonsomsin2009impacts}. The relationship between fuel consumption and road grade is nonlinear, with higher grades (+5\% and above) resulting in increased fuel consumption due to power enrichment events. Conversely, lower grades (-4\% and below) lead to less fuel consumption as gravitational force assists the vehicle's force~\citep{boriboonsomsin2009impacts, liu2018modeling}. Emissions increase more on uphill drives, particularly with air conditioning on and varying vehicle loads~\citep{cicero1997effects}. \\ 
\hline
Road type & The type of road and its composition influence fuel economy. Road types differ in materials used for construction, altering friction levels and serving distinct purposes (e.g., arterial roads versus highways)~\citep{burgess2003parametric}. Notably, about 35\% of the fuel energy allotted to mechanical power to overcome friction is for overcoming rolling friction in the tire-road contact~\citep{holmberg2012global}.\\ 
\hline
Speed limit & For vehicles, there is usually a fuel-optimal speed point. While these points are dependent on vehicle characteristics and other confounding factors such as the slope of the road, and weather, studies suggest that 60-90 km/h could yield benefits~\citep{el2005comparative}. The optimal speed point for fuel efficiency for electric vehicles is lower than for internal combustion engine vehicles and falls between 40-60 km/h~\citep{fafoutellis2020eco}. Therefore, setting speed limits higher or lower than generally fuel-optimal speed points of vehicles could increase emissions and fuel consumption. \\ 
\hline
Number of lanes & The literature lacks a comprehensive understanding of the influence of the number of lanes on eco-driving benefits. Nevertheless, it is anticipated that an increase in lanes and dedicated turn lanes will alter lane-changing dynamics, thereby affecting eco-driving behaviors. \\ 
\hline
Turn lane configuration & There is a deficiency in the literature on precise understanding of the impact of turn lane configurations on eco-driving. However, it is expected that it will affect the overall benefits. The presence of turn lanes increases the frequency of lane changes, and it also leads to alterations in traffic signal timing plans, contributing to changes in eco-driving behaviors. \\ 
\hline
Lane length & Changes in the lane lengths can have an impact on the benefits as they affect the capacity of the road and the possibility of platooning~\citep{chowdhury2022estimation}, which in turn impacts how well the eco-driving can be performed at the fleet level. However, there is a gap in the literature on this front.  \\ 
\hline
Traffic signal timing & By looking ahead, idling time at intersections could be reduced by decelerating earlier and more smoothly (releasing the throttle and using the engine brake rather than the foot brake) and avoiding unnecessary accelerating and hard braking again, which saves fuel during both driving and idling~\citep{huang2018eco}. \cite{6832481} show a 61\% increase in fuel economy in a motivating case study with fixed-time traffic signals but only a 16\% increase across 1000 multisignal simulations and 6\% for actuated signals. Dynamic speed advice using real-time signal data shows a 12-14\% reduction in energy consumption and emissions~\citep{mandava2009arterial}. Traffic light-to-vehicle communication (TLVC) can yield up to 22\% benefits at a single intersection with a single vehicle, but this advantage reduces to 8\% in a road network scenario with multiple vehicles~\citep{tielert2010impact}. \\ 
\hline
Pedestrians & Pedestrians' impact on emissions and fuel consumption is understudied. If pedestrians follow road rules and avoid jaywalking, there is minimal influence. However, jaywalking pedestrians often disrupt eco-driving strategies, prompting vehicles to adjust their approaches.   \\ 
\hline
Cyclists & The impact of cyclists is understudied. However, the behaviors of cyclists can also influence eco-driving behaviors, specifically when the cyclist occupies the same lane as the vehicle, which entails the vehicle adjusting its behavior for low-speed cyclists in front. \\ 
\hline
\end{tabular}
\caption{Summary of the known impact of each factor on eco-driving emission benefits (2/4 tables).}
\label{tab:eco-driving-factors-table2}
\end{table*}
\clearpage

\begin{table*}[!ht]
\centering
\begin{tabular}{ |p{3cm}|p{12cm}| } 
\hline
\textbf{Factor} & \textbf{Description}  \\
\hline
Nearby intersections & When eco-driving at signalized intersections, the vehicle flow to one intersection is the outflow of another intersection. This changes the vehicle inflow process, which is often assumed as uniform~\citep{jayawardana2022learning}. To capture this effect, some studies model experimental environments beyond the control zones to include uncontrolled vehicle feeding zones~\citep{lichtle2021fuel}.  \\ 
\hline
Temperature & Temperature affects the optimal operating temperature of engines, thereby influencing vehicle performance. Research indicates that a decrease in temperature by 1 °C leads to a fuel consumption increase of 0.38 ± 0.079\%~\citep{beusen2009using}. Neglecting temperature's influence in eco-driving analyses can misrepresent benefits. For instance, eco-driving courses may not sustain their fuel-saving effects over time due to seasonal temperature variations~\citep{degraeuwe2013corrigendum}. Furthermore, considering temperature effects is crucial in assessing emission levels and policies for hybrid electric vehicles.~\cite{zahabi2014fuel} noted that hybrid electric vehicle fuel economy is more sensitive to temperature changes by highlighting that if the temperature goes from 8°C to 16°C, the fuel economy rate of hybrid electric vehicles reduces by 3\%, while for conventional vehicles, it's only 1\%.\\ 
\hline
Humidity & The study by \cite{choi2010moves} found that, as for the sensitivity of emissions to humidity, high temperatures (above 75°F) affect HC and CO emissions, and temperatures above 25°F affect NOx emissions. They also find that gasoline vehicles are more affected by humidity compared to diesel vehicles. \\ 
\hline
Weather conditions & Weather conditions, including rain, snow, and wind, have a substantial impact on eco-driving by affecting road conditions and vehicle operation. Wet roads increase fuel consumption due to higher rolling resistance caused by water or slush~\citep{mpg_guide}. Rain and snow change the characteristics of the road surface and affect the vehicle rolling resistance. The basis of this resistance is a phenomenon called “aquaplaning”. Maintaining the same speed on the wet road could increase fuel consumption by over 30\%~\citep{karlsson2012energy}. Snow and ice also lead to increased fuel consumption due to reduced traction and slower driving\citep{fueleconomy_website}. Headwinds and crosswinds worsen aerodynamic drag, reducing fuel efficiency by approximately 13\% for every 10mph increase in wind~\citep{mpg_guide}.\\ 
\hline
Seasonal effects & 
Seasonal effects refer to the collective impact of temperature, humidity, and weather conditions. Research indicates that during winter, hybrid electric vehicles experience a roughly 20\% reduction in fuel efficiency compared to the summer, while this effect is not prominent for standard gasoline sedans~\citep{zahabi2014fuel}. Disregarding these seasonal effects in eco-driving analyses can distort the perceived advantages. For instance, eco-driving programs may not maintain their fuel-saving benefits consistently due to fluctuations in seasonal temperatures~\citep{degraeuwe2013corrigendum}.\\ 
\hline
Vehicle age distribution & Older vehicles usually have wear and tear in the engine and other components, which could lead to increased emissions~\citep{beydoun2006vehicle, anilovich1996survey}. Newer vehicles are designed with emission reduction technologies such as enhanced eco-mode, which can reduce emissions and fuel consumption~\citep {lee2023effect}.  \\ 
\hline
Vehicle type distribution & The model of the vehicle affect the benefits, as discussed earlier. Furthermore, the distribution of these vehicle types (or, more specifically, models) in a city is essential in the city-level analysis.   \\ 
\hline
Driver mix distribution & Human driver behaviors vary widely due to factors like demographics, travel time, vehicle type, and travel purpose~\citep{treiber2013traffic}. \cite{ma2020energetic} illustrate that eco-driving in a diverse human-driven platoon can reduce fuel consumption by 11.93\%. \cite{li2023physics} further show that the energy
consumption of mixed traffic flows can be reduced by 4.83\% - 9.16\% on average under diverse human behaviors, indicating the benefits are still significant even when the diversity is high. \\  
\hline
Fuel type distribution & The choice of fuel for vehicles have varying effects on eco-driving. Diesel-powered engines offer better fuel efficiency and greater low-end torque compared to gasoline engines of similar size~\citep{fueleconomy_website}. Diesel engines produce fewer greenhouse gas emissions, but there are significant worries regarding the potential health impacts of their particulate matter emissions~\citep{diesel_gasoline}. \\ 
\hline
\end{tabular}
\caption{Summary of the known impact of each factor on eco-driving emission benefits (3/4 tables).}
\label{tab:eco-driving-factors-table3}
\end{table*}
\clearpage

\begin{table*}[!ht]
\centering
\begin{tabular}{ |p{3cm}|p{12cm}| } 
\hline
\textbf{Factor} & \textbf{Description}  \\
\hline
Semi-autonomous vehicle penetration & Penetration level significantly influence eco-driving benefits. At lower penetration levels, benefits are limited as only a few vehicles practice eco-driving. Conversely, higher penetration yields greater benefits. \cite{jayawardana2022learning} demonstrate that 100\% autonomous vehicle penetration can lead to an 18\% reduction in fuel consumption and a 25\% reduction in CO$_2$ emissions. Even at 25\% penetration, a significant portion (at least 50\%) of total fuel and emission reduction benefits can be achieved. \cite{garcia2017modeling} found that more eco-drivers reduce emissions in low or medium-demand scenarios, but in high-demand with many eco-drivers, emissions increase. Smoother traffic flow aimed at reducing congestion can unexpectedly raise overall emissions.\\
\hline
Idling & Idling leads to zero fuel efficiency (0 km/L)~\citep{sanguinetti2017many}. In the U.S., idling results in the waste of approximately 30 billion liters of fuel annually, half of which is attributed to personal vehicles\citep{idling_report}. Mitigating idling can substantially reduce emissions and fuel consumption by up to 20\%\citep{demir2011comparative}, akin to the impact of removing 5 million vehicles from the road in terms of fuel savings and emissions reduction\citep{idling_report}.\\ 
\hline
Speeding & Vehicles typically have a fuel-optimal speed point. Fuel consumption and emission rates per-unit distance are optimum in the range of 60–90 km/h~\citep{el2005comparative}. For electric vehicles, the fuel-efficient speed range is lower, typically between 40-60km/h~\citep{fafoutellis2020eco}. \cite{rakha2003impact} show that vehicle fuel consumption rate is, in fact, more sensitive to cruise-speed levels than to vehicle stops. Using cruise control
when possible is commonly recommended for eco-driving~\citep{sivak2012eco}. \\ 
\hline
Acceleration & Keeping smooth transitions between various vehicle operating modes without hard accelerations are usually advised. The intensity and frequency of accelerations lead to an increase in fuel consumption~\citep{wang2008modelling}. Shifting up gears quickly also has a considerable impact. Choosing the right shift point and driving the vehicle using the highest gear possible are effective methods to reduce fuel consumption~\citep{beusen2009using}. According to
US Department of Energy, aggressive driving can lower the fuel economy by 15–30\% in highways and 10–40\% in stop-and-go traffic~\citep{fueleconomy_website}.\\ 
\hline
Deceleration & Although deceleration could reduce fuel consumption, hard decelerations and alternating between deceleration and acceleration frequently could increase fuel consumption. Applying engine brake (without changing down through gears) for smooth deceleration and minimizing the use of foot brake where appropriate is recommended for eco-driving~\citep{birrell2014analysis}. \\ 
\hline
Pulse and glide & In pulse and glide, the vehicle first accelerates to a high speed and then freely reduces the speed to a lower speed (costing). Studies show that this strategy can bring fuel reduction benefits, specifically for electric vehicles and hybrid vehicles, due to their ability to regenerate energy during braking~\citep{eo2018development}. Also,~\cite{hall2019reduced} show that fuel consumption can be reduced by up to 8.1\% compared to a constant speed driving strategy with pulse and glide. \\ 
\hline
Lane changing & Lane changing, especially by non-eco-driving vehicles in partial penetration settings, can reduce the benefit levels as they may interfere with eco-driving behaviors. \cite{yang2016eco} show that fuel-saving can be negative under low penetration of eco-driving vehicles because of extensive lane changing by non-eco-driving vehicles.\\ 
\hline
\end{tabular}
\caption{Summary of the known impact of each factor on eco-driving emission benefits (4/4 tables).}
\label{tab:eco-driving-factors-table4}
\end{table*}

\begin{algorithm}
    \caption{AADT Imputation} 
    \label{graphsimp}
    \hspace*{\algorithmicindent} \textbf{Input:} \\ 
    \hspace*{\algorithmicindent}\hspace*{\algorithmicindent} \texttt{G(V, E)} denotes the directed graph representing the road network where E is the set of road segments and V is the set of connections between road segments.  \\
    \hspace*{\algorithmicindent}\hspace*{\algorithmicindent}\texttt{U} denotes the set of edges where AADT data is available. The existing flow data is $f(e)$ for $e\in U$. \\
    \hspace*{\algorithmicindent}\hspace*{\algorithmicindent}\texttt{$S_1$} denotes the set of source nodes in the graph. \\
    \hspace*{\algorithmicindent}\hspace*{\algorithmicindent}\texttt{$S_2$} denotes the set of sink nodes in the graph. \\
    \hspace*{\algorithmicindent}\hspace*{\algorithmicindent}\texttt{Des} returns the child nodes of a given node. \\
    \hspace*{\algorithmicindent}\hspace*{\algorithmicindent}\texttt{Pre} returns the parent nodes of a given node.\\
    \hspace*{\algorithmicindent} \textbf{Output} \\
    \hspace*{\algorithmicindent}\hspace*{\algorithmicindent} $\hat{\mathcal{X}}$ Optimal flow assignment
    \begin{algorithmic}[1]
        \State $\hat{\mathcal{X}} = \arg\min_{\mathcal{X}} \sum_{e\in U} (\mathcal{X}(e) - f(e))^2$
        \State s.t. $\sum_{e'\in \texttt{Des}(e)} \mathcal{X}(e') - \sum_{e' \in \texttt{Pre}(e)}\mathcal{X}(e') = 0 \hfill \forall e \in E\setminus \{S_1 \cup S_2\}$
        \State \textbf{return} $\hat{\mathcal{X}}$
    \end{algorithmic} 
\end{algorithm}

To differentiate the inflows between peak and off-peak hours, we use recommended conversion rates from AADT to peak and off-peak traffic flow rates~\citep{precisiontraffic_2014}. In particular, we assume 5.5\% of AADT is due to an off-peak hour while 8.4\% of AADT is due to a peak hour. We also assume these conversions are applicable across all three cities.

Furthermore, traffic flow at intersections can be linked, as the flow of vehicles at one intersection can impact neighboring ones, specifically at peak hours. To address this within our modeling, we introduce the concept of \textit{ghost intersections} by modeling neighboring one-hop intersections without active control of vehicles for eco-driving while referring to the intersection where the eco-driving controllers are active as \textit{control active intersection}.

Modeling different traffic signal timing plans poses a combinatorial number of possibilities (with ghost intersections). Thus, we exhaustively search through the fixed-time traffic signal control plans for the control active intersections after fixing the ghost intersection plans to the default SUMO program. In practice, traffic signal control plans are seldom optimal, and thus, our results can be seen as a lower bound on the potential for eco-driving. 

In modeling pedestrians and cyclists, we assume their adherence to traffic regulations, which eliminates disruptions to vehicular traffic. We assume that pedestrians will use marked crosswalks rather than jaywalking, and cyclists will make use of designated bike lanes that do not overlap with vehicular traffic lanes.

\textbf{Emission models.} 
Accurate emission models are needed to quantify the emissions of vehicle trajectories produced by the micro-simulations. The Motor Vehicle Emission Simulator (MOVES) from the US Environmental Protection Agency is the industry-standard emission simulator, but it is prohibitively expensive to use in our case. MOVES takes roughly three minutes per scenario while we deal with a million scenarios. Thus, we create fast neural surrogate emission models to imitate MOVES, following and extending the work by \cite{sanchez2022learning,sanchez2022learningJMLR}. 

Our goal is to leverage MOVES as an emission data source and train a set of fast surrogate models via supervised learning to imitate MOVES. We follow the guidelines of ~\cite{sanchez2022learning,sanchez2022learningJMLR} and define a vehicle trajectory $\tau$ as a sequence of velocities $v_t$ at each time step $t$. We define a baseline trajectory $\tau_b$ with a constant speed over $n$ time steps and a controlled trajectory $\tau_c$ of length $n+1$, identical to $\tau_b$ except for an additional time step with an extra velocity term. Importantly, $\tau_b[0:n] = \tau_c[0:n]$ holds. We use the tuple $<\tau_b, \tau_c>$ to represent both trajectories.

We use MOVES to calculate CO$_2$ emissions for each trajectory tuple, factoring in vehicle type, vehicle age, road grade, temperature, humidity, and vehicle fuel type. By subtracting baseline trajectory emissions from controlled trajectory emissions, instantaneous emissions which depend on velocity and acceleration introduced in the extra time step of the controlled trajectory, can be obtained. Repeated iterations yield a dataset linking emissions to velocity, acceleration, temperature, humidity, and road grade under set conditions.

We use this dataset to train surrogate emission models categorized by vehicle type (passenger cars, trucks, buses), fuel types (gasoline, diesel), and vehicle age (1 to 10 years in increments of 1 year and 10+ years as a representative age category of all older vehicles). \cite{sanchez2022learning,sanchez2022learningJMLR} use decision trees as surrogate models. For better accuracy, our surrogate emission models are implemented as neural networks featuring a single hidden layer with 32 neurons. 

Occasionally, these learned models may yield negative emission predictions due to fitting errors. Therefore, we adjust the final emission value as $e^* = \max(0, n_e)$ where $n_e$ is the predicted emission. Additionally, we find that caching emission values in memory speeds up the evaluation process. In total, our suite comprises 88 instantaneous emission models, each defined by specific vehicle type, fuel type, and age, with inputs of instantaneous velocity, acceleration, road grade, temperature, and humidity.

We further use surrogate emission models to simplify our scenario modeling efforts by capturing the cumulative effects of factors such as vehicle maintenance, vehicle model, vehicle characteristics, road types, aerodynamics drag, air conditioning, and vehicle load without modeling specific individual effects of them. In other words, we rely on the accuracy of MOVES data to represent the collective average effect of these factors. This simplification is reasonable since these factors are often beyond our control, and optimizing for specific conditions has limited practical significance.

\textbf{Human driver mix distributions.} For the foreseeable future, human-driven vehicles will remain prevalent. To model human drivers in our simulations, we use the Intelligent Driver Model (IDM)~\citep{Treiber2000CongestedTS}. IDM is a widely accepted car-following model that can produce realistic traffic waves. IDM calculates a vehicle's acceleration using Equation~\ref{IDM}, with desired velocity $v_0$, space headway $s_0$, time headway $T$, maximum acceleration $\alpha$, and comfortable braking deceleration $\beta$. The velocity difference with the leading vehicle is denoted as $\Delta v(t)$, and $\delta$ is a constant.

\begin{equation}
a(t)=\alpha \left[1-\left(\frac{v(t)}{v_{0}}\right)^{\delta}-\left(\frac{s^{*}(v(t), \Delta v(t)}{s(t)}\right)^{2}\right]
\label{IDM}
\end{equation}

\begin{equation}
s^{*}(v(t), \Delta v(t))=s_{0}+\max \left(0, v(t) T+\frac{v(t) \Delta v(t)}{2 \sqrt{\alpha \beta}}\right)
\end{equation}

For simulation accuracy, precise calibration of parameters $v_0$, $s_0$, $T$, $\alpha$, and $\beta$ is crucial. Different regions may exhibit varying driving behaviors, such as American drivers versus British drivers. Thus, calibrating the IDM parameters is critical for accurate human driver modeling.

Our IDM model calibration goals are threefold. We aim to align it with real-world human driving behavior at US signalized intersections, tailoring the five IDM parameters for human-like trajectories. We also need separate IDM models for distinct vehicle types (e.g., cars, buses, trucks) due to their influence on driving style. Lastly, we aim to establish a range of IDM models reflecting diverse driver behaviors and aggressiveness as observed in real-world driving.

For this purpose, we employ Bayesian inference and a Gaussian process-based approach proposed by ~\cite{zhang2022bayesian} and leverage real-world arterial driving data from CitySim~\citep{zheng2022citysim}. For $\forall(t, d) \in\{(t, d)\}_{t=t_0, d=1}^{t_0+(T-1) \Delta t, D}$ where $d$ represents the index for each driver $d \in \{1,\cdots,D\}$ and $t$ represents the timestamp, we have,

\begin{subequations}
\label{idm-bayesian}
\begin{gather}
\ln ({\theta}) \sim \mathcal{N}\left({\mu}_0, {\Sigma}_0\right) \in \mathbb{R}^5 \\
\ln \left(\sigma_\epsilon\right) \sim \mathcal{N}\left(\mu_\epsilon, \sigma_1\right) \in \mathbb{R} \\
v_d^{(t+\Delta t)} \mid {i}_d^{(t)}, {\theta} \stackrel{\text{i.i.d.}}{\sim} \mathcal{N}\left(\mathcal{F}_{\mathrm{IDM}}\left({i}_d^{(t)} ; {\theta}\right),\left(\sigma_\epsilon \Delta t\right)^2\right) \in \mathbb{R}
\end{gather}
\end{subequations}

Here, ${\theta}=\left[v_0, s_0, T, \alpha, \beta\right] \in \mathbb{R}^5$ is the IDM model parameters where $\mathcal{N}(\mu ,\sigma)$ represents a Gaussian distribution with a mean of $\mu$ and standard deviation of $\sigma$. Independent and identically distributed is indicated by $i.i.d$.
Furthemore, we denote the inputs at time $t$ as a vector $i_d^{(t)} = [s^{(t)}_d, v^{(t)}_d, \Delta v^{(t)}_d], \forall t \in \{t_0, \cdots, t_0 + (T-1)\Delta t \}$ where $s^{(t)}_d$ is the headway, $v^{(t)}_d$ is the velocity and $\Delta v^{(t)}_d$ is the relative velocity of driver $d$ and its leading vehicle at time $t$. We further define a function $\mathcal{F}_{IDM}(\cdot)$ that updates the ego vehicle’s speed at $t + \Delta t$ using Equation~\ref{IDM} and Equation~\ref{speed_update} where $\Delta t$ is the step size. 

\begin{equation}
    v^{(t + \Delta t)} = v^{(t)} + a^{(t)} \Delta t
    \label{speed_update}
\end{equation}

The formulation in Equations~\ref{idm-bayesian} generates a population-level joint distribution of IDM parameters. As vehicle type influences driving behavior, we build distinct joint distributions for each vehicle type. This requires separate human-driving trajectory datasets for every vehicle type.

CitySim~\citep{zheng2022citysim} is a drone-based vehicle trajectory dataset that provides vehicle trajectories along arterial roads, but it lacks explicit vehicle type labels. Therefore, we use the bounding box length of vehicles to distinguish cars from buses and trucks. While this identifies cars, it does not allow us to differentiate between trucks and buses. Thus, we assume a common IDM parameter distribution for trucks and buses. This is reasonable as both are heavy-weight vehicles with similar behavior at signalized intersections. After modeling these joint distributions, we utilize them to sample human drivers for the micro-simulations.

\textbf{Other factor.}
We source vehicle age, fuel type, and vehicle type distributions from the MOVES databases~\citep{epa_moves}. These distributions represent the national distributions. Once sourced, we assign each simulation vehicle with an age, fuel type, and vehicle type drawn from these distributions, which will subsequently used in emission estimates through surrogate emission models.

We use the US National Centers for Environmental Information ~\citep{NCEI} as our data source for temperature and humidity values. We also make the informed decision to model seasons and weather conditions as changes in temperature and humidity. Therefore, we do not explicitly model the impact of weather events like rain or snow on driving behavior but capture the effect of these events on vehicle engine efficiency. Driving behavior changes due to weather events may not be prominent at low-speed limits typically observed at intersections. However, we find that temperature and humidity changes during weather events have significantly impacted vehicle engine efficiency. Thus, we adopt this informed modeling strategy. 

\subsection{Intersection count distribution}
\label{int-count}
\begin{figure}
\includegraphics[width=1.0\textwidth]{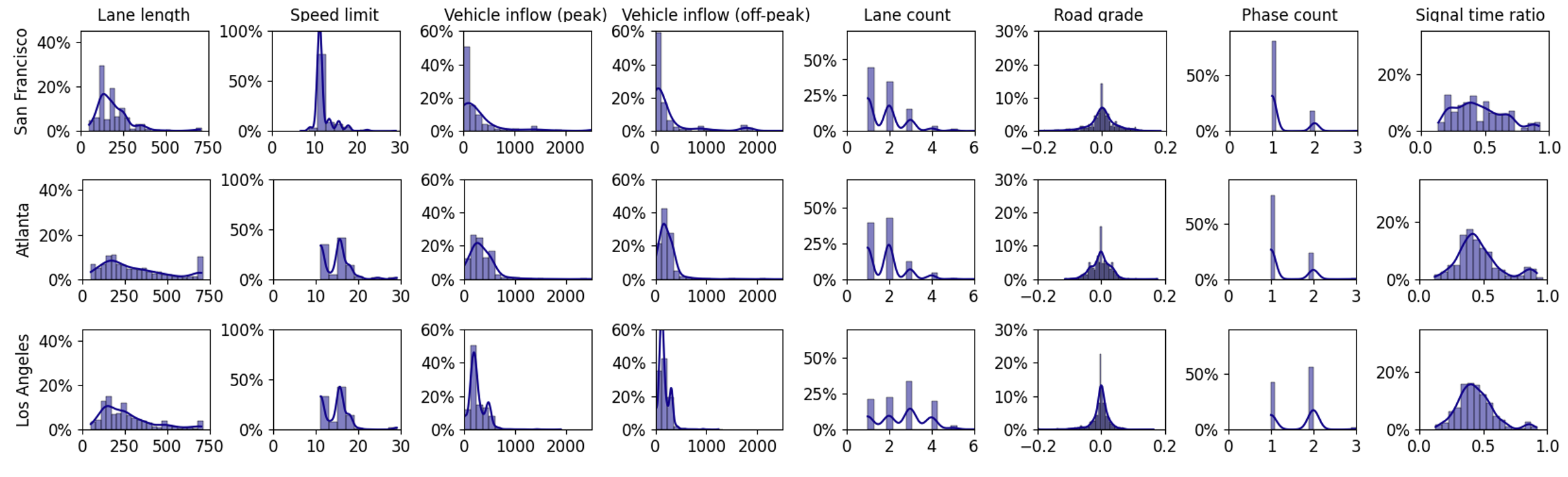}
\caption{\textbf{City feature distributions.} Comparison of feature distributions of the incoming approaches of the three cities. The y-axis is the percentage, while each x-axis is the respective feature range. Lane length is measured in meters, speed limit is in meters per second, vehicle inflows are vehicles per hour, and road grade is in percent grade. Phase count denotes how many different traffic signal phases are applicable for a given incoming approach. The signal time ratio denotes the ratio between approach-related phase times and total cycle time.}
\label{fig:city_feature_distributions}
\end{figure}

In Table~\ref{table:int_counts}, we list the number of intersections considered for scenario modeling in each city. In Figure~\ref{fig:city_feature_distributions}, we plot the feature distributions of each city.

\begin{table}[h]
\centering
\begin{tabular}{ |c|c|} 
\hline
City & Intersection Count \\
\hline
\hline
Atlanta & 621 \\
\hline
Los Angeles & 4276 \\
\hline
San Francisco & 1213 \\
\hline
\hline
Total & 6110 \\
\hline
\end{tabular}
\caption{Intersection count distribution within the three cities.}
\label{table:int_counts}
\end{table}

\subsection{Scenario variations} 
\label{sceanrio_count}
Based on our scenario modeling efforts, we comprehensively analyze one million traffic scenarios. As a high-level breakdown, we first look at 0.2 million scenarios by considering different combinations of guided semi-autonomous vehicle penetration levels (10\%, 20\%, 50\%, and 100\%), four seasons (summer, fall, spring, and winter), and driving hours (peak and off-peak) in the 6,011 signalized intersections in the selected three cities. Subsequently, we look at another 0.4 million scenarios by varying guided semi-autonomous vehicle penetration levels, weather conditions (rain, clear), and driving hours for all four seasons. Furthermore, we analyze another 0.4 million scenarios by considering different combinations of guided semi-autonomous vehicle penetration levels, four seasons, driving hours, and electric and hybrid vehicles in the 6,011 signalized intersections in the selected three cities.

\subsection{MT-DRL training environments}
\label{appendix_training_environments}

We define a synthetic environment distribution for training MT-DRL policies. Each training environment is defined based on uniform distributions of a selected set of factors as given in Table~\ref{table:training-environments}. In total, we utilize a total of 3,840 synthetic training environments per partition.

\begin{table}[h]
\centering
\begin{tabular}{ |c|c|c|c| } 
\hline
Factor & Min & Max & samples  \\
\hline
\hline
inflow & 200veh/h & 400 veh/h & 4  \\ 
\hline
lane length & 50m & 750m & 4  \\ 
\hline
speed limit & 6.5m/s & 30m/s & 4  \\ 
\hline
green phase duration & 15s & 35s & 3  \\ 
\hline
red phase duration & 30s & 55s & 4  \\ 
\hline
traffic signal offset & 0 & 1 & 5  \\ 

\hline
\end{tabular}
\caption{Factors and factor valuations defining synthetic training environment distribution.}
\label{table:training-environments}
\end{table}

\subsection{Hyperparameters}
\label{hyperparameters}

All hyperparameter configurations used for training the MT-DRL policies are given in Table~\ref{tab:params}. 

\begin{table*}[h!]
    \centering
    \begin{tabular}{|c|c||c|c|}
        \hline
        parameter & value & parameter & value \\
        \hline
        \hline
        PPO policy clip & $0.03$ & PPO initial KL divergence coef & $0.1$ \\
        PPO value clip & $3$ & PPO KL target & $0.02$ \\
        PPO value loss coefficient & $1$ & PPO entropy coefficient & $0.005$ \\
        gradient update steps & $10$ & epochs & $5000$ \\
        number of episodes per epoch & 10 & batch size & 40\\
        episode horizon & $1500$ & learning rate & $0.0001$ \\
        optimizer & $Adam$ & discounting factor & $0.99$ \\
        Adam weight decay & $0.97$ & Adam betas & $(0.9, 0.999)$ \\
        neural network layers & $4$ & activation function & $tanh$ \\
        weight initialization & $orthogonal$ & simulation warmup steps & 50\\
        width of a neural network layer & $256$ & simulation step size & $0.5s$ \\
        \hline
    \end{tabular}
    \caption{MT-DRL training-related hyperparameter configuration}
    \label{tab:params}
\end{table*}

\subsection{Overall effectiveness - peak vs off-peak hours}
\label{peak-off-peak-analysis}

In Table~\ref{tab:peak-off-peak-effectiveness}, we present the emission benefit percentages in peak and off-peak hours. As can be seen, peak hours at most adoption levels yield slightly better emission benefits except in 100\% adoption, in which they are closely related. 

\begin{table}[h!]
\centering
\setlength{\tabcolsep}{4pt} % adjust column separation
\renewcommand{\arraystretch}{1.2} % adjust row separation
\small % reduce font size
\begin{tabular}{|l|cc|cc|cc|cc|}
\hline
\textbf{City} & \multicolumn{2}{c|}{\textbf{10}} & \multicolumn{2}{c|}{\textbf{20}} & \multicolumn{2}{c|}{\textbf{50}} & \multicolumn{2}{c|}{\textbf{100}} \\
 & \textbf{P (\%)} & \textbf{OP (\%)} & \textbf{P (\%)} & \textbf{OP (\%)} & \textbf{P (\%)} & \textbf{OP (\%)}& \textbf{P (\%)} & \textbf{OP (\%)} \\
\hline
Atlanta & 5.5540 & 4.6228 & 7.9499 & 6.5693 & 14.5445 & 12.7584 & 21.2136 & 21.2587 \\
Los Angeles & 4.6385 & 3.6571 & 6.0912 & 5.1146 & 10.3615 & 9.1247 & 15.0320 & 15.0428 \\
San Francisco & 6.3749 & 6.3833 & 7.2342 & 7.1573 & 9.5765 & 8.8940 & 10.8552 & 11.2618 \\
\hline
\end{tabular}
\caption{Emission saving percentages in peak and off-peak hours in each city. \textbf{P} stands for peak hours and \textbf{OP} stand for off-peak hours.}
\label{tab:peak-off-peak-effectiveness}
\end{table}

\subsection{Correlation analysis of factors}
\label{stat-test-factor-correlation}

In Table~\ref{alt-cor-sig}, \ref{sf-cor-sig}, \ref{la-cor-sig}, we present the correlation values and their statistical significance for Atlanta, San Francisco, and Los Angeles, respectively. We formulate the null hypothesis and alternative hypothesis as follows. 

\textbf{Null Hypothesis}: The population correlation coefficient is not significantly different from zero.

\textbf{Alternate Hypothesis}: The population correlation coefficient is significantly different from zero.

We conduct a p-value-based analysis using a significance level of $\alpha = 0.05$ and apply a two-tailed test for all correlations. 

\begin{table*}
\centering
\begin{tabular}{|l|ll|ll|ll|ll|}
\hline
Feature & \multicolumn{2}{c|}{10\%} & \multicolumn{2}{c|}{20\%} & \multicolumn{2}{c|}{50\%} & \multicolumn{2}{c|}{100\%} \\
        & Cor. & Sig. & Cor. & Sig. & Cor. & Sig. & Cor. & Sig. \\
\hline
\hline
      Temperature &    -0.00 &          No &    -0.00 &          No &    -0.00 &          No &      0.00 &           No \\
      Speed limit &     0.06 &         Yes &     0.14 &         Yes &     0.23 &         Yes &      0.28 &          Yes \\
            Road grade &     0.02 &         Yes &     0.03 &         Yes &     0.04 &         Yes &      0.02 &          Yes \\
Signal ratio &    -0.06 &         Yes &    -0.03 &         Yes &    -0.04 &         Yes &      0.04 &          Yes \\
Vehicle inflow &    -0.02 &         Yes &    -0.01 &         Yes &    -0.05 &         Yes &     -0.12 &          Yes \\
      Phase count &    -0.03 &         Yes &    -0.00 &          No &    -0.05 &         Yes &     -0.03 &          Yes \\
         Lane count &    -0.05 &         Yes &    -0.03 &         Yes &    -0.04 &         Yes &      0.03 &          Yes \\
      Lane Length &     0.09 &         Yes &     0.18 &         Yes &     0.33 &         Yes &      0.47 &          Yes \\
         Humidity &     0.01 &          No &     0.01 &          No &     0.01 &          No &     -0.00 &           No \\
\hline
\end{tabular}
\caption{Pearson correlation values and statistical significance of different factors on emission benefits in Atlanta. \textit{Cor.} denotes the correlation value and \textit{Sig.} denotes the statistical significance.}
\label{alt-cor-sig}
\end{table*}

\begin{table*}
\centering
\begin{tabular}{|l|ll|ll|ll|ll|}
\hline
Feature & \multicolumn{2}{c|}{10\%} & \multicolumn{2}{c|}{20\%} & \multicolumn{2}{c|}{50\%} & \multicolumn{2}{c|}{100\%} \\
        & Cor. & Sig. & Cor. & Sig. & Cor. & Sig. & Cor. & Sig. \\
\hline
\hline
      Temperature &     0.00 &          No &     0.00 &          No &     0.00 &          No &      0.00 &           No \\
      Speed limit &    -0.04 &         Yes &     0.01 &         Yes &     0.09 &         Yes &      0.16 &          Yes \\
            Road grade &     0.00 &          No &     0.01 &          No &     0.04 &         Yes &      0.05 &          Yes \\
Signal ratio &    -0.06 &         Yes &    -0.03 &         Yes &    -0.00 &          No &      0.04 &          Yes \\
Vehicle inflow &    -0.06 &         Yes &    -0.07 &         Yes &    -0.09 &         Yes &     -0.10 &          Yes \\
      Phase count &    -0.03 &         Yes &    -0.02 &         Yes &    -0.03 &         Yes &     -0.04 &          Yes \\
         Lane count &    -0.07 &         Yes &    -0.05 &         Yes &    -0.05 &         Yes &     -0.00 &           No \\
      Lane length &     0.01 &          No &     0.05 &         Yes &     0.16 &         Yes &      0.28 &          Yes \\
         Humidity &    -0.00 &          No &    -0.00 &          No &     0.00 &          No &     -0.00 &           No \\
\hline
\end{tabular}
\caption{Pearson correlation values and statistical significance of different factors on emission benefits in San Francisco. \textit{Cor.} denotes the correlation value and \textit{Sig.} denotes the statistical significance.}
\label{sf-cor-sig}
\end{table*}

\begin{table*}
\centering
\begin{tabular}{|l|ll|ll|ll|ll|}
\hline
Feature & \multicolumn{2}{c|}{10\%} & \multicolumn{2}{c|}{20\%} & \multicolumn{2}{c|}{50\%} & \multicolumn{2}{c|}{100\%} \\
        & Cor. & Sig. & Cor. & Sig. & Cor. & Sig. & Cor. & Sig. \\
\hline
\hline
      Temperature &    -0.00 &         Yes &    -0.00 &          No &    -0.00 &          No &      0.00 &           No \\
      Speed limit &     0.04 &         Yes &     0.14 &         Yes &     0.23 &         Yes &      0.30 &          Yes \\
            Road grade &     0.00 &          No &     0.01 &         Yes &     0.01 &         Yes &      0.02 &          Yes \\
Signal ratio &    -0.06 &         Yes &    -0.04 &         Yes &    -0.03 &         Yes &      0.01 &          Yes \\
Vehicle inflow &    -0.10 &         Yes &    -0.13 &         Yes &    -0.14 &         Yes &     -0.17 &          Yes \\
      Phase count &    -0.02 &         Yes &    -0.02 &         Yes &    -0.07 &         Yes &     -0.08 &          Yes \\
         Lane count &    -0.03 &         Yes &     0.01 &         Yes &    -0.03 &         Yes &      0.01 &          Yes \\
      Lane length &     0.05 &         Yes &     0.14 &         Yes &     0.29 &         Yes &      0.46 &          Yes \\
         Humidity &     0.00 &          No &     0.00 &          No &     0.00 &          No &     -0.00 &           No \\
\hline
\end{tabular}
\caption{Pearson correlation values and statistical significance of different factors on emission benefits in Los Angeles. \textit{Cor.} denotes the correlation value and \textit{Sig.} denotes the statistical significance.  }
\label{la-cor-sig}
\end{table*}

\subsection{Safety assessment}
\label{safety_assessment}

To assess the safety implications, we employ surrogate safety measures derived from the Surrogate Safety Assessment Model (SSAM) developed by the US Federal Highway Administration~\citep{FHWA-SSAM}. SSAM is a software tool designed to automatically detect, categorize, and evaluate events of traffic conflicts in vehicle trajectory data. In this study, we collect vehicle trajectories by simulating control policies at 100 randomly chosen intersections across the three cities. Subsequently, we utilize SSAM to conduct a safety assessment based on the trajectories. 

As the safety surrogate measures, we utilize the following five surrogate measures. 

\begin{itemize}
    \item \textbf{TTC}: The minimum time-to-collision value observed during the conflict. In general, the higher the TTC, the safer it is. 
    \item \textbf{PET}: The minimum post-encroachment time observed during the conflict. Post-encroachment time refers to the interval starting from the moment the first vehicle vacated a position until the subsequent arrival of the second vehicle at the same position. In general, the higher the PET, the safer it is. 
    \item \textbf{MaxS}: The maximum speed of either vehicle throughout the conflict. A higher MaxS value indicates a more severe conflict, while a lower value indicates a less critical or potentially safer situation.
    \item \textbf{DeltaS}: Quantifies the variation in speeds between the two vehicles. If $v_1$ and $v_2$ denote the velocity vectors of the first and second vehicles, respectively, DeltaS is defined as DeltaS = || $v_1$ - $v_2$ ||. A higher value indicates a more severe conflict, while a lower value indicates a less critical or potentially safer situation. 
    \item \textbf{DR}: The initial deceleration rate of the second vehicle. A higher value indicates a more severe conflict, while a lower value indicates a less critical or potentially safer situation. 
\end{itemize}

\subsection{Eco-driving and alternatives}
\label{alternatives_and_eco_driving}

We contextualize eco-driving in the presence of other decarbonizing strategies. Mainly, we look at two strategies: vehicle electrification and hybrid vehicle adoption. Below, we discuss how we conduct these analyses. 

\textbf{Vehicle electrification and adoption of hybrid vehicles:} To estimate the emissions impact of vehicle electrification and the adoption of hybrid vehicles for comparison with eco-driving adoption, we utilize alternative vehicle adoption projections from the US Energy Information Administration (EIA) 2023 Energy Outlook's reference case~\citep{eia2023outlook}. This projects electric vehicle (EV) and hybrid adoption into 2050, accounting for existing policies, consumer preferences, macroeconomic factors, technological constraints, the current legal environment, and more. 
The projections are in terms of vehicle miles traveled (VMT) by vehicle type. Electric vehicles include 100-, 200-, and 300-mile EVs; hybrids include 20- and 50-mile plug-in gasoline hybrids and electric-gasoline hybrids; internal combustion engine (ICE) vehicles include gasoline, diesel, and ethanol-flex fuel vehicles. 

Carbon dioxide emissions for each vehicle are sourced from~\citep{miotti2016personal}. This accounts for both fuel production and use, including the emissions produced via current electricity production in the state where a city is located. To reflect the possibility of technology improvements over our period of interest, we select values from the least-emitting vehicle in each category: the 2021 automatic-transmission Mitsubishi Mirage (ICE), the 2021 Toyota Prius Prime (Hybrid), and the 2020 electric Hyundai Ioniq (Electric).

Emission reductions are expressed as a fractional reduction for the pooled period from 2024-2050, relative to a hypothetical scenario in which all projected VMTs are from ICE vehicles. Eco-driving adoption is assumed to be uniform across vehicle types, and results are reported in terms of cumulative average adoption rates over the projection period. For example, a linear adoption from 0\% to 100\% over the duration would correspond to the results shown at the 50\% cumulative eco-driving adoption rate level. Nonlinear eco-driving adoption can be similarly mapped to a cumulative average rate and found on the plots. The 100\% level therefore represents a hypothetical upper bound in which all vehicles immediately adopt eco-driving at the period's beginning and continue doing so through 2050. The full range of eco-driving adoption is shown to allow the reader to consider how a variety of adoption schemes might affect emissions improvements. 

The results indicate that eco-driving yields additional benefits relative to a future with hybrid vehicle adoption or electrification alone. In many cases, the projected emissions cuts from eco-driving are comparable to or greater than the projected benefits from hybrid vehicle adoption and electrification for the period analyzed, even at relatively low eco-driving adoption rates.

\subsection{Additional assumptions}
\label{assumptions_limitations}

Here, we list the additional assumptions we make that are not mentioned in the relevant sections of the work, as well as assumptions that warrant further details. 

We assume recommended conversion rates from AADT to peak and off-peak traffic flow rates~\citep{precisiontraffic_2014}. Specifically, we use 5.5\% of AADT represents an off-peak hour, and 8.4\% represents a peak hour flow. These conversions are assumed to be consistent across all three cities, disregarding factors like weather and driver behavior differences. Additionally, we assume peak hours are from 6 am to 10 am and 4 pm to 8 pm. Off-peak hours are from 10 am to 4 pm and 8 pm to 1 am. These time ranges were used to calibrate peak and off-peak hour ratios and scale the overall emission benefits. 

To calibrate traffic signal timings, we use a default rule-based timing strategy provided by the SUMO simulator for all ghost intersections. Subsequently, we fine-tune the signal timing at the control active intersection through an exhaustive search within a predefined timing range of 15-45 seconds, with increments of 5 seconds.

We calibrate city-level parameters based on US-level details, including vehicle type distribution, fuel type distribution, and vehicle age distribution. Additionally, we assume that distributions such as vehicle type distribution are not influenced by seasonal effects. 

When modeling weather effects, we assume the weather effects have minimum impact on the driving behavior specifically at signalize intersections where driving speeds are relatively low. We however model impact of temperature and humidity changes when weather changes. While, in reality, severe weather conditions like heavy snow can cause driving behavior changes, other weather changes do not make much difference in driving at the intersection level. 

When designing the observation space of each POMDP, we assume that the semi-autonomous vehicles are equipped with onboard sensors or vehicle-to-vehicle communication (V2V) to gather information about nearby vehicles (immediate leading and following vehicles in the ego-lane and adjacent lanes to the left and right of the vehicle). We also assume that the vehicles can receive Signal Phase and Timing (SPaT) information through infrastructure-to-vehicle communication (I2V) or cloud-based communication up to 750m from the intersection. 

When modeling ghost intersections, we use synthetic ghost intersections instead of modeling ghost intersections as real-world 1-hop neighboring intersections. Since the role of ghost intersections is to feed the vehicles to control active intersections in a realistic fashion, we believe this modeling simplification is reasonable. 

In modeling intersections, we do not model unprotected left turns. Rather, we assume every turn is protected. 

In modeling trucks, we model light-duty trucks and not heavy-duty trucks, as within city regions, the light-duty trucks are more prominent. 

\putbib[main] 
\end{bibunit}

\end{document}